\documentclass[aps,prd,twocolumn,superscriptaddress,nofootinbib]{revtex4-2}

\usepackage{amsmath}
\usepackage{amssymb}
\usepackage{graphicx}
\usepackage{float}
\usepackage{multirow}
\usepackage[pdftex,breaklinks,colorlinks,
linkcolor=Blue,
citecolor=teal,
anchorcolor=red,
urlcolor=cyan,
pdfencoding=auto]{hyperref}
\usepackage[dvipsnames]{xcolor}
\usepackage{array}
\usepackage{dcolumn}
\usepackage{mathtools}
\usepackage{orcidlink}
\usepackage[normalem]{ulem}
\usepackage{mathrsfs}
\usepackage[scr=boondoxo]{mathalpha}
\usepackage{bm}

\usepackage{accents}

\newcommand{\e}{\varepsilon}
\newcommand{\ee}{\epsilon}
\newcommand{\E}{\sigma}
\newcommand{\trunctwo}{+\mathcal{O}(\varepsilon^2)}
\newcommand{\truncthree}{+\mathcal{O}(\varepsilon^3)}

\usepackage{cancel}

\begin{document}

\title{Post-adiabatic waveform-generation framework for asymmetric precessing binaries}

\author{Josh Mathews\,\orcidlink{0000-0002-5477-8470}}
\affiliation{Department of Physics, National University of Singapore, 21 Lower Kent Ridge Rd, Singapore 119077}

\author{Adam Pound\,\orcidlink{0000-0001-9446-0638}}
\affiliation{School of Mathematical Sciences and STAG Research Centre, University of Southampton, Southampton, United Kingdom, SO17 1BJ}

\begin{abstract}
Recent years have seen rapid progress in calculations of gravitational waveforms from asymmetric compact binaries containing spinning secondaries. Here we outline a complete waveform-generation scheme, through first post-adiabatic order (1PA) in gravitational self-force theory, for generic secondary spin and generic (eccentric, precessing) orbital configurations around a generic Kerr primary. We emphasize the utility of a Fermi-Walker frame in parametrizing the secondary spin, and we analyse precession and nutation effects in the spin-orbit dynamics. We also explain convenient gauge choices within the waveform-generation scheme, and the gauge invariance of the resulting waveform. Finally, we highlight that, thanks to recent results due to Grant and Witzany et al., all relevant spin effects at 1PA order can now be computed without evaluating local self-forces or torques.  
\end{abstract}

\maketitle

\tableofcontents

\section{Introduction}
\label{sec:introduction}

Gravitational self-force (SF) theory originally grew out of the need to model extreme-mass-ratio inspirals (EMRIs) for the planned gravitational-wave detector LISA~\cite{Mino:1996nk}. After decades of progress~\cite{Barack:2018yvs}, it has now matured into a practical framework for building fast, accurate gravitational waveform models~\cite{Katz:2021yft, Isoyama:2021jjd,Wardell:2021fyy,Nasipak:2023kuf}. This development is exemplified by the FastEMRIWaveforms (FEW) software package, which can generate LISA-length EMRI waveforms on a timescale of $\sim 10$ms~\cite{Chua:2020stf, Katz:2021yft, michael_l_katz_2023_8190418, Speri:2023jte, BHPToolkit, Chapman-Bird:2025xtd}. Moreover, while it was traditionally motivated by EMRIs, with mass ratios as extreme as $\sim 1:10^7$~\cite{Babak:2017tow}, SF theory exhibits good agreement with fully nonlinear numerical relativity simulations even at mass ratios $\sim 1:10$~\cite{LeTiec:2011dp, LeTiec:2011bk, LeTiec:2013uey,vandeMeent:2020xgc,Warburton:2021kwk,Wardell:2021fyy,Albertini:2022rfe,Albalat:2022lfz,Ramos-Buades:2022lgf, NavarroAlbalat:2022tvh, Warburton:2024xnr}, and it is now recognized as one of the most viable methods of directly modelling intermediate-mass-ratio inspirals~\cite{LISAConsortiumWaveformWorkingGroup:2023arg} as well as an important tool for calibrating effective models that cover the entire binary parameter space~\cite{Damour:2009sm,Nagar:2022fep,vandeMeent:2023ols}.

In the context of compact-binary models, SF theory is based on an asymptotic expansion of the spacetime metric in powers of the binary mass ratio $\epsilon\equiv m_2/m_1$, where $m_1$ and $m_2$ are the masses of the heavier, primary and lighter, secondary object, respectively. The zeroth order in this expansion represents the Kerr spacetime of the primary black hole as if it were in isolation, with the secondary treated as a source of small perturbations on that background. 

This expansion for small mass ratios means that SF models are adapted to  asymmetric binaries in which the primary is significantly heavier than the secondary. Currently, other waveform models have limited accuracy for such systems~\cite{Hu:2022rjq,Jan:2023raq,Dhani:2024jja}, even in  the range of mass ratios that can be observed by present-day ground-based detectors (such as the system GW191219\_163120, with mass ratio $\approx 1:26$~\cite{KAGRA:2021vkt}). Self-force models are also naturally adapted to another challenging area of parameter space: highly precessing systems in which the two objects' spin vectors rapidly precess around the orbital angular momentum. Although they are perturbative in the mass ratio, SF models can account for the primary object's spin, and its precession around the orbital angular momentum, non-perturbatively because the expansion is performed around the fully nonlinear background Kerr spacetime of the primary.

Systems exhibiting both of these features---high mass asymmetry and spin precession---are currently subject to the largest modeling uncertainties~\cite{Dhani:2024jja,MacUilliam:2024oif}. Discrepancies between waveform models have led to substantial parameter biases even for mildly asymmetric precessing binaries~\cite{Hannam:2021pit}. For EMRIs specifically, spin precession is likely to be ubiquitous~\cite{LISA:2022yao}, which has motivated the EMRI modeling community's long-term goal of developing SF models for generic, precessing orbital configurations around Kerr black holes~\cite{LISA:2017pwj}. The need for improved models of less extreme precessing binaries provides additional motivation for this goal.

It is also widely expected that meeting the accuracy requirements of future gravitational-wave detectors \cite{LISA:2017pwj,Colpi:2024xhw,TianQin:2015yph,Gong:2021gvw, Luo:2019zal,Kawamura:2011zz, Punturo:2010zz, Reitze:2019iox,Li:2024rnk} will require carrying the SF expansion to second perturbative order in the mass ratio~\cite{LISAConsortiumWaveformWorkingGroup:2023arg,Burke:2023lno}, necessitating second-order calculations for generic configurations on a Kerr background. Moreover, at second order in SF theory, the spin of the secondary object enters~\cite{Pound:2012dk} and contributes significantly to the waveform phase~\cite{Mathews:2021rod}. Over the past several years, this importance of the secondary spin has spurred rapid progress toward SF waveform models for spinning secondaries on generic orbits around spinning primaries~\cite{Ruangsri:2015cvg,Harms:2015ixa,Harms:2016ctx,Lukes-Gerakopoulos:2017vkj,Warburton:2017sxk,Witzany:2018ahb,Witzany:2019nml,vandeMeent:2019cam,Akcay:2019bvk,Zelenka:2019nyp,Piovano:2020zin,Piovano:2021iwv,Compere:2021kjz,Mathews:2021rod,Skoupy:2021asz,Skoupy:2022adh,Drummond:2022efc,Drummond:2022xej,Drummond:2023loz,Drummond:2023wqc,Witzany:2023bmq,Skoupy:2023lih,Ramond:2022vhj,Ramond:2024ozy,Ramond:2024sfp,Skoupy:2024jsi,Grant:2024ivt,Piovano:2024yks, Witzany:2024ttz,Skoupy:2024uan, Skoupy:2025nie}. Our aim in this paper is to describe a complete waveform-generation framework for that generic scenario, focusing on the impact of the secondary's spin.

\subsection{Multiscale waveform generation with a nonspinning secondary}

Typically, in modern SF calculations and waveform models, the small-mass-ratio expansion is formulated as a multiscale expansion based on the quasi-periodic behavior of asymmetric binaries~\cite{Hinderer:2008dm,Miller:2020bft,Pound:2021qin}. This multiscale expansion enables rapid waveform generation by allowing one to pre-compute all the necessary, expensive inputs for the waveform in an offline step; the online waveform generation then comprises a cheap, fast evolution through pre-computed data~\cite{Katz:2021yft}.

If we neglect the secondary's spin, then generic inspiraling orbits around Kerr black holes are approximately tri-periodic, with three independent, slowly evolving frequencies $\Omega^i=(\Omega^r,\Omega^\theta,\Omega^\phi)$~\cite{Drasco:2005kz}. Each frequency has an associated phase $\mathring\psi^i=\int \Omega^i\, dt$ describing the phase of the secondary's radial, azimuthal, or polar motion. Calculations are performed in the background Kerr geometry with the primary's spin axis orthogonal to the equatorial plane $\theta=\pi/2$, such that spin-orbit precession manifests itself as precession of the secondary's orbital plane (meaning $\Omega^\theta\neq\Omega^r$). 

The multiscale expansion is adapted to this tri-periodicity, leading to waveforms with the form of slowly varying mode amplitudes multiplying tri-periodic phase factors~\cite{Pound:2021qin,Hughes:2021exa}:
\begin{multline}\label{multiscale waveform}
    h = \sum_{\bm{k}\in\mathbb{Z}^3}\Bigl[\e \mathring h^{(1)}_{\bm{k}}(\mathring\pi_i,\theta,\phi)
    +\e^2 \mathring h^{(2)}_{\bm{k}}(\mathring\pi_i,\delta m_1,\delta \chi_1,\theta,\phi)\\
    +{\cal O}(\e^3)\Bigr]e^{-ik_i\mathring\psi^i},
\end{multline}
where $(\theta,\phi)$ are angles on the sphere at future null infinity, $\bm{k}=(k_r,k_\theta,k_\phi)$ are integers, $\e\equiv1$ is a bookkeeping parameter used to count powers of the mass ratio $\ee$, and $\mathring\pi_i$ are three independent, slowly evolving orbital parameters (such as suitably defined eccentricity, semi-latus rectum, and orbital inclination). $\delta m_1$ and $\delta \chi_1$ are slowly evolving corrections to the mass and spin of the primary, describing the effect of absorption through its horizon~\cite{Miller:2020bft}. The waveform's time dependence is contained in its dependence on the binary's variables $\mathring\psi^i$, $\mathring\pi_i$,  $\delta m_1$, and $\delta \chi_1$, which are governed by simple ordinary differential equations~\cite{Pound:2021qin}:
\begin{align}
    \frac{d\mathring\psi^i}{dt} &= \Omega^i(\mathring\pi_k),\label{psidot}\\
    \frac{d\mathring\pi_i}{dt} &= \e\left[F^{(0)}_i(\mathring\pi_k) + \e F^{(1)}_i(\mathring\pi_k,\delta m_1,\delta\chi_1)+{\cal O}(\e^2)\right]\!,\label{Jdot}
\end{align}
and 
\begin{align}
\frac{d\delta m_1}{dt} &= \e\dot{\cal E}^{(1)}_{\cal H}(\mathring\pi_i)+{\cal O}(\e^2),\label{mdot}\\  
\frac{d\delta \chi_1}{dt} &= \e\dot{\cal L}^{(1)}_{\cal H}(\mathring\pi_i)+{\cal O}(\e^2),\label{Sdot}
\end{align}
where $\dot{\cal E}^{(1)}_{\cal H}$ and $\dot{\cal L}^{(1)}_{\cal H}$ are the leading-order fluxes of energy and angular momentum into the primary's horizon. We use rings over variables that cleanly separate slow evolution from rapid oscillations, avoiding oscillations in the mode amplitudes $\mathring h^{(n)}_{\bm{k}}$ and in the right-hand side of Eqs.~\eqref{psidot} and \eqref{Jdot}.

In practical models, all of the inputs, $\mathring h^{(n)}_{\bm{k}}$, $F^{(n)}_i$, $\dot{\cal E}^{(n)}_{\cal H}$, and $\dot{\cal L}^{(n)}_{\cal H}$, are pre-computed on a grid of $\mathring\pi_i$ values.\footnote{The background spin must also be included as an axis on this parameter-space grid, while the background mass only provides an overall scale. The corrections $\delta m_1$ and $\delta\chi_1$ do not need to be included as axes because $h^{(2)}_{\bm{k}}$ and $F^{(1)}_i$ are linear in these corrections, meaning one can pre-compute coefficients of $\delta m_1$ and $\delta\chi_1$  as functions of $\mathring\pi_i$ and of the background spin.} Waveforms are then rapidly generated by solving Eqs.~\eqref{psidot}--\eqref{Sdot} and summing the modes in Eq.~\eqref{multiscale waveform}. Equations~\eqref{psidot}--\eqref{Sdot} can be solved rapidly because all dependence on the phases $\mathring\psi^i$ has been eliminated through the choice of variables $(\mathring\psi^i,\mathring\pi_i)$, avoiding the need to resolve oscillations on the orbital time scales $\sim 2\pi/\Omega^i$~\cite{VanDeMeent:2018cgn}.

In Eqs.~\eqref{psidot}--\eqref{Sdot}, numeric labels in parentheses denote the post-adiabatic (PA) order at which the term enters the evolution of the orbital phases~\cite{Hinderer:2008dm,Pound:2021qin}, and  $F^{(n)}_{i}$ are referred to as $n$PA forcing functions.  
To understand this order counting, note   Eqs.~\eqref{psidot}--\eqref{Sdot} imply that the orbital phases, and hence the waveform phase~\cite{Warburton:2024xnr}, have an asymptotic expansion of the form
\begin{equation}\label{phases}
\mathring\psi^i = \frac{1}{\e}\left[\mathring\psi_{(0)}^i(\e t)+\e \mathring\psi_{(1)}^i(\e t)+{\cal O}(\e^2)\right]\!.
\end{equation}
Although this expansion has limited accuracy compared to a direct solution of Eqs.~\eqref{psidot}--\eqref{Sdot}~\cite{Wardell:2021fyy}, it provides a simple estimate of the impact of any given term in the dynamics: $n$PA forcing functions directly contribute to the $n$PA phase $\mathring\psi_{(n)}^i$. 

In a 1PA model, the residual phase error in Eq.~\eqref{phases} scales as ${\cal O}(\e)$, showing that 1PA models should achieve gravitational-wave phase errors much smaller than 1 radian for mass ratios much smaller than unity. 
Detailed analyses for binaries with a nonspinning secondary~\cite{Hinderer:2008dm,Pound:2021qin} show what is required to achieve this 1PA accuracy. The 0PA forcing function $F^{(0)}_i$ can be computed from the dissipative self-force exerted by the first-order (linear in $\ee$) metric perturbation $h^{(1)}_{\alpha\beta}$, while $F^{(1)}_i$ requires the complete (conservative and dissipative) first-order self-force as well as the dissipative self-force exerted by the \emph{second}-order (quadratic in $\ee$) metric perturbation $h^{(2)}_{\alpha\beta}$. 

Fully relativistic, generic 0PA waveforms were first generated in Ref.~\cite{Hughes:2021exa}, building on decades of progress~\cite{Hughes:2001jr, Glampedakis:2002ya,Mino:2003yg,Drasco:2005kz, Hughes:2005qb, Fujita:2009us, Fujita:2009bp, Sago:2015rpa, vandeMeent:2017bcc, Isoyama:2018sib, Fujita:2020zxe, Isoyama:2021jjd}. 
Efforts are now underway to extend those results to build a fast 0PA model with complete coverage of the parameter space. 

Some 1PA effects, particularly the effects of the conservative first-order self-force, have been thoroughly studied~\cite{Barack:2018yvs} and have been computed for a small sample of generic orbits~\cite{vandeMeent:2017bcc}. However, due to the challenge of computing $h^{(2)}_{\alpha\beta}$, until recently the only complete 1PA waveform model was restricted to the simplest case of quasicircular orbits around a nonspinning primary black hole~\cite{Wardell:2021fyy,Burke:2023lno}. We report extensions of that model in a series of companion papers~\cite{Mathews:2025txc,Honet:2025gge,Honet:2025lmk}.

\subsection{Addition of the secondary spin}

Recent work has gone a long way toward incorporating a spinning secondary into the type of framework laid out above. See, for example, Refs.~\cite{Skoupy:2023lih,Drummond:2023wqc,Piovano:2024yks} for nearly complete treatments. In this paper we consolidate and extend those studies, presenting a unified framework for multiscale waveform generation with a spinning secondary. Such a framework is greatly simplified  by the fact that at 1PA order, the  secondary spin $\chi_2$ only enters linearly; this is because, for a compact object, spin scales as mass squared. Concretely, we show the secondary spin only enters the 1PA waveform in the following ways:
\begin{enumerate}
    \item The second-order amplitudes, $h^{(2)}_{\bm{k}}$, in the waveform~\eqref{multiscale waveform} are modified by the addition of a linear-in-$\chi_2$ term $h^{(2\text{-}\chi_2)}_{\bm{k}}$, and this term is multiplied by an additional phase factor $e^{-iq\mathring{\psi}_s}$, where $\mathring{\psi}_s$ denotes a precession angle of the secondary spin and $q=0,\pm1$. However, this modification to $h^{(2)}_{\bm{k}}$ can probably be neglected in most circumstances because it only contributes  $\mathcal{O}(\e)$ to the waveform phase.
    \item The orbital frequencies $\Omega^i$ in Eq.~\eqref{psidot} are modified by a linear-in-$\chi_2$ correction $\Omega^i_{(1\text{-}\chi_2)}$. This represents the \emph{conservative} effect of the secondary spin.  
    \item The 1PA forcing functions $F^{(1)}_i$ are modified by the addition of a linear-in-$\chi_2$ correction $F_i^{(1\text{-}\chi_2)}$. This represents the \emph{dissipative} effect of the secondary spin.
\end{enumerate}
However, there is gauge freedom in the multiscale expansion, which can be used to move 1PA contributions between $F^{(1\text{-}\chi_2)}_i$ and $\Omega^i_{(1\text{-}\chi_2)}$. This freedom can be used to eliminate $\Omega^i_{(1\text{-}\chi_2)}$, for example, altering $F^{(1\text{-}\chi_2)}_i$ in the process. As part of our analysis, we characterize this freedom and the invariance of the 1PA waveform.

Conservative 1PA effects of the secondary spin have been well studied (e.g., in~\cite{Witzany:2019nml,Drummond:2022efc,Drummond:2022xej, Drummond:2023wqc,Piovano:2024yks,Skoupy:2024uan}), and the frequency corrections $\Omega^i_{(1\text{-}\chi_2)}$ were derived by Witzany~\cite{Witzany:2018ahb,Piovano:2024yks, Witzany:2024ttz}. Dissipative 1PA effects have also received much attention (e.g., in~\cite{Akcay:2019bvk,Piovano:2020zin, Skoupy:2021asz,Skoupy:2022adh,Skoupy:2023lih,Skoupy:2024jsi,Piovano:2024yks}) but are not yet complete.  Recent work has highlighted the possibility of obtaining all relevant dissipative effects by computing the rates of change of quantities that would be constant for a test body~\cite{Skoupy:2023lih}: spin-corrected versions of the energy, angular momentum, and Carter constant, along with a fourth quantity known as the Rüdiger constant~\cite{doi:10.1098/rspa.1981.0046,c839080c-5d3d-34d9-a247-38e3c109da14,Compere:2021kjz}. The spin-corrected energy and angular momentum satisfy balance laws, which allow one to compute  their evolution (at linear order in $\chi_2$) from asymptotic fluxes at infinity and the horizon~\cite{Akcay:2019bvk}.  

Grant~\cite{Grant:2024ivt} recently showed that the 1PA evolution of the spin-corrected Carter constant can also be obtained from asymptotic fluxes, extending the analogous, classic formulas~\cite{Sago:2005fn,Isoyama:2018sib} for the 0PA evolution of the Carter constant. Finally, Ref.~\cite{Skoupy:2024jsi} showed that the evolution of the Rüdiger constant is redundant at 1PA order. This is because it depends functionally on the other quasi-conserved quantities and a component of the secondary's spin, commonly labeled $\chi_\parallel$, which they have shown does not evolve at 1PA order.

Our analysis solidifies and extends Ref.~\cite{Skoupy:2024jsi}'s conclusion that the evolution of $\chi_\parallel$ can be neglected at 1PA order. We also show how to incorporate Grant's evolution equation for the spin-corrected Carter constant into the complete waveform-generation scheme. The essential step here is to combine Grant's formulas with those of Witzany et al.~\cite{Witzany:2024ttz} for the conservative sector of the linear-in-spin dynamics and those of Isoyama et al.~\cite{Isoyama:2018sib} for asymptotic fluxes.

\subsection{Outline and conventions}

In Sec.~\ref{sec:sf} we review SF theory with a spinning compact secondary.\footnote{The spin of the primary body is automatically assimilated at the level of the metric in SF theory.} In Sec.~\ref{sec:spin} we highlight a convenient parameterization of the secondary's spin and derive equations for its precession and nutation. We show, in particular, that the spin's two independent components (parallel and orthogonal to the orbital angular momentum) are exactly constant at linear order in the spin.

In Sec.~\ref{sec:generic} we present the multiscale expansion describing the evolving inspiral and metric of a spinning secondary on a generic orbit in a Kerr background spacetime. We elucidate the gauge freedom in this expansion, among other things. In Sec.~\ref{sec:waveform generation} we summarize the waveform-generation scheme. We analyze the gauge invariance of the waveform and characterize the impact of the secondary spin precession. In Sec.~\ref{sec:flux balance}, we describe how the results of Grant~\cite{Grant:2024ivt}, Witzany et al.~\cite{Witzany:2024ttz}, and Isoyama et al.~\cite{Isoyama:2018sib} can be combined to obtain a complete, practical prescription for the calculation of 1PA spin effects in the dynamics. In Sec.~\ref{sec:conclusions} we summarise our findings, outline the path to complete 1PA waveforms, and discuss other avenues for future extensions. 

Our analysis in Secs.~\ref{sec:generic}--\ref{sec:flux balance} applies away from orbital resonances~\cite{Hinderer:2008dm,Pound:2021qin,Lynch:2023gpu}, while Secs.~\ref{sec:sf} and \ref{sec:spin} apply generically, both away from and across resonances.

We work in geometric units with $G=c=1$. We denote the individual masses as $m_i$ with $m_1\gg m_2$. We use $\e\equiv 1$ as a counting parameter of the small mass ratio $\ee\equiv m_2 / m_1$. $\chi_i$ are the dimensionless spins $S_i/m_i^2$, where $S_i$ are the spin angular momenta of the two bodies. For astrophysical compact objects, $S_i\sim m_i^2$ and so in the small-mass-ratio expansion $S_1\sim\e^0$ and  $S_2\sim \e^2$. For each dimensionless spin this implies $\chi_i\sim \e^0$. If the body is a Kerr black hole, then we have the more precise restriction $0\leq|\chi_i|\leq1$.
 
\section{Self-force theory with a spinning secondary}
\label{sec:sf}

In this section we review the formulation of SF theory with a spinning secondary at second order in perturbation theory. We will ultimately use a multiscale formulation of the problem in later sections, but in this section for simplicity we adopt the self-consistent formulation~\cite{Pound:2009sm} as extended in Ref.~\cite{Miller:2020bft}. 

\subsection{Field equations and effective metric}

The Einstein field equations governing the compact binary's metric ($\textbf{g}_{\mu\nu}$) are
\begin{equation}
\label{eq:FE}
G_{\mu \nu}(\textbf{g})=8\pi T_{\mu \nu},
\end{equation}
where $T_{\mu\nu}$ is an effective stress-energy tensor for the secondary object, described below. Taking the usual black hole perturbation theory approach, we solve the field equations by expanding the binary's metric in powers of $\e$:
\begin{equation}
\label{eq:metric}
\textbf{g}_{\mu \nu}=g_{\mu \nu}+\e h_{\mu \nu}^{(1)}+\e^2 h_{\mu \nu}^{(2)} \truncthree,
\end{equation}
where $g_{\mu \nu}$ is the spacetime of the primary black hole as if it were in isolation, and $h_{\mu \nu}^{(n)}$ is the $n^{\text{th}}$-order metric perturbation. 

Since it absorbs gravitational radiation during the binary's evolution, the primary black hole itself slowly evolves. This can be accounted for in one of two ways. The first option is to take $g_{\mu \nu}$ to be a Kerr metric with time-dependent mass and spin parameters. In this first approach, the Einstein tensor of the background metric, $G_{\mu\nu}(g)$, is small but nonzero. The second option is to split the mass and spin into constant terms ($m_1^{(0)}$ and $\chi_1^{(0)}$) and small, dynamical corrections ($\delta m_1$ and $\delta\chi_1$); $g_{\mu\nu}$ is then taken to be a Kerr metric with mass $m_1^{(0)}$ and dimensionless spin $\chi_1^{(0)}$, and the perturbations $h^{(n)}_{\mu\nu}$ then include terms that are $n^{\text{th}}$ order in $\delta m_1$ and $\delta\chi_1$. Both approaches, which yield equivalent total metrics, are described in Sec.~II of Ref.~\cite{Miller:2020bft}. Here we adopt the second approach, which implies $G_{\mu\nu}(g)=0$ and allows us to exploit the exact Killing symmetries of the fixed Kerr background. We return to the treatment of the primary in Ref.~\cite{Mathews:2025txc}, where we specialize to the case of a slowly spinning primary with $\chi^{(0)}_1=0$ and $\delta\chi_1\neq0$.

Because the secondary compact object is much smaller than the curvature scale of $g_{\mu\nu}$, it can be represented in the field equations as a `gravitational skeleton', meaning a point-particle stress-energy tensor $T_{\mu\nu}$ equipped with the object's multipole moments. More concretely, through second order in $\e$, the secondary object can be described as a spinning point particle under the pole-dipole approximation~\cite{Mathews:2021rod,Pound:2012dk, Upton:2021oxf}. For a material body whose size is much smaller than the curvature scale of $g_{\mu\nu}$, this pole-dipole approximation follows from a Mathisson-Dixon multipole expansion of the body's physical stress-energy tensor~\cite{Mathisson:1937zz,Mathisson:2010opl,Dixon:1970zz} (as extended by Harte to the physical case of a gravitating material body~\cite{Harte:2011ku,Harte:2014wya}). More generally, for black holes as well as material bodies, the pole-dipole approximation has been derived via matched asymptotic expansions~\cite{Gralla:2008fg,Pound:2012dk,Upton:2021oxf}. In either approach, the monopole and dipole terms in the stress-energy are
\begin{equation}
\label{eq:pole-dipole}
 T^{\alpha \beta}=\e T_{(m)}^{\alpha \beta} + \e^2 T_{(d)}^{\alpha \beta} + \mathcal{O}(\e^3),
\end{equation}
where $T_{(m)}^{\alpha \beta}$ is a mass-monopole term and $T_{(d)}^{\alpha \beta} $ is a spin-dipole term; quadrupole and higher moments would appear at higher orders in $\e$. Explicitly, the two contributions are
\begin{subequations}
\begin{align}
\label{eq:SEmono}
T_{(m)}^{\alpha \beta}&=m_2\int d \hat\tau' \, \frac{\delta^{4}\left[x^{\mu}- z^{\mu}(\tau' )\right]}{\sqrt{-\hat g'}} \hat u^{\alpha}(\tau') \hat u^{\beta}(\tau' ),\\
\label{eq:SEdipole}
T_{(d)}^{\alpha \beta} &= (m_2)^2\, \hat \nabla_{\rho} \int d\hat \tau'  \, \frac{\delta^{4}\left[x^{\mu}- z^{\mu}(\tau' )\right]}{\sqrt{-\hat g'}} \hat u^{(\alpha}(\tau' ) {\hat S}^{\beta)\rho}(\tau'),
\end{align}
\end{subequations}
where $\delta^{4} $ is the four-dimensional Dirac delta function, $z^\mu$ is the object's effective center-of-mass worldline, $\hat S^{\alpha\beta}$ is the object's dimensionless (mass-normalized) spin tensor, and primes are used to indicate evaluation at $z^\mu(\tau')$. 

Importantly, the stress-energy terms~\eqref{eq:SEmono} and \eqref{eq:SEdipole} take the form of a spinning particle in a certain \emph{effective} vacuum metric $\hat g_{\alpha\beta}$ rather than in the external background $g_{\alpha\beta}$. The proper time $\hat\tau$, four-velocity $\hat u^\alpha\equiv \frac{dz^\alpha}{d\hat\tau}$, metric determinant $\hat g$, and covariant derivative $\hat\nabla_\alpha$ are all defined from $\hat g_{\alpha\beta}$. The effective metric itself is defined by subtraction of suitably defined, singular self-fields $h^{{\rm S}(n)}_{\alpha\beta}$ from the physical metric. Given the physical metric from Eq.~\eqref{eq:metric}, which we can write as
\begin{equation}
\textbf{g}_{\mu \nu}=g_{\mu \nu}+\sum_{n\geq0} \e^n h_{\mu \nu}^{(n)},
\end{equation}
the effective metric is then
\begin{equation}
\hat g_{\mu \nu}=g_{\mu \nu}+ h_{\mu \nu}^{\mathrm{R}},
\end{equation}
in which we have defined
\begin{equation}
h_{\mu \nu}^{\mathrm{R}}= \sum_{n\geq0} \e^n \left( h_{\mu \nu}^{(n)}-h_{\mu \nu}^{\mathrm{S}(n)}\right).
\end{equation}
The appropriate singular/self-field $h_{\mu \nu}^{\mathrm{S}(2)}$ for a spinning compact secondary was derived in Ref.~\cite{Pound:2012dk}; we refer to Ref.~\cite{Mathews:2021rod} for further discussion.

With the expansions of the metric and stress-energy tensor, we can rewrite Eq.~\eqref{eq:FE} through second order as
\begin{multline}\label{eq:EFE expanded}
    \e\delta G_{\mu\nu}(h^{(1)}) + \e^2\delta G_{\mu\nu}(h^{(2)})\\ = 8\pi\left(\e T^{(m)}_{\mu\nu}+\e^2 T^{(d)}_{\mu\nu}\right) \\
    -\e^2\delta^2 G_{\mu\nu}(h^{(1)},h^{(1)}) +{\cal O}(\e^3),
\end{multline}
where $\delta G_{\mu\nu}$ is the linearized Einstein tensor and $\delta^2 G_{\mu\nu}$ is the quadratic term in the expansion of the Einstein tensor. We will not require more explicit expressions, but they can be found in Sec.~II of Ref.~\cite{Spiers:2023mor}, for example. 

Equation~\eqref{eq:EFE expanded} can be divided into a sequence of equations for each $h^{(1)}_{\mu\nu}$ as described in Sec.~II of Ref.~\cite{Miller:2020bft}. However, this is nontrivial due to the presence of the evolving mass and spin corrections $(\delta m_1,\delta\chi_1)$ and because the trajectory $z^\mu$ and spin $\hat S^{\alpha\beta}$ satisfy $\e$-dependent dynamical equations. We will ultimately adopt a multiscale expansion that cleanly splits Eq.~\eqref{eq:EFE expanded} into hierarchical equations. To enable that expansion, we next turn to the dynamical equations for $z^\mu$ and $\hat S^{\alpha\beta}$.

\subsection{MPD-Harte equations of motion}

Through second order in $\e$, the secondary behaves as a test body in the effective metric $\hat g_{\mu\nu}$~\cite{Pound:2012nt,Harte:2011ku,Pound:2017psq}, obeying the Mathisson-Papapetrou-Dixon (MPD) test-body equations~\cite{Mathisson:1937zz,Papapetrou:1951pa,Dixon:1974xoz}. The secondary's quadrupole moment enters these equations at $\mathcal{O}(\e^2)$, but we allow ourselves to neglect its effects because they are purely conservative (at least for a small Kerr black hole~\cite{Ramond:2024ozy,Rahman:2021eay}) and therefore only influence the waveform at 2PA order. For the same reason, we neglect terms that are quadratic or higher order in the particle's spin, which we denote ${\cal O}(s^2)$. We hence assume the pole-dipole approximation, under which the MPD equations in  $\hat g_{\mu\nu}$ read
\begin{subequations}
\label{eq:MPDhat}
\begin{align}
\label{eqn:MPD1hat}
\frac{\hat D \hat u^{\alpha}}{d \hat\tau} &=-\frac{m_2}{2 } \hat R^{\alpha}{}_{\beta \gamma \delta} \hat u^{\beta} \hat S^{\gamma \delta}+ {\cal O}(s^2), \\
\label{eqn:MPD2hat}
\frac{\hat D \hat S^{\gamma \delta}}{d \hat\tau} &= {\cal O}(s^2),
\end{align}
\end{subequations}
where $\hat D/d\hat \tau \equiv \hat u^\alpha\hat\nabla_\alpha$.  
We refer to these as the MPD-Harte equations since Harte extended them (subject to some caveats~\cite{Mathews:2021rod}) to gravitating material bodies rather than only test bodies in a fixed background. Reference~\cite{Mathews:2021rod} contains a critical assessment of their validity in our context.

In Eq.~\eqref{eq:MPDhat} we have  imposed the Tulczyjew-Dixon spin supplementary condition (SSC),
\begin{equation}
\label{eq:TDssc}
\hat p^{\alpha}\hat S_{\alpha \beta} = 0,
\end{equation}
where here (and throughout this paper) indices on hatted quantities are lowered with the effective metric. Recall we have defined $\hat S_{\alpha \beta}$ as the mass-normalised effective spin tensor of the secondary, such that
\begin{equation}\label{eq:chi2=hatS.hatS}
(\chi_2)^2=\frac{1}{2}\hat S_{\alpha \beta}\hat S^{\alpha \beta}.
\end{equation}
The quantity $\hat p^{\mu}$ is the secondary's linear momentum in the effective spacetime. We may also define an effective (dimensionless) spin vector,
\begin{equation}
\hat{S}^{\mu} = -\frac{1}{2} \hat \epsilon^{\mu}{}_{\alpha \beta \gamma} \hat u^{\alpha} \hat{S}^{\beta \gamma},
\end{equation}
with inverse relation
\begin{equation}
\label{eq:SpinTensorFromVector}
\hat{S}^{\mu \nu}=- \hat \epsilon^{\mu \nu \alpha \beta}\hat{S}_{\alpha}\hat u_{\beta}.
\end{equation}
Note that the hat on the Levi-Civita tensor is important as it indicates dependence on the metric determinant of $\hat g_{\alpha \beta}$ as opposed to the metric determinant of $g_{\alpha \beta}$ or $\textbf{g}_{\alpha \beta}$. The Tulczyjew-Dixon SSC implies that $\hat p^\alpha=m_2 \hat u^\alpha+{\cal O}(s^2)$, and $\hat u^{\alpha} \hat S_{\alpha}=0$, where the higher-order spin terms only contribute conservative effects at $\mathcal{O}(\e^2)$ and hence may be neglected. Given our definitions and Eq.~\eqref{eq:MPDhat}, the spin vector must satisfy the parallel transport equation
\begin{equation}
\label{eq:PTeqn}
\frac{\hat D\hat{S}^{\alpha}}{d \hat \tau}={\cal O}(s^2).
\end{equation}

It is more practical to re-express the equations of motion in Eq.~\eqref{eq:MPDhat} in terms of the \emph{background} metric and its regular perturbations, expanding in powers of $\e$. The equations of motion become those of a self-accelerated and self-torqued spinning body in the background spacetime~\cite{Mathews:2021rod}:
\begin{subequations}
\label{eq:selfforceEOMs}
\begin{align}
\frac{D u^{\mu}}{d \tau} &= -\frac{1}{2} P^{\mu\nu}(g_\nu{}^\lambda - h^{\mathrm{R}\, \lambda}_\nu)\left(2 h_{\lambda \rho ; \sigma}^{\mathrm{R}}-h_{\rho \sigma ; \lambda}^{\mathrm{R}}\right) u^{\rho} u^{\sigma}\nonumber\\
&\quad -\frac{m_2}{2}R^{\mu}{}_{\alpha \beta \gamma}\left(1-\frac{1}{2}h^{\mathrm{R}}_{\rho\sigma}u^\rho u^\sigma\right)u^{\alpha}\hat S^{\beta \gamma} \nonumber\\
&\quad +\frac{m_2}{2} P^{\mu\nu}(2h^{\mathrm{R}}_{\nu(\alpha;\beta)\gamma}-h^{\mathrm{R}}_{\alpha\beta;\nu\gamma})u^{\alpha}\hat S^{\beta \gamma}\nonumber\\
&\quad+\mathcal{O}(\e^3,s^2),\label{eq:selfforceorbit}\\
\frac{D\hat S^{\mu\nu}}{d \tau} &= u^{(\rho}\hat S^{\sigma)[\mu}g^{\nu]\lambda}\left(2 h_{\lambda \rho ; \sigma}^{\mathrm{R}}-h_{\rho \sigma ; \lambda}^{\mathrm{R}}\right)  + \mathcal{O}(\e^2,s^2), \label{eq:selfforcespin}
\end{align}
\end{subequations}
where $D/d\tau\equiv u^\mu\nabla_\mu$, $u^\mu\equiv dz^\mu/d\tau$,  $P^{\mu\nu}\equiv g^{\mu\nu}+u^\mu u^\nu$, $\tau$ is the proper time as measured in $g_{\mu\nu}$, and $\nabla_\mu$ and semicolons both denote the covariant derivative compatible with $g_{\mu\nu}$.

In deriving Eq.~\eqref{eq:selfforceEOMs} we have used the normalization conditions 
\begin{equation}\label{eq:u normalisations}
\hat g_{\alpha \beta}\hat u^{\alpha} \hat u^{\beta}=-1=g_{\alpha \beta}u^{\alpha} u^{\beta},
\end{equation}
or equivalently, $\hat g_{\alpha \beta}\hat p^{\alpha} \hat p^{\beta}=-m_2^2 +\mathcal{O}(s^2) = g_{\alpha \beta}p^{\alpha} p^{\beta}$. By writing $\hat u^{\alpha} =\frac{d\tau}{d\hat\tau}u^{\alpha}$ in the normalization conditions, one finds
\begin{equation}
\frac{d\tau}{d\hat\tau}=\sqrt{1-h^{\rm R}_{\alpha\beta}u^\alpha u^\beta},
\end{equation}
or equivalently we may write 
\begin{equation}\label{eq:hatu=u+du}
\hat u^{\alpha}=u^{\alpha}+\delta u^{\alpha}
\end{equation}
with
\begin{equation}
\delta u^{\alpha}= \left(1-\sqrt{1-h^{\rm R}_{u u}}\right)u^{\alpha}\approx \frac{1}{2}h^{\rm R}_{u u}u^{\alpha},
\end{equation}
where the indices replaced with $u$ indicate contraction with the background four-velocity. 

We emphasise that Eq.~\eqref{eq:hatu=u+du} does not represent an expansion of the four-velocity around a zeroth-order value. $\hat u^\alpha$ and $u^\alpha$ are both tangent to the same self-accelerated curve $z^\alpha$; they differ only in the choice of parameter  ($\hat\tau$ or $\tau$) along that curve.

\section{Precession and nutation of the secondary spin}
\label{sec:spin}

In this section we present a convenient parametrization of the secondary's spin degrees of freedom. By decomposing the spin vector in a local, Fermi-Walker transported frame along the particle's worldline, we show its components in this frame are exactly constant. The spin's precession then corresponds to the local frame's rotation relative to a second frame. 

\subsection{Decomposition in a Fermi-Walker tetrad}

Along the particle's worldline, we construct a tetrad $\{\hat e^\alpha_0,\hat e^\alpha_A\}$ ($A=1,2,3$) that is orthonormal in the effective metric $\hat g_{\alpha\beta}$:
\begin{subequations}
\begin{align}
\hat g_{\alpha\beta}\hat e^\alpha_0 \hat e^\beta_0 &= -1,\\
\hat g_{\alpha\beta}\hat e^\alpha_0 \hat e^\beta_A &= 0,\label{eq:ehat0.ehatA}\\
\hat g_{\alpha\beta}\hat e^\alpha_A \hat e^\beta_B &= \delta_{AB}.\label{eq:ehatA.ehatB}
\end{align}
\end{subequations}
We take the zeroth leg of the tetrad to be the four-velocity, $\hat e^\alpha_0=\hat u^\alpha$, which is propagated along the worldline according to
\begin{equation}
    \frac{\hat D \hat e^\alpha_0}{d\hat\tau} = \hat a^\alpha.
\end{equation}
Here $\hat a^\alpha\equiv \frac{\hat D \hat u^\alpha}{d\hat\tau}$ is the covariant acceleration in $\hat g_{\alpha\beta}$, given by Eq.~\eqref{eqn:MPD1hat}. We take the spatial triad $\hat e^\alpha_A$ to be Fermi-Walker transported along the worldline, meaning~\cite{Poisson:2011nh}
\begin{equation}\label{eq:F-W transport}
    \frac{\hat D\hat e^\alpha_A}{d\hat\tau} = \hat a_A \hat u^\alpha,
\end{equation}
where $\hat a_A\equiv\hat e^\beta_A \hat a_\beta$. The resulting tetrad $\{\hat e^\alpha_0,\hat e^\alpha_A\}$ represents the particle's local frame in the effective metric $\hat g_{\alpha\beta}$. Decomposed in this frame, the spin vector reads
\begin{equation}
\hat S^{\alpha} = \hat S^0 \hat e_0^{\alpha}+\hat S^{A}\hat e_A^{\alpha}. 
\end{equation}

We can show almost immediately that the tetrad components $\hat S^0$ and $\hat S^A$ are constant (up to spin-squared terms). The Tulczyjew-Dixon SSC~\eqref{eq:TDssc} implies
\begin{equation}\label{eq:hatS0=0}
    \hat S^0 = 0.
\end{equation}
The parallel-transport equation~\eqref{eq:PTeqn} then implies
\begin{equation}
    \frac{d\hat S^A}{d\hat\tau}\hat e^\alpha_A + \hat S^A \hat a_A\hat u^\alpha = {\cal O}(s^2).
\end{equation}
Contracting this equation with the four-velocity yields
\begin{equation}\label{eq:hatS.hata=0}
    \hat S^A \hat a_A = {\cal O}(s^2),
\end{equation}
while contracting with a member of the spatial triad yields
\begin{equation}\label{eq:dhatSdtau=0}
    \frac{d\hat S^A}{d\hat\tau} = {\cal O}(s^2).
\end{equation}
Equation~\eqref{eq:hatS.hata=0} is trivially satisfied because $\hat a_\alpha$ is proportional to the spin. 
Equation~\eqref{eq:dhatSdtau=0} shows that the spin's nonzero degrees of freedom are constant.

Given the trajectory $z^\alpha$, one can always construct the triad $\hat e^\alpha_A$ by finding a triad orthogonal to $\hat u^\alpha$ at a single point on the trajectory and then solving the Fermi-Walker transport equation~\eqref{eq:F-W transport}. We do so in effect by expressing the Fermi-Walker transported tetrad as a perturbed version of a commonly used background tetrad: the Marck tetrad~\cite{Marck}.

\subsection{Marck tetrad}

The Marck tetrad $\{e^\alpha_0,e_A^{\alpha}\}$ forms a basis of solutions to the parallel transport equation along geodesics of Kerr spacetime~\cite{Marck,vandeMeent:2019cam,Witzany:2019nml}. As we describe below, it is not Fermi-Walker transported in the background metric $g_{\alpha\beta}$, but the violation is ${\cal O}(\e,s)$, meaning we will be able to write the Fermi-Walker tetrad as a small deformation of the Marck tetrad:
\begin{subequations}    
\begin{align}
\hat e^\alpha_0 = e^\alpha_0+\delta e^\alpha_0 + {\cal O}(\e^2),\\
\hat e_A^{\alpha} = e_A^{\alpha}+\delta e^\alpha_A + {\cal O}(\e^2).
\end{align}
\end{subequations}

The Marck tetrad legs on $z^\mu$ can be constructed from the four-velocity $u^\alpha$ and the Killing-Yano tensor $Y_{\alpha\beta}$:
\begin{subequations}
\begin{align}
e_0^{\alpha} &= u^{\alpha},\\
e_1^{\alpha}&=\cos \psi_s\, \E_1^{\alpha} + \sin\psi_s\, \E_2^{\alpha},\label{eq:e1}\\
e_2^{\alpha} &=\cos \psi_s\, \E_2^{\alpha} - \sin\psi_s\, \E_1^{\alpha},\label{eq:e2}\\
e_3^{\alpha} &= Y^\alpha_{\ \;\beta} u^{\beta}/\sqrt{K^{(0)}},\label{eq:e3}
\end{align}
\end{subequations}
where $\psi_s$ is a spin-precession phase discussed below,
\begin{subequations}
\label{eq:sigmatetrad}
\begin{align}
\E_0^\alpha &= u^\alpha,\\
\E_1^\alpha &= \epsilon^{\alpha \beta \gamma \delta}\E^0_\beta \E^2_\gamma \E^3_\delta,\\
\E_2^\alpha &= \frac{1}{N}P^{\alpha \beta}K_{\beta \gamma}u^{\gamma},\\
\E_3^\alpha &= e_3^{\alpha},
\end{align}
\end{subequations}
$K_{\mu\nu}=Y_{\mu}{}^\rho Y_{\nu\rho}$ is the Killing tensor, $K^{(0)}= K^{\alpha \beta}u_{\alpha}u_{\beta}$ and $N\equiv-\sqrt{P^{\alpha \beta}K_{\beta \gamma}u^{\gamma}P_{\alpha}{}^{ \delta}K_{\delta \lambda}u^{\lambda}}$.
One can straightforwardly check that $\{e^\alpha_0,e^\alpha_A\}$ is orthonormal in the background metric, satisfying 
\begin{subequations}
\begin{align}
g_{\alpha\beta}e^\alpha_0 e^\beta_0 &= -1,\\
g_{\alpha\beta}e^\alpha_0 e^\beta_A &= 0,\\
g_{\alpha\beta}e^\alpha_A e^\beta_B &= \delta_{AB}
\end{align}
\end{subequations}
at all points along $z^\alpha$. 

If $z^\alpha$ were a geodesic of the Kerr background, the Marck tetrad would be parallel propagated (with respect to $g_{\alpha\beta}$) along it. 
The Fermi-Walker transport equation reduces to the parallel transport equation at zeroth order in acceleration, and the corresponding basis of solutions reduces to the Marck tetrad. Equivalently, we can say the Marck tetrad represents the local frame of an inertial observer (freely falling in the Kerr background) whose trajectory is instantaneously co-moving with $z^\alpha$.
In the remainder of this section, we assess the tetrad's failure to be Fermi-Walker propagated along the accelerated trajectory, and we explain the interpretation of its precession.

Given $e^\alpha_0=u^\alpha$, we first note the tetrad leg $e^\alpha_3$ defined from Eq.~\eqref{eq:e3} is trivially orthogonal to $e^\alpha_0$ because $Y_{\alpha\beta}$ is antisymmetric. We can loosely interpret the quantity
\begin{equation}\label{eq:l def}
l^\alpha\equiv Y^\alpha_{\ \;\beta}u^\beta 
\end{equation}
as the particle's leading-order specific orbital angular momentum (noting $l^\alpha l_\alpha = K^{(0)}$) \cite{c839080c-5d3d-34d9-a247-38e3c109da14}. If the spin of either body is co-directional with $l^\alpha$, it will not exhibit precession. $l^\alpha$ plays a role analogous to the orbital angular momentum vector in post-Newtonian theory, and $\sqrt {K^{(0)}}$ has the units of a specific angular momentum. $e^\alpha_3$ is then a unit vector in the direction of this angular momentum. If we project the Killing-Yano tensor onto the intermediary tetrad $\{\sigma^\alpha_0,\sigma^\alpha_A\}$, we find only three independent components,
\begin{subequations}
\label{eq:KYtetrad}
\begin{align}
Y_{\alpha\beta}\E_0^\alpha \E_3^\beta=&-\sqrt{K^{(0)}},\\
Y_{\alpha\beta}\E_1^\alpha \E_2^\beta=&\frac{\mathcal{Z}}{\sqrt{K^{(0)}}},\\
Y_{\alpha\beta}\E_2^\alpha \E_3^\beta=&-\frac{N}{\sqrt{K^{(0)}}},
\end{align}
\end{subequations}
having introduced the Killing-Yano scalar, which in Kerr spacetime takes the value $\mathcal{Z}=r a \cos\theta$. By definition,
\begin{equation}
\frac{D \E_3^\alpha}{d\tau}=\frac{1}{\sqrt{K^{(0)}}}Y^\alpha_{\ \;\beta}a^\beta-\frac{1}{K^{(0)}}\E_3^\alpha (K_{\delta \gamma}u^\gamma a^\delta),
\end{equation}
where $a^\alpha$ is the covariant acceleration in $g_{\alpha\beta}$, given by Eq.~\eqref{eq:selfforceorbit}. Substituting Eq.~\eqref{eq:KYtetrad} yields
\begin{align}
    \frac{D \E_3^\alpha}{d\tau}&= (a^\beta \E^3_\beta) u^\alpha+\frac{\mathcal{Z}}{K^{(0)}}(a^\beta \E^2_\beta) \E_1^\alpha\nonumber\\
    &\quad-\frac{1}{K^{(0)}}\left(\mathcal{Z}a^\beta \E^1_\beta +N a^\beta \E^3_\beta\right) \E_2^\alpha.\label{eq:DE3dtau}
\end{align}
The terms proportional to $\sigma^\alpha_A$ on the right-hand side represent the failure of $\sigma^\alpha_3$ (and therefore of $e^\alpha_3$) to be Fermi-Walker transported along the trajectory. From the formula above, we see that if the trajectory is confined to the equatorial plane, $\theta=\pi/2$ with $a^\theta=0$, then $\E_3^\alpha$ is automatically Fermi-Walker transported.

At leading order, the secondary spin's precession corresponds to rigid rotation of the remaining two legs, $e^\alpha_1$ and $e^\alpha_2$, around $l^\alpha$. To understand this, first consider an arbitrary pair of mutually orthonormal vectors $\E^\alpha_1$ and $\E^\alpha_2$ that are also orthonormal to $e^\alpha_0$ and $e^\alpha_3$. The triad $\E^\alpha_A$ (with $\E^\alpha_3\equiv e^\alpha_3$) necessarily satisfies
\begin{equation}\label{eq:generic propagation}
\frac{D\E^\alpha_A}{d\tau} = (\E^\alpha_A a_\alpha) u^\alpha + \omega_{A}^{\ \,B}\E^\alpha_B
\end{equation}
for some antisymmetric $\omega_{AB}=-\omega_{BA}$, which represents the angular velocity of the triad relative to the particle's natural Fermi-Walker frame. Here and elsewhere, triad indices are raised with $\delta^{AB}$ and lowered with $\delta_{AB}$. Equation~\eqref{eq:generic propagation} (together with $Du^\alpha/d\tau=a^\alpha$) is the most general form of propagation of any tetrad $\{u^\alpha,\E^\alpha_A\}$ along an accelerated worldline; this can be seen by writing $\frac{D\E^\alpha_A}{d\tau}$ as a linear combination of $u^\alpha$ and $\E^\alpha_B$, contracting it with $u_\alpha$ or $\E^C_\alpha$, and using the identities
\begin{equation}
    u_\alpha \frac{D\E^\alpha_A}{d\tau} = -a_\alpha \E^\alpha_A \quad\text{and}\quad \E^B_\alpha \frac{D\E^\alpha_A}{d\tau} = -\E_A^\alpha \frac{D\E_\alpha^B}{d\tau},
\end{equation}
which follow from orthonormality. Such a calculation also shows
\begin{equation}\label{eq:omegaAB}
    \omega_A^{\ \,B} = \frac{D\E^\alpha_A}{d\tau}\E^B_\alpha.
\end{equation}

Any two triads that are orthogonal to $u^\alpha$ must be related to each other by a rigid rotation. Hence, starting from a generic $\E^\alpha_A=(\E^\alpha_1,\E^\alpha_2,e^\alpha_3)$, we can construct the last two Marck tetrad legs using the rotation in Eqs.~\eqref{eq:e1} and \eqref{eq:e2}.
More compactly, we write the Marck triad as $e^\alpha_A = R_A^{\ \,B} \E^\alpha_B$, where
\begin{equation}\label{eq:RAB}
R_A^{\ \,B}=\begin{pmatrix}\cos\psi_s & \sin\psi_s & 0\\
-\sin\psi_s & \cos\psi_s & 0\\
0 & 0 & 1\end{pmatrix} \equiv \bm{R}_z(\psi_s)
\end{equation}
is the rotation matrix.

The covariant derivative of $e^\alpha_A$ along $z^\mu$ is
\begin{equation}\label{eq:Dedtau}
    \frac{De^\alpha_A}{d\tau} = \frac{dR_A^{\ \,B}}{d\tau}\E^\alpha_B + R_A^{\ \,B} \frac{D\E^\alpha_B}{d\tau}. 
\end{equation}
First contracting this with $u_\alpha$, and using $u_\alpha e^\alpha_A = 0 = u_\alpha \E^\alpha_A$, we find 
\begin{equation}
     u_\alpha\frac{De^\alpha_A}{d\tau} = - a_A = -R_A^{\ \,B} a_\alpha\E^\alpha_B.
\end{equation}
Next contracting Eq.~\eqref{eq:Dedtau} with $\E_{C\alpha}$ and using $(R_A^{\ \,B})^{-1}=R^B_{\ \,A}$, we find 
\begin{equation}\label{eq:rotation equation}
    R^B_{\ \,A}\frac{dR_{BC}}{d\tau} = - \frac{D\E_A^{\alpha}}{d\tau}\E_{C\alpha} = -\omega_{AC}+\mathcal{O}(\e,s). 
\end{equation}
The left-hand side can be immediately evaluated and the equation reduced to
\begin{equation}\label{eq:psidot AB}
    \begin{pmatrix}0 & 1 & 0\\
-1 & 0 & 0\\
0 & 0 & 0\end{pmatrix}\frac{d\psi_s}{d\tau} = - \omega_{AB}+\mathcal{O}(\e,s),
\end{equation}
which implies
\begin{equation}\label{eq:psidot}
\frac{d \psi_s}{d \tau} =-\omega_{12}+\mathcal{O}(\e,s)= \omega_{21}+\mathcal{O}(\e,s).
\end{equation}
In other words, the dyad $\{\E^\alpha_1,\E^\alpha_2\}$ precesses around $l^\alpha$ with an angular frequency $\omega_{21}$. Note that Eq.~\eqref{eq:psidot AB} also requires $\omega_{31}=\mathcal{O}(\e)=\omega_{32}$, but this is satisfied by virtue of Eq.~\eqref{eq:DE3dtau}.

The spin precession frequency $\Omega_s$ will be an appropriate orbit average of $\omega_{21}$. We note this frequency depends on \emph{which} dyad $\{\E^\alpha_1,\E^\alpha_2\}$ we begin with. Our derivation of Eq.~\eqref{eq:psidot} is valid for any choice of this dyad; we have not made use of the specific choice in Eq.~\eqref{eq:sigmatetrad}. If $\{\E^\alpha_1,\E^\alpha_2\}$ is chosen to have fixed orientation relative to a suitably stationary frame in the Kerr background~\cite{Bini:2017slb}, then the Marck tetrad legs precess around $l^\alpha$ with a frequency $\omega_{21}$ when measured in the stationary frame.

\subsection{Fermi-Walker tetrad}

Our goal is to find a Fermi-Walker tetrad $\{\hat e^\alpha_0,\hat e^\alpha_A\}$ such that each of its legs reduces to the corresponding Marck tetrad leg when $h^{\rm R}_{\alpha\beta}$ and $S^\alpha$ vanish. 
We achieve this by first constructing a perturbation to the triad $\E^\alpha_A$ and then finding the necessary rotation to eliminate the frame's angular velocity. 

Given $\hat e_0^{\alpha}=\hat u^{\alpha}=u^\alpha +\delta u^\alpha$,
we construct a triad
\begin{equation}
\hat\E_A^{\alpha}=\E_A^{\alpha}+\delta \E_A^{\alpha}+\mathcal{O}(\e^2),
\end{equation}
satisfying the same orthonormality conditions~\eqref{eq:ehat0.ehatA} and~\eqref{eq:ehatA.ehatB} as $\hat e^\alpha_A$. Writing $\delta \E_A^{\alpha}$ as a linear combination of $u^\alpha$ and $\E^\alpha_A$ and using the fact that $\delta u^\alpha$ is parallel to $u^\alpha$, we immediately find that the orthonormality conditions imply
\begin{equation}
\label{eq:tetrad_shift}
\delta \E_A^{\alpha} = u^{\alpha}  h^{\mathrm{R}}_{\beta \gamma}u^\beta \E_A^{\gamma} -\frac{1}{2}\E^{B\alpha}h^{\mathrm{R}}_{\beta \gamma}\E_B^{\beta}\E_A^{\gamma}.
\end{equation}
This frame has an angular velocity
\begin{equation}\label{eq:omegaABhat}
    \hat\omega_A^{\ \,B} = \frac{\hat D\hat\E^\alpha_A}{d\hat\tau}\hat\E^B_\alpha
\end{equation}
relative to a Fermi-Walker frame.

From the triad $\hat\E^\alpha_A$, we now construct our Fermi-Walker triad as 
\begin{equation}\label{eq:ehat=REhat}
\hat e^\alpha_A \equiv \hat R_A{}^B\hat\E^\alpha_B,
\end{equation}
in analogy with the background triad, where $\hat R_A{}^B$ is a perturbed version of the rotation matrix~\eqref{eq:RAB}.
We can enforce that $\hat e^\alpha_A$ is Fermi-Walker propagated by following the same steps as for the background tetrad, arriving again at Eq.~\eqref{eq:rotation equation} (with hats on all quantities). We rewrite that equation in terms of Euler angles $(\psi_s,\vartheta_s,\varphi_s)$, decomposing the rotation matrix in the form
\begin{equation}
\label{eq:EulerRotation}
    \hat{\bm{R}} = \bm{R}_z(\varphi_s)\bm{R}_x(\vartheta_s)\bm{R}_z(\psi_s),
\end{equation}
where $\bm{R}_z(\alpha)$ is a rotation around the `$z$' axis, as defined in Eq.~\eqref{eq:RAB}, and 
\begin{equation}
\label{eq:Rx}
    \bm{R}_x( \vartheta_s) \equiv \begin{pmatrix}1 & 0 & 0 \\ 
    0 & \cos  \vartheta_s & \sin \vartheta_s \\
    0 & -\sin \vartheta_s & \cos  \vartheta_s
    \end{pmatrix}
\end{equation}
is a rotation around the `$x$' axis. The angles are defined in the ranges $ \psi_s, \varphi_s\in(0,2\pi)$ and $ \vartheta_s\in (0,\pi)$.

In terms of these angles, the hatted version of Eq.~\eqref{eq:rotation equation} reduces to
\begin{subequations}\label{eq:d/dt Euler angles}
\begin{align}
\frac{d \psi_s}{d\hat\tau} &= \hat\omega_{21} -\left(\hat\omega_{31} \cos \psi_s + \hat\omega_{32}  \sin \psi_s\right)\cot \vartheta_s,\label{eq:dpsihat/dtauhat}\\
\frac{d \vartheta_s}{d\hat\tau} &= \hat\omega_{32} \cos \psi_s-\hat\omega_{31} \sin \psi_s,\\
\frac{d \varphi_s}{d\hat\tau} &= \left(\hat\omega_{31} \cos \psi_s+\hat\omega_{32} \sin \psi_s\right)\csc  \vartheta_s.
\end{align}
\end{subequations}
These are the standard evolution equations for the Euler angles describing rigid-body motion, given the angular-velocity tensor $\hat\omega_{AB}$~\cite{Euler-angles}. When $ \vartheta_s\rightarrow0$, the rotation matrix in Eq.~\eqref{eq:Rx} becomes the identity matrix. While Eq.~\eqref{eq:d/dt Euler angles} is singular in that limit, the total rotation in Eq.~\eqref{eq:EulerRotation} remains regular as it goes to $\bm{R}_z( \psi_s+ \varphi_s)$ and the singular terms clearly cancel in the evolution of the sum of the two angles.

The angular velocity $\hat\omega_{AB}$ may be expressed in terms of its background equivalent and $h^{\rm R}_{\alpha\beta}$ as
\begin{equation}
\hat\omega_{AB}=\dfrac{d\tau}{d\hat\tau}\left(\omega_{AB}+\delta \omega_{AB} \right).
\end{equation}
Explicit evaluation of Eq.~\eqref{eq:omegaABhat} yields
\begin{equation}\label{eq:domegaAB}
\delta\omega_{AB}=-h^{\mathrm{R}}_{\alpha \beta;\gamma}u^\alpha\E_A^{[\beta}\E_B^{\gamma]}+{\cal O}(\e^2,s).
\end{equation}
Thus, in Eq.~\eqref{eq:d/dt Euler angles} we can write the angular velocity components as
\begin{subequations}
\begin{align}
    \hat\omega_{21} &= \dfrac{d\tau}{d\hat\tau}\left[\omega_{21} + \delta\omega_{21} + {\cal O}(\e^2,s)\right],\\
    \hat\omega_{3B} &= \delta\omega_{3B} + {\cal O}(\e^2,s),
\end{align}
\end{subequations}
using the fact that $\omega_{31}=\mathcal{O}(s)=\omega_{32}$. We are neglecting the $\mathcal{O}(s)$ contributions to the angular velocity since an $\mathcal{O}(s)$ violation of the Fermi-Walker transport equation does not affect Eq.~\eqref{eq:dhatSdtau=0}. The secondary's spin contributions would be straightforward to include perturbatively by folding them into the definition of $\delta\omega_{AB}$. 

Given this form of the angular velocity, we seek a solution to Eq.~\eqref{eq:d/dt Euler angles} of the form
\begin{subequations}\label{eq:Euler angle ansatz}    
\begin{align}
     \psi_s &= \tilde\psi_s - \delta\varphi \cot\vartheta_0 + {\cal O}(\e^2) ,\label{eq:psi NIT}\\
     \vartheta_s &= \vartheta_0 + \delta\vartheta + {\cal O}(\e^2),\\
     \varphi_s &= \delta\varphi\csc\vartheta_0 + {\cal O}(\e^2),    
\end{align}
\end{subequations}
where $\delta\vartheta$ and $\delta\varphi$ are linear in $h^{\rm R}_{\alpha\beta}$. The $\delta\varphi$ term in Eq.~\eqref{eq:psi NIT} accounts for the oscillatory terms in Eq.~\eqref{eq:dpsihat/dtauhat}. $\vartheta_0$ is an arbitrary constant designed to avoid the singularity of the Euler angles at $\vartheta_s=0$; we will ultimately take the limit $\vartheta_0\to0$ to ensure our perturbed tetrad reduces to the unperturbed one in the $\ee\to0$ limit. To facilitate taking the $\vartheta_0\to0$ limit, we have factored out singular functions of $\vartheta_0$, allowing us to work with variables $\tilde\psi_s$ and $\delta\varphi$ that are smooth at $\vartheta_0=0$.

Substituting the ansatz into Eq.~\eqref{eq:d/dt Euler angles}, we quickly find
\begin{align}
    \frac{d\tilde\psi_s}{d\tau} &= \omega_{21} + \delta\omega_{21},\label{eq:dpsi_s eqn}\\
    \frac{d\delta\vartheta}{d\tau} &= \delta\omega_{32} \cos\tilde\psi_s-\delta\omega_{31} \sin\tilde\psi_s,\label{eq:dtheta eqn}\\    
    \frac{d\delta\varphi}{d\tau} &=  \delta\omega_{31} \cos\tilde\psi_s+\delta\omega_{32} \sin\tilde\psi_s.\label{eq:dphi eqn}
\end{align}
Note that unlike the original variable $ \psi_s$, the new variable $\tilde\psi_s$ has a non-oscillatory rate of change.\footnote{More precisely, $\frac{d\tilde\psi_s}{d\tau}$ is independent of $\tilde\psi_s$. It can still have oscillatory dependence on the \emph{orbital} phases.} It represents the mean precession angle, the secularly growing piece of the original phase $ \psi_s$;  Eq.~\eqref{eq:psi NIT} can hence be interpreted as a near-identity averaging transformation~\cite{VanDeMeent:2018cgn}. 

The angle $\delta\vartheta$ represents nutation of the Fermi-Walker frame. Given the form of the nutation equation~\eqref{eq:dtheta eqn}, we can factor out its precession dependence with an ansatz
\begin{equation}\label{eq:dtheta ansatz}
    \delta\vartheta = \delta\vartheta_c \cos\tilde\psi_s + \delta\vartheta_s \sin\tilde\psi_s, 
\end{equation}
where $\delta\vartheta_c$ and $\delta\vartheta_s$ are independent of $\tilde\psi_s$.\footnote{This ansatz would need to be modified when including the secondary spin since $\delta\omega_{AB}$ would also depend on $\tilde\psi_s$ via precession terms in the spin-curvature coupling force.} This reduces Eq.~\eqref{eq:dtheta eqn} to coupled differential equations for the nutation degrees of freedom $\delta\vartheta_c$ and $\delta\vartheta_s$:
\begin{align}
\label{eq:dtheta_c eqn}
    \frac{d\delta\vartheta_c}{d\tau} - \omega_{12}\delta\vartheta_s &= -\delta\omega_{23},\\
\label{eq:dtheta_s eqn}
    \frac{d\delta\vartheta_s}{d\tau} + \omega_{12}\delta\vartheta_c &= \delta\omega_{13}.   
\end{align}
We can also straightforwardly check that Eq.~\eqref{eq:dphi eqn} can be satisfied with an ansatz in terms of these same functions $\delta\vartheta_c$ and $\delta\vartheta_s$:
\begin{equation}\label{eq:dphi value}
    \delta\varphi = -\delta\vartheta_s \cos\tilde\psi_s + \delta\vartheta_c \cos\tilde\psi_s. 
\end{equation}

The necessary conditions for the parallel transport of the spin vector in the effective metric are finally boiled down to the evolution of the precession phase variable $\tilde \psi_s$ and two nutation variables, $\delta \vartheta_c$ and $\delta \vartheta_s$. Their evolution is determined by Eq.~\eqref{eq:dpsi_s eqn}, Eq.~\eqref{eq:dtheta_c eqn}, and Eq.~\eqref{eq:dtheta_s eqn}, respectively. See Appendix~\ref {sec:NutationSol} for a sketch of the solution to Eqs.~\eqref{eq:dtheta_c eqn} and~\eqref{eq:dtheta_s eqn}.

\subsection{Summary}\label{sec:summary}

We now return to the spin vector itself,
\begin{equation}\label{eq:hatS=hatSA.ehat}
\hat S^\alpha = \hat S^A \hat e^\alpha_A.
\end{equation}
Without loss of generality, we set 
\begin{equation}
\hat S^1=\chi_{\perp},\quad \hat S^2 = 0, \quad \hat S^3=\chi_{\parallel}. 
\end{equation}
The choice $\hat S^2 = 0$ corresponds to a choice of origin for the precession angle $\tilde\psi_s$; see Ref.~\cite{Isoyama:2021jjd} for discussion of analogous freedom in the nonspinning case. By construction, according to Eq.~\eqref{eq:dhatSdtau=0}, the two spin magnitudes are constants:
\begin{align}\label{eq:spinmagsevolveEff}
\frac{d \chi_\perp}{d\tau} &= 0 = \frac{d \chi_\parallel}{d\tau},
\end{align}
including during orbital resonances.
These spin components are hence freely specifiable constants describing the particle, akin to its mass $m_2$.

We now substitute Eqs.~\eqref{eq:ehat=REhat}, \eqref{eq:Euler angle ansatz}, \eqref{eq:dtheta ansatz}, and \eqref{eq:dphi value} into Eq.~\eqref{eq:hatS=hatSA.ehat} and take the limit $\vartheta_0\to0$. The result is 
\begin{equation}
\hat S^{\alpha}=S^{\alpha}+\e \delta S^{\alpha}\trunctwo
\end{equation}
where the two terms are parameterized as
\begin{subequations}\label{eq:S and dS}
\begin{align}
S^{\alpha}&=\chi_{\parallel}\E_3^{\alpha}+\chi_{\perp}\cos\tilde\psi_s\,\E_1^{\alpha}+\chi_{\perp}\sin\tilde\psi_s\,\E_2^{\alpha},\\
\delta S^{\alpha}&=\chi_{\parallel}\delta\E_3^{\alpha}+\chi_\parallel \delta \vartheta_s \, \E^\alpha_1 - \chi_\parallel\delta \vartheta_c\, \E^\alpha_2\nonumber\\
&\quad +\chi_\parallel  (\delta\vartheta_c \sin \tilde\psi_s -\delta\vartheta_s \cos\tilde\psi_s)\sigma^\alpha_3 \nonumber\\
&\quad  +\chi_{\perp}\cos\tilde\psi_s\,\delta\E_1^{\alpha}+\chi_{\perp}\sin\tilde\psi_s\,\delta\E_2^{\alpha},
\end{align}
\end{subequations}
with $\delta\E_A^{\alpha}$ as defined in Eq.~\eqref{eq:tetrad_shift}.
As a consequence of the normalisations of the tetrad legs, we have
\begin{equation}
\hat S^{\alpha}\hat S_{\alpha}= S^{\alpha}S_{\alpha}=\chi^2=\chi_\parallel^2+\chi_\perp^2.
\end{equation}

The parameterization~\eqref{eq:S and dS} has cleanly separated oscillatory and non-oscillatory dependence on the precession phase $\tilde\psi_s$. If we average over that phase, we obtain
\begin{subequations}\label{eq:<S>}
\begin{align}
\left\langle S^{\alpha}\right\rangle_{\tilde\psi_s}&=\chi_{\parallel}\E_3^{\alpha},\label{eq:<S> leading}\\
\left\langle \delta S^{\alpha}\right\rangle_{\tilde\psi_s} &=\chi_{\parallel}\left(\delta\E_3^{\alpha}+\delta \vartheta_s \, \E^\alpha_1 - \delta \vartheta_c\, \E^\alpha_2\right).
\end{align}
\end{subequations}

Finally expressing the spin tensor as $\hat S^{\alpha \beta} = S^{\alpha \beta} + \e \delta S^{\alpha \beta}+\mathcal{O}(\e^2)$, then by Eq.~\eqref{eq:SpinTensorFromVector} we have 
\begin{subequations}\label{eq:deltasab}
\begin{align}
S^{\mu \nu} &= -\epsilon^{\mu \nu \alpha \beta}S_{\alpha}u_{\beta},\\
\delta S^{\alpha \beta} &= \epsilon^{\alpha \beta}{}_{\gamma \lambda}u^{\gamma}\delta S^{\lambda}+\frac{1}{2}P^{\gamma \lambda} h^{\mathrm{R}}_{\gamma \lambda} S^{\alpha \beta}-2 h^{\mathrm{R} \, [\beta}_{\lambda}S^{\alpha] \lambda}.
\end{align}
\end{subequations}

In the next section we utilize our results here to develop a complete 1PA expansion of the orbital motion and Einstein field equations. The spin magnitudes $\chi_\parallel$ and $\chi_\perp$ are exactly constant, but the nutation angles $\delta\vartheta_c$ and $\delta\vartheta_s$ will generally evolve. In principle, their evolution contributes to the dissipative self-force at the same order as the leading linear-in-spin energy and angular momentum fluxes and must be taken into account. However, in Sec.~\ref{sec:gauge choices} and Appendix~\ref{sec:localforces}, we show that  the nutation equations do not need to be explicitly solved to compute the inspiral (and waveform) at 1PA order.

\section{Multiscale expansion for generic orbits in Kerr spacetime}
\label{sec:generic}

In the Introduction we recalled the structure of the multiscale expansion for a nonspinning secondary on a generic orbit in Kerr spacetime. Here, using our results from the previous section, we derive the extension to the case of a spinning secondary. The goal is to solve the Einstein field equations~\eqref{eq:EFE expanded} coupled to the equations of motion~\eqref{eq:selfforceEOMs}. 

Throughout this section, for visual clarity we suppress functional dependence on the constant parameters $m_1^{(0)}$, $\chi_1^{(0)}$, $m_2$, $\chi_\parallel$, and $\chi_\perp$.

\subsection{Orbital motion}\label{sec:orbital motion}

As summarized in Sec.~\ref{sec:summary}, we have reduced the equation for the secondary spin, Eq.~\eqref{eq:selfforcespin}, to evolution equations for a precession angle and two nutation angles, Eqs.~\eqref{eq:dpsi_s eqn},~\eqref{eq:dtheta_c eqn}, and~\eqref{eq:dtheta_s eqn}. We now wish to similarly reduce the equation of orbital motion~\eqref{eq:selfforceorbit} to evolution equations of the form~\eqref{psidot} and \eqref{Jdot}. We adopt Boyer-Lindquist coordinates $(t,x^i)$, with spatial coordinates $x^i=(r,\theta,\phi)$.

We first rewrite Eq.~\eqref{eq:selfforceorbit} in terms of our solution for the spin vector:
\begin{align}
\frac{D u^{\mu}}{d \tau} &= -\frac{1}{2} P^{\mu\nu}(g_\nu{}^\lambda - h^{\mathrm{R}(1)\, \lambda}_\nu)\left(2 h_{\lambda \rho ; \sigma}^{\mathrm{R}}-h_{\rho \sigma ; \lambda}^{\mathrm{R}}\right) u^{\rho} u^{\sigma}\nonumber\\
&\quad -\frac{m_2}{2}R^{\mu}{}_{\alpha \beta \gamma}\left(1-\frac{1}{2}h^{\mathrm{R}(1)}_{\rho\sigma}u^\rho u^\sigma\right)u^{\alpha} S^{\beta \gamma} \nonumber\\
&\quad -\frac{m_2}{2}R^{\mu}{}_{\alpha \beta \gamma}u^{\alpha}\delta S^{\beta \gamma} \nonumber\\
&\quad +\frac{m_2}{2} P^{\mu\nu}\left(2h^{\mathrm{R}(1)}_{\nu(\alpha;\beta)\gamma}-h^{\mathrm{R}(1)}_{\alpha\beta;\nu\gamma}\right)u^{\alpha}S^{\beta \gamma}\nonumber\\
&\quad+\mathcal{O}(\e^3,s^2),\label{eq:selfforceorbit full}
\end{align}
where $S^{\alpha\beta}$ and $\delta S^{\alpha\beta}$ are given by Eq.~\eqref{eq:deltasab} with Eq.~\eqref{eq:S and dS}. To mesh with the multiscale expansion of the Einstein equations and the waveform generation scheme, it is convenient to use $t$, rather than $\tau$ as the parameter along the worldline, in which case Eq.~\eqref{eq:selfforceorbit full} becomes
\begin{align}\label{eq:selfforceorbit t}
    \frac{D \dot z^{\mu}}{dt} = a^\mu/(u^t)^2 +\kappa \dot z^\mu,
\end{align}
where an overdot denotes $d/dt$, $a^\mu$ is the right-hand side of Eq.~\eqref{eq:selfforceorbit full}, $\kappa \equiv -\frac{1}{u^t}\frac{du^t}{dt}$, and $g_{\alpha\beta}u^\alpha u^\beta=-1$ implies
\begin{equation}\label{eq:ut}
u^t = [-(g_{tt}+2g_{ti}\dot z^i+g_{\alpha\beta}\dot z^i\dot z^j)]^{-1/2}.
\end{equation}

Equation~\eqref{eq:selfforceorbit t} can be recast in terms of slow and fast variables following Ref.~\cite{Pound:2021qin}, starting from osculating geodesics~\cite{Pound:2007th,Gair:2010iv} and then performing averaging transformations~\cite{VanDeMeent:2018cgn}. We do not belabor the details as this type of procedure is thoroughly illustrated elsewhere~\cite{VanDeMeent:2018cgn,Lynch:2021ogr,Lynch:2023gpu}; see, in particular, Ref.~\cite{Drummond:2023wqc}, which carried out an analysis very similar to the one we do here but omitted dissipative ${\cal O}(\e^2)$ terms in the equations of motion. 

The analysis in Ref.~\cite{Pound:2021qin} is valid for generic accelerated orbits in Kerr, and we will make extensive use of equations in it. However, our treatment differs from Ref.~\cite{Pound:2021qin} in that we work with $t$ as our time parameter from the beginning, while  Ref.~\cite{Pound:2021qin} begins with a complete analysis in terms of Mino time $\lambda$ and then transforms to variables based on $t$.\footnote{We also make several changes of notation from Ref.~\cite{Pound:2021qin}: $p^i\to \pi_i$, $\varphi_i\to \mathring\psi^i$, $p^i_\varphi\to\mathring\pi_i$, and $\mathscr{P}^\alpha\to\varpi_I$, along with various less noteworthy alterations.} Differential equations in terms of the two variables are related by the factor 
\begin{equation}\label{eq:f_t}
\frac{dt}{d\lambda}\equiv\mathscr{f}_t, 
\end{equation}
given in Eq.~(205) of Ref.~\cite{Pound:2021qin}.

\subsubsection{Osculating geodesic equations}
As in Ref.~\cite{Pound:2021qin}, we first introduce quasi-Keplerian orbital elements $\pi_i=\{p,e,\iota\}$ and phases $\psi^i = \{\psi^r,\psi^\theta,\phi\}$, with which we parameterize the Boyer-Lindquist trajectory $z^i$ and velocity $\dot z^i$ as follows:
\begin{subequations}\label{eq:z(psi,p)}
\begin{align}
    r(\psi^i,\pi_i) &= \frac{p\, m_1}{1+e \cos\psi^r},\\
    \cos\theta(\psi^i,\pi_i) &= \sin\iota \cos\psi^\theta,
\end{align}
\end{subequations}
and 
\begin{equation}\label{eq:zdot(psi,p)}
\dot z^i(\psi^j,\pi_j) = \frac{\partial z^i}{\partial \psi^j}\omega^j_{(0)}(\psi^k,\pi_k).
\end{equation}
Here $\omega^j_{(0)}$ is the expression for $\dot\psi^j$ on a Kerr geodesic, given by
\begin{equation}
\omega^j_{(0)} = \frac{d\lambda}{dt}\mathscr{f}_j = \mathscr{f}_j/\mathscr{f}_t, 
\end{equation}
with $\mathscr{f}_t$, $\mathscr{f}_r$, $\mathscr{f}_\theta$, and $\mathscr{f}_\phi$ as given in Eqs.~(205), (216), (217), and (206) of Ref.~\cite{Pound:2021qin}, respectively.

The parameters $p$, $e$, and $\iota$ are referred to as osculating orbital elements; they correspond to the eccentricity, semi-latus rectum, and maximum inclination of the geodesic that is instantaneously comoving with the accelerated orbit. If the motion were geodesic, we would have $d\pi_i/dt=0$ and $d\psi^i/dt=\omega^i_{(0)}$, while for the accelerated motion we have $d\pi_i/dt\neq0$ and $d\psi^i/dt\neq\omega^i_{(0)}$. We can also simply think of  Eqs.~\eqref{eq:z(psi,p)} and \eqref{eq:zdot(psi,p)} as a coordinate transformation on the particle's orbital phase space, $(z^i,\dot z^i)\mapsto (\psi^i,\pi_i)$. The full list of slowly evolving binary parameters we denote with $\varpi_I=\{\pi_i, \delta m_1, \delta \chi_1\}$.

Applying the chain rule $d/dt = \dot\psi^i\partial_{\psi^i} + \dot\varpi_I\partial_{\varpi_I}$ on the left-hand side of Eq.~\eqref{eq:zdot(psi,p)}, we obtain the kinematical equation
\begin{align}
    \frac{\partial z^i}{\partial \psi_j}\left(\dot\psi^j-\omega^j_{(0)}\right)+\frac{\partial z^i}{\partial \pi_j}\dot \pi_j &= 0.\label{eq:osculating 1}
\end{align}
Substituting Eqs.~\eqref{eq:ut}--\eqref{eq:zdot(psi,p)} into Eq.~\eqref{eq:selfforceorbit t} and appealing to the same chain rule, we obtain the dynamical equation
\begin{align}\label{eq:osculating 2}
    \frac{\partial \dot z^i}{\partial \psi^j}\left(\dot\psi^j-\omega^j_{(0)}\right) + \frac{\partial \dot z^i}{\partial \pi_j}\dot \pi_j &= a^i/(u^t)^2+(\kappa-\kappa_{(0)})\dot z^i,
\end{align}
where 
\begin{equation}
    \kappa-\kappa_{(0)} = -\frac{1}{u^t}\left[\frac{\partial u^t}{\partial\psi^i}(\dot\psi^i-\omega^i_{(0)})+\frac{\partial u^t}{\partial \pi_j}\dot \pi_j\right]. 
\end{equation}
In obtaining Eq.~\eqref{eq:osculating 2} we have used the Kerr geodesic equation in the form $\frac{\partial \dot z^i}{\partial \psi^j}\omega^j_{(0)}+\Gamma^i_{\beta\gamma}\dot z^\beta \dot z^\gamma=\kappa_{(0)}\dot z^i$. Note we have restricted to spatial components of the equation of orbital motion because $z^t=t$ and $\dot z^t=1$ trivially.

We can rearrange Eqs.~\eqref{eq:osculating 1} and \eqref{eq:osculating 2} to obtain equations for $\dot\psi^i$ and $\dot \pi_i$, while the evolution of the primary's mass and spin is described by Eq.~\eqref{mdot} and Eq.~\eqref{Sdot} respectively. Given the form of the acceleration~\eqref{eq:selfforceorbit full}, these equations take the form
\begin{align}
    \frac{d\psi^i}{dt} &= \omega^i_{(0)}(\psi^j,\pi_j)+\e\omega^i_{(1)}(\psi^j,\varpi_J,\tilde\psi_s)+{\cal O}(\e^2),\label{eq:dpsi/dt}\\
    \frac{d\pi_i}{dt} &= \e f^{(0)}_i(\psi^j,\varpi_J,\tilde\psi_s) + \e^2 f^{(1)}_i(\psi^j,\varpi_J,\tilde\psi_s)+{\cal O}(\e^3).\label{eq:dpi/dt}
\end{align}
Here the `frequency' corrections $\omega^i_{(1)}$ and forcing functions $f^{(n)}_i$ are linear combinations of the acceleration components, which we write as
\begin{align}
    \omega^i_{(1)}(\psi^i,\varpi_i,\tilde\psi_s) &= A^i_{\ j}(\psi^i,\pi_i) a^j_{(1)}(\psi^i,\varpi_I,\tilde\psi_s), \label{eq:omega1=Aa}\\
    f^{(n)}_i(\psi^i,\varpi_i,\tilde\psi_s) &= B_{ij}(\psi^i,\pi_i) a^j_{(n+1)}(\psi^i,\varpi_I,\tilde\psi_s),\label{eq:fn=Ba}
\end{align}
with the right-hand side of Eq.~\eqref{eq:selfforceorbit full} expanded as
\begin{align}
    a^i &= \e a^i_{(1)}(\psi^i,\varpi_I,\tilde\psi_s)+\e^2 a^i_{(2)}(\psi^i,\varpi_I,\tilde\psi_s) 
+{\cal O}(\e^3,s^2).\label{eq:acceleration expansion}
\end{align}
The coefficients $A^i_{\ j}$ and $B_{ij}$ can be read off Eqs.~(289)--(291) and (293) in Ref.~\cite{Pound:2021qin}, after dividing those equations by $dt/d\lambda=\mathscr{f}_t$.

Equations~\eqref{eq:dpsi/dt} and \eqref{eq:dpi/dt} are coupled to the spin precession phase, whose evolution equation~\eqref{eq:dpsi_s eqn} we can write as
\begin{equation}
    \frac{d\tilde\psi_s}{dt} = \omega_{s(0)}(\psi^j,\pi_j)+\e \omega_{s(1)}(\psi^j,\varpi_J) + {\cal O}(\e^2), \label{eq:dpsistilde/dt}
\end{equation}
where
\begin{align}
    \omega_{s(0)} = \frac{\omega_{21}}{u^t},\quad\text{and}\quad
    \omega_{s(1)} = \frac{\delta\omega_{21}}{u^t}\label{eq:omegas1}
\end{align}
with $\delta\omega_{21}$ given in Eq.~\eqref{eq:domegaAB}.

It is also useful to introduce geodesic action angles, which we denote $\mathring\psi^i_{(0)}$. For geodesics, these satisfy 
\begin{equation}\label{eq:geodesic action-angles}
\frac{d\mathring\psi^i_{(0)}}{dt}=\Omega^i_{(0)}(\pi_i) \qquad\text{(geodesic case)},
\end{equation}
growing exactly linearly in time. The frequencies $\Omega^i_{(0)}$ are an appropriate average of the `frequencies' $\omega^i_{(0)}$~\cite{Pound:2021qin}, such that $\mathring\psi^i_{(0)}$ represents the non-oscillatory part of $\psi^i$. Such angle variables and their frequencies were first derived in Ref.~\cite{Schmidt:2002qk}. We review them, and appeal to them in deriving some of our results, in Appendix~\ref{app:action-angles}.

\subsubsection{Averaging transformation}

We now transform to new variables $\{\mathring\psi^i,\mathring \varpi_I,{\mathring\psi}_s\}$ that contain no oscillations in their rates of change, meaning they satisfy equations of the form
\begin{subequations}\label{eq:ringed equations}
\begin{align}
    \frac{d\mathring\psi^i}{dt} &= \Omega^i_{(0)}(\mathring \pi_j)+\e\Omega^i_{(1)}(\mathring \varpi_J)+{\cal O}(\e^2),\label{eq:dpsitilde/dt}\\
    \frac{d\mathring \pi_i}{dt} &= \e F^{(0)}_i(\mathring \pi_j) + \e^2 F^{(1)}_i(\mathring \varpi_J)+{\cal O}(\e^3),\label{eq:dptilde/dt}\\
    \frac{d{\mathring{\psi}}_s}{dt} &= \Omega_{s(0)}(\mathring \pi_j)+\e\Omega_{s(1)}(\mathring \varpi_J)+{\cal O}(\e^2).\label{eq:dpsisttilde/dt}
\end{align}
\end{subequations}
The new variables are related to the old ones by an averaging transformation of the form
\begin{subequations}\label{eq:averaging transformation}
\begin{align}
    \psi^i &= \mathring\psi^i + \Delta\psi^i(\mathring\psi^j,\mathring \pi_j) + \e \delta \psi^i(\mathring\psi^j,\mathring \varpi_J,{\mathring\psi}_s) + {\cal O}(\e^2),\label{eq:psi to psitilde}\\
    \pi_i &= \mathring \pi_i + \e\delta \pi_i(\mathring\psi^j,\mathring\varpi_J,\mathring\psi_s) + {\cal O}(\e^2),\label{eq: pi to pitilde}\\
    \tilde\psi_s &= \mathring{\psi}_s + \Delta\tilde\psi_s(\mathring\psi^j,\mathring \pi_j) + \e \delta \psi_s(\mathring\psi^j,\mathring \varpi_J) + {\cal O}(\e^2),\label{eq:psis to psistilde}
\end{align}
\end{subequations}
where all functions are $2\pi$-periodic in $\mathring\psi^j$ and $\mathring\psi_s$.\footnote{If $\delta m_1$ and $\delta \chi_1$ are defined via integrals over the primary's perturbed horizon, they will generally contain oscillatory phase dependence which may be removed in the averaging transformation by defining the new variables $\mathring{\delta m_1}$ and $\mathring{\delta \chi_1}$. In practice, when identifying the corrections to the primary's parameters at the level of the field equations or by integrating the average horizon fluxes in Eqs.~\eqref{mdot} and \eqref{Sdot}, we work directly with the non-oscillatory parameters. Thus in this work it is sufficient to take $\mathring{\delta m_1}=\delta m_1$ and $\mathring{\delta \chi_1}=\delta \chi_1$.} The zeroth-order terms in this transformation are precisely the transformation to the geodesic action angles $\mathring\psi^i_{(0)}$ reviewed in Appendix~\ref{app:action-angles}.

The functions $\Delta\psi^i$, $\delta\psi^i$, $\delta \pi_i$, $\Delta\tilde\psi_s$, and $\delta\psi_s$ are chosen to remove oscillations from the equations of motion. They, along with the functions on the right-hand sides of the equations of motion~\eqref{eq:ringed equations}, can be determined (up to a residual freedom that we discuss below) by substituting Eq.~\eqref{eq:averaging transformation}, with Eq.~\eqref{eq:ringed equations}, into Eqs.~\eqref{eq:dpsi/dt}, \eqref{eq:dpi/dt}, and~\eqref{eq:dpsistilde/dt}. 

After these substitutions, the leading-order terms in Eqs.~\eqref{eq:dpsi/dt}, \eqref{eq:dpi/dt}, and~\eqref{eq:dpsistilde/dt} become  
\begin{subequations}\label{eq:0PA transformation eqns}
\begin{align}
    \Omega^i_{(0)}+\Omega^j_{(0)}\frac{\partial\Delta \psi^i}{\partial\mathring\psi^j} &= \omega^i_{(0)}(\psi^j_{(0)},\mathring\pi_j),\label{eq:psi0PA}\\
    \hspace{-5pt}F^{(0)}_i + \Omega^j_{(0)}\frac{\partial\delta \pi_i}{\partial\mathring\psi^j}  + \Omega_{s(0)}\frac{\partial\delta \pi_i}{\partial\mathring\psi_s}  &= f^{(0)}_i(\psi^j_{(0)},\mathring\varpi_j,\tilde\psi_{s(0)}),\label{eq:pi0PA}\\ \Omega_{s(0)}+\Omega^j_{(0)}\frac{\partial\Delta \tilde\psi_s}{\partial\mathring\psi^j} &= \omega_{s(0)}(\psi^j_{(0)},\mathring\pi_j), \label{eq:psis0PA}   
\end{align}
\end{subequations}
where
\begin{align}
\psi^i_{(0)}&\equiv\mathring\psi^i+\Delta\psi^i(\mathring\psi^j,\mathring\pi_j),\label{eq:psi0}\\
\tilde\psi_{s(0)}&\equiv\mathring\psi_s+\Delta\tilde\psi_s(\mathring\psi^j,\mathring\pi_j),\label{eq:psis0}
\end{align}
are the zeroth-order terms in Eq.~\eqref{eq:psi to psitilde} and~\eqref{eq:psis to psistilde}.

Averaging Eq.~\eqref{eq:0PA transformation eqns} over angles reveals that the leading term in each of the equations of motion~\eqref{eq:ringed equations} is simply the average of the corresponding term in Eqs.~\eqref{eq:dpsi/dt}, \eqref{eq:dpi/dt}, and~\eqref{eq:dpsistilde/dt}:
\begin{subequations}\label{eq:0PA terms}  
\begin{align}
    \Omega^i_{(0)}(\mathring \pi_j) &= \left\langle\omega^i_{(0)}(\psi^j_{(0)},\mathring \pi_j)\right\rangle,\label{eq:Omega0}\\
    F_i^{(0)}(\mathring \pi_j) &= \left\langle f_{i}^{(0)}(\psi^j_{(0)},\mathring \varpi_j,\tilde\psi_{s(0)})\right\rangle,\label{eq:F0}\\
    \Omega_{s(0)}(\mathring \pi_j) &= \left\langle\omega_{s(0)}(\psi^j_{(0)},\mathring \pi_j)\right\rangle, \label{eq:Omegas0}
\end{align}
\end{subequations}
where 
\begin{equation}\label{eq:angle average}
    \langle\cdot\rangle \equiv \frac{1}{(2\pi)^4}\oint \cdot \;d^4\mathring\psi
\end{equation}
is the average over $\mathring\psi^i$ and $\mathring\psi_s$. Note that $\mathring\psi^\phi$ does not appear in the equations of motion, meaning $\langle\cdot\rangle$ reduces to $\frac{1}{(2\pi)^3}\oint \cdot \,d\mathring\psi^rd\mathring\psi^\theta d\mathring\psi_s$ in the above expressions. We also note that the 0PA forcing function $F^{(0)}_i$ is independent of $\chi_2$, $\delta m_1$, and $\delta\chi_1$ despite the fact that $f^{(0)}_i$ depends on these quantities. This is because the first-order MPD force and the linear force due to $\delta m_1$ and $\delta\chi_1$ are purely conservative. We remind the reader of this property in more detail in Appendix~\ref{app:conservative/dissipative}.  

Equations~\eqref{eq:0PA terms} eliminate the non-oscillatory parts of Eqs.~\eqref{eq:dpsi/dt}, \eqref{eq:dpi/dt}, and~\eqref{eq:dpsistilde/dt}. The oscillatory parts then determine the functions $\Delta\psi^i$, $\delta\pi_i$, and $\Delta\tilde\psi_s$:
\begin{subequations}\label{eq:leading oscillations}
\begin{align}
    \Omega^j_{(0)}\frac{\partial\Delta \psi^i}{\partial\mathring\psi^j} &= \omega^i_{(0)} - \left\langle\omega^i_{(0)}\right\rangle,\label{eq:Dpsi}\\
    \Omega^j_{(0)}\frac{\partial\delta \pi_i}{\partial\mathring\psi^j}+\Omega_{s(0)}\frac{\partial\delta \pi_i}{\partial\mathring\psi_s} &= f^{(0)}_i - \left\langle f^{(0)}_i\right\rangle,\label{eq:dpi}\\
    \Omega^j_{(0)}\frac{\partial\Delta \tilde\psi_s}{\partial\mathring\psi^j} &= \omega_{s(0)} - \left\langle\omega_{s(0)}\right\rangle. \label{eq:Dpsis}
\end{align}
\end{subequations}

At the first subleading order, the averaged part of Eqs.~\eqref{eq:dpsi/dt}, \eqref{eq:dpi/dt}, and~\eqref{eq:dpsistilde/dt}, with Eq.~\eqref{eq:averaging transformation} and~\eqref{eq:ringed equations} substituted, yields the first subleading terms in Eqs.~\eqref{eq:dpsi/dt}, \eqref{eq:dpi/dt}, and~\eqref{eq:dpsistilde/dt}: 
\begin{subequations}\label{eq:1PA terms}
\begin{align}
    \Omega^i_{(1)} &= \left\langle\omega^i_{(1)}\right\rangle + \left\langle\delta \pi_j\frac{\partial\omega^i_{(0)}}{\partial\mathring \pi_j}+\delta \psi^j \frac{\partial\omega^i_{(0)}}{\partial\psi^j_{(0)}}\right\rangle \nonumber\\
    &\quad - F^{(0)}_j\frac{\partial\left\langle \Delta\psi^i\right\rangle}{\partial\mathring\pi_j},\label{eq:Omega1}\\
    F_i^{(1)} &= \left\langle f_i^{(1)}\right\rangle + \left\langle\delta \pi_j\frac{\partial f_i^{(0)}}{\partial\mathring \pi_j}+\delta \psi^j \frac{\partial f_i^{(0)}}{\partial\psi^j_{(0)}}\right\rangle\nonumber\\
    &\quad - F^{(0)}_j\frac{\partial\left\langle \delta\pi_i\right\rangle}{\partial\mathring\pi_j},\label{eq:F1}\\ 
    \Omega_{s(1)} &= \left\langle\omega_{s(1)}\right\rangle + \left\langle\delta \pi_j\frac{\partial\omega_{s(0)}}{\partial\mathring \pi_j}+\delta \psi^j \frac{\partial\omega_{s(0)}}{\partial\psi^j_{(0)}}\right\rangle\nonumber\\
    &\quad - F^{(0)}_j\frac{\partial\bigl\langle \Delta\tilde\psi_s\bigr\rangle}{\partial\mathring\pi_j},\label{eq:Omegas1}
\end{align}
\end{subequations}
where functions of $(\psi^i,\pi_i)$ on the right-hand side are evaluated at $(\psi^i_{(0)},\mathring\pi_i)$.

Equations~\eqref{eq:1PA terms} require $\delta\psi^i$, the subleading term in the transformation~\eqref{eq:psi to psitilde}.
This function is determined by the complete (not averaged) $\mathcal{O}(\e)$ part of Eq.~\eqref{eq:dpsi/dt}:
\begin{multline}
    \Omega^i_{(1)} + \Omega^j_{(1)}\frac{\partial\Delta \psi^i}{\partial\mathring\psi^j} 
    + F^{(0)}_j\frac{\partial\Delta \psi^i}{\partial\mathring\pi_j} + \Omega^j_{(0)}\frac{\partial\delta \psi^i}{\partial\mathring\psi^j}+\Omega_{s(0)}\frac{\partial\delta \psi^i}{\partial\mathring\psi_s}\\ 
    = \omega^i_{(1)}  +\delta \pi_j\frac{\partial\omega^i_{(0)}}{\partial\mathring \pi_j}+\delta \psi^j \frac{\partial\omega^i_{(0)}}{\partial\psi^j_{(0)}}.\label{eq:dpsi eqn}
\end{multline}
Equation~\eqref{eq:dpsi eqn} is complicated by the fact that $\Omega^i_{(1)}$ depends on $\delta\psi^i$ through Eq.~\eqref{eq:Omega1}. However, it admits a series solution for $\delta\psi^i$ of the form (325) in Ref.~\cite{Pound:2021qin}. Alternatively, we can find $\delta\psi^i$ using the geodesic action angles $\mathring\psi^i_{(0)}$ as a stepping stone. We present that method in Appendix~\ref{app:action-angles}.

Finally, the (omitted) order-$\e^2$ term in the transformation~\eqref{eq: pi to pitilde} and the order-$\e$ term in the transformation~\eqref{eq:psis to psistilde} are responsible for satisfying the oscillatory part of Eqs.~\eqref{eq:dpi/dt} and~\eqref{eq:dpsistilde/dt} at first subleading order. However, these terms in the transformations are not explicitly required at 1PA order because (unlike $\delta\psi^i$) they do not contribute to the evolution equations~\eqref{eq:ringed equations} at the given orders, nor (again, unlike $\delta\psi^i$) do they enter the field equations through second order.

Note that the criterion of removing oscillations only determines the oscillatory parts of the functions $\Delta\psi^i$, $\delta\pi_i$, and $\Delta\psi_s$ (and their higher-order analogues); their averages $\langle \Delta\psi^i\rangle$, $\langle \Delta\pi_i\rangle$, and $\langle \Delta\psi_s\rangle$ do not enter into equations such as~\eqref{eq:leading oscillations}. Hence, these non-oscillatory pieces of the transformation can be chosen as arbitrary functions of $\mathring\pi_i$ (or of $\mathring\varpi_J$ in the case of $\langle \Delta\pi_i\rangle$). However, such a choice affects the subleading terms in the equations of motion~\eqref{eq:dpsi/dt}, \eqref{eq:dpi/dt}, and~\eqref{eq:dpsistilde/dt}. Concretely, we see from Eqs.~\eqref{eq:1PA terms} that the non-oscillatory functions have the following contribution to the subleading frequency corrections and forcing functions:
\begin{subequations}\label{eq:1PA gauge transformation}
\begin{align}
    \Delta\Omega^i_{(1)} &= \left\langle\delta \pi_j\right\rangle\frac{\partial \Omega^i_{(0)}}{\partial\mathring\pi_j} - F^{(0)}_j\frac{\partial\left\langle \Delta\psi^i\right\rangle}{\partial\mathring\pi_j},\label{eq:DOmega1}\\
    \Delta F_i^{(1)} &= \left\langle\delta \pi_j\right\rangle\frac{\partial F_i^{(0)}}{\partial\mathring\pi_j}  - F^{(0)}_j\frac{\partial\left\langle \delta\pi_i\right\rangle}{\partial\mathring\pi_j},\\ 
    \Delta\Omega_{s(1)} &= \left\langle\delta \pi_j\right\rangle\frac{\partial\Omega_{s(0)}}{\partial\mathring\pi_j} - F^{(0)}_j\frac{\partial\bigl\langle \Delta\tilde\psi_s\bigr\rangle}{\partial\mathring\pi_j},\label{eq:DOmegas1}    
\end{align}
\end{subequations}
where we have used $\left\langle \langle \delta\pi_j\rangle \frac{\partial}{\partial\mathring\pi_j}\cdot\right\rangle = \langle \delta\pi_j\rangle\frac{\partial}{\partial\mathring\pi_j}\left\langle \cdot\right\rangle$ and Eqs.~\eqref{eq:0PA terms}. We return to this residual gauge freedom in later sections.

\subsection{Field equations}\label{sec:field equations}

In the multiscale expansion of the field equations, all time dependence is encoded in a dependence on the mechanical phase-space variables, such that $h_{\alpha\beta}=h_{\alpha\beta}(\mathring\psi^i,\mathring \varpi_I,\mathring\psi_s,x^i,\e)$~\cite{Pound:2021qin,Miller:2023ers}. The expansion at small $\e$ then becomes
\begin{equation}\label{eq:h multiscale}
    h_{\alpha\beta} = \e \mathring h^{(1)}_{\alpha\beta}(\mathring\psi^i,\mathring \varpi_I,x^i)+\e^2 \mathring h^{(2)}_{\alpha\beta}(\mathring\psi^i,\mathring \varpi_I,{\mathring\psi}_s,x^i)+\ldots,
\end{equation}
where all functions are $2\pi$-periodic in $\mathring\psi^i$ and $\mathring\psi_s$. 
The coefficients here are not identical to those in Eq.~\eqref{eq:metric}. Instead, each coefficient in Eq.~\eqref{eq:metric} must be re-expanded in multiscale form,
\begin{equation}
    h^{(n)}_{\alpha\beta}(t,x^i,\e) = \sum_{k=0}^\infty \e^{k}h^{(n,k)}_{\alpha\beta}(\mathring\psi^i(t,\e),\mathring\varpi_I(t,\e),\mathring\psi_s(t,\e),x^i),
\end{equation}
such that
\begin{align}
    \mathring h^{(1)}_{\alpha\beta}(\mathring\psi^i,\mathring \varpi_I,x^i) &= h^{(1,0)}_{\alpha\beta}(\mathring\psi^i,\mathring \varpi_I,x^i),\\
    \mathring h^{(2)}_{\alpha\beta}(\mathring\psi^i,\mathring \varpi_I,\mathring\psi_s,x^i) &= h^{(2,0)}_{\alpha\beta}(\mathring\psi^i,\mathring \varpi_I,\mathring\psi_s,x^i) \nonumber\\
    &\quad + h^{(1,1)}_{\alpha\beta}(\mathring\psi^i,\mathring \varpi_I,\mathring\psi_s,x^i),
\end{align}
and so on.\footnote{For simplicity we take $t$ to be Boyer-Lindquist time. However, our analysis throughout this paper applies for any choice of time coordinate $t$; for times other than Boyer-Lindquist, the only change is the formula for $\mathscr{f}_t$ in Eq.~\eqref{eq:f_t}. As long as the time coordinate reduces to Boyer-Lindquist time along the particle's trajectory, then even Eq.~\eqref{eq:f_t} remains unchanged. The multiscale expansion is usually formulated in terms of hyperboloidal time~\cite{Miller:2020bft,Miller:2023ers} for reasons explained in Refs.~\cite{Miller:2020bft,Cunningham:2024dog}, and hyperboloidal coordinates also bring other advantages~\cite{PanossoMacedo:2022fdi}.} 

When substituting the expansion~\eqref{eq:h multiscale} into the field equations, we apply the chain rule
\begin{subequations}\label{eq:chain rule}
    \begin{align}
    \frac{\partial}{\partial t} &= \frac{d\mathring\psi^i}{dt}\frac{\partial}{\partial\mathring\psi^i} + \frac{d\mathring\varpi_I}{dt}\frac{\partial}{\partial\mathring\varpi_I} +  \frac{d\mathring\psi_s}{dt}\frac{\partial}{\partial\mathring\psi_s} \\
    &= \Omega^i_{(0)}\frac{\partial}{\partial\mathring\psi^i} + \Omega_{s(0)}\frac{\partial}{\partial\mathring\psi_s} \nonumber\\
    &\quad + \e\left(\Omega^i_{(1)}\frac{\partial}{\partial\mathring\psi^i} + F^{(0)}_I \frac{\partial}{\partial\mathring\varpi_I} + \Omega_{s(1)}\frac{\partial}{\partial\mathring\psi_s}\right) + \mathcal{O}(\e^2),
\end{align}
\end{subequations}
having denoted $F^{(0)}_I=\{F^{(0)}_i, \dot{\cal E}^{(1)}_{\cal H},\dot{\cal L}^{(1)}_{\cal H}\}$. 
This implies expansions
\begin{align}
    \delta G_{\alpha\beta}(\mathring h^{(n)}) &= \delta G^{(0)}_{\alpha\beta}(\mathring h^{(n)}) + \e \delta G^{(1)}_{\alpha\beta}(\mathring h^{(n)}) + \mathcal{O}(\e^2),\\
    \delta^2 G_{\alpha\beta}(\mathring h^{(1)},\mathring h^{(1)}) &= \delta^2 G^{(0)}_{\alpha\beta}(\mathring h^{(1)},\mathring h^{(1)}) + \mathcal{O}(\e).
\end{align}
We can further decompose the field equations into the Fourier domain by expanding the metric perturbations in discrete Fourier series,
\begin{align}
    \mathring h^{(1)}_{\alpha\beta} &= \sum_{\bm{k}\in\mathbb{Z}^3}\mathring h^{(1,\bm{k})}_{\alpha\beta}(\mathring \varpi_I,x^i)e^{-i\mathring\psi_{\bm{k}}},\label{eq:h1 Fourier}\\
    \mathring h^{(2)}_{\alpha\beta} &=\sum_{\bm{k}\in\mathbb{Z}^3}\sum_{q\,=-1}^{+1} \mathring h^{(2,\bm{k},q)}_{\alpha\beta}(\mathring \varpi_I,x^i)e^{-i\mathring\psi_{\bm{k}}-iq\mathring\psi_s},\label{eq:h2 Fourier}
\end{align}
with 
\begin{equation}
    \bm{k} = k_i = (k_r,k_\theta,k_\phi)
\end{equation}
and
\begin{equation}
\mathring\psi_{\bm{k}}\equiv k_i \mathring\psi^i.
\end{equation}
Note the $i$ appearing in the exponential is the imaginary number, not to be confused with a spatial index. The precession phase only enters with mode numbers $q=0,\pm1$ because it only appears in the field equations through the sines and cosines in Eq.~\eqref{eq:S and dS}. Given these Fourier expansions, the chain rule~\eqref{eq:chain rule} becomes
\begin{align}
    \hspace{-5pt}\left(\frac{\partial}{\partial t}\right)_{\!\bm{k},q} 
    &= -i\bigl(k_i\Omega^i_{(0)} +q\,\Omega_{s(0)}\bigr) \nonumber\\
    &\quad + \e\left[F^{(0)}_I \frac{\partial}{\partial\mathring\varpi_I}-i\bigl(k_i\Omega^i_{(1)}+q\, \Omega_{s(1)}\bigr)\right] + \mathcal{O}(\e^2).\label{eq:ddt modes}
\end{align}
The operators $\delta^kG^{(n)}_{\alpha\beta}$ correspondingly become operators on individual Fourier coefficients $\mathring h^{(1,\bm{k})}_{\alpha\beta}$ and $\mathring h^{(2,\bm{k},q)}_{\alpha\beta}$:
\begin{align}
    \delta G^{(n)}_{\alpha\beta}(\mathring h^{(1)}) &= \sum_{\bm{k}\in\mathbb{Z}^3}\delta G^{(n,\bm{k})}_{\alpha\beta}(\mathring h^{(1,\bm{k})})e^{-i\mathring\psi_{\bm{k}}},\\ 
    \delta G^{(0)}_{\alpha\beta}(\mathring h^{(2)}) &= \sum_{\bm{k}\in\mathbb{Z}^3}\sum_{q\,=-1}^1\delta G^{(0,\bm{k},q)}_{\alpha\beta}(\mathring h^{(2,\bm{k},q)})e^{-i\mathring\psi_{\bm{k}}-iq\mathring\psi_s},
\end{align}
and
\begin{multline}
    \delta^2 G^{(0)}_{\alpha\beta}(\mathring h^{(1)},\mathring h^{(1)}) \\
    = \sum_{\bm{k}\in\mathbb{Z}^3}\sum_{\bm{k}'\in\mathbb{Z}^3} \delta^2 G^{(0,\bm{k},\bm{k}')}_{\alpha\beta}(\mathring h^{(1,\bm{k}')},\mathring h^{(1,\bm{k}'')})e^{-i\mathring\psi_{\bm{k}}}
\end{multline}
with $\bm{k}''=\bm{k}-\bm{k}'$. The linear operator $\delta G^{(0,\bm{k})}_{\alpha\beta}$ is identical to the familiar linearized Einstein tensor in the frequency domain, with frequency $\omega^{(0)}_{\bm{k}}=k_i\Omega^i_{(0)}$. We refer to  Refs.~\cite{Miller:2020bft,Miller:2023ers} for more detailed explorations of multiscale expansions of the Einstein field equations.

To fully expand the field equations, we must also expand the stress-energy tensor. 
We write the monopole term~\eqref{eq:SEmono} as
\begin{equation}\label{eq:SEmono multiscale}
    T^{(m)}_{\alpha \beta}(\mathring\psi^i,\mathring\varpi_I,\mathring\psi_s,x^i,\e) = m_2 \frac{\hat g_{\alpha\mu}\hat g_{\beta\nu}\dot z^\mu \dot z^\nu}{\sqrt{-\hat g_{\rho\sigma}\dot z^\rho \dot z^\sigma}} \frac{\delta^3(x^i-z^i)}{\sqrt{-\hat g}}
\end{equation}
after evaluating the integral and using
\begin{equation}
\frac{dt}{d\hat\tau} = \frac{1}{\sqrt{-\hat g_{\alpha\beta}\dot z^\alpha \dot z^\beta}}.
\end{equation}
In Eq.~\eqref{eq:SEmono multiscale}, the effective metric, trajectory, and velocity are
\begin{align}
\hat g_{\alpha\beta} &= g_{\alpha\beta}(x^i)+\e \mathring h^{{\rm R}(1)}_{\alpha\beta}(\mathring\psi^i,\mathring\varpi_I,x^i) +\mathcal{O}(\e^2),\\
z^i &= z^i_{(0)}(\psi^i_{(0)},\mathring\pi_i) + \e z^i_{(1)}(\psi^i_{(0)},\mathring\varpi_I,\mathring\psi_s) + \mathcal{O}(\e^2),\label{eq:z multiscale}\\
\dot z^i &= v^i_{(0)}(\psi^i_{(0)},\mathring\pi_i) + \e v^i_{(1)}(\psi^i_{(0)},\mathring\varpi_I,\mathring\psi_s) + \mathcal{O}(\e^2),\label{eq:zdot multiscale}
\end{align}
where $\psi^i_{(0)}$ is given in Eq.~\eqref{eq:psi0}. The coefficients $z^i_{(n)}$ and $v^i_{(n)}$ are obtained from Eqs.~\eqref{eq:z(psi,p)} and~\eqref{eq:zdot(psi,p)} with the expansions~\eqref{eq:averaging transformation}. Concretely, at leading order,
\begin{subequations}
\begin{align}
    r_{(0)} &= \frac{\mathring p\, m_1}{1+\mathring e \cos\psi^r_{(0)}},\\
    \cos\theta_{(0)} &= \sin\mathring\iota \cos\psi^\theta_{(0)},
\end{align}
\end{subequations}
$\phi_{(0)}=\psi^\phi_{(0)}$, and
\begin{equation}
    v^i_{(0)} = \frac{\partial z^i_{(0)}}{\partial \psi^j_{(0)}}\omega^j_{(0)}(\psi^k_{(0)},\mathring\pi_k).
\end{equation}
We emphasize that this leading-order trajectory is not a geodesic of the background spacetime; it would only be a geodesic if $\mathring\pi_i$ were constant and $\mathring\psi^i$ were exactly linear in $t$. At the next order,
\begin{align}
z^i_{(1)} &= \delta\psi^j\frac{\partial z^i_{(0)}}{\partial \psi^j_{(0)}} + \delta \pi_i\frac{\partial z^i_{(0)}}{\partial \mathring\pi_j},\label{eq:z1}\\
v^i_{(1)} &= \delta\psi^j\frac{\partial v^i_{(0)}}{\partial \psi^j_{(0)}} + \delta \pi_i\frac{\partial v^i_{(0)}}{\partial \mathring\pi_j}.\label{eq:v1}
\end{align}
We also have the trivial identities 
\begin{equation}
v^t_{(0)}=1 \quad \text{and} \quad v^t_{(n>0)}=0.
\end{equation}

These expansions of $\hat g_{\mu\nu}$, $z^i$, and $\dot z^i$ imply
\begin{multline}
    T^{(m)}_{\alpha \beta} = \e T^{(m,0)}_{\alpha \beta}(\mathring\psi^i,\mathring\pi_i,x^i) \\
    + \e^2 T^{(m,1)}_{\alpha \beta}(\mathring\psi^i,\mathring\varpi_I,\mathring\psi_s,x^i) + \mathcal{O}(\e^3)
\end{multline}
with
\begin{align}
    T^{(m,0)}_{\alpha \beta} &= m_2 \frac{ g_{\alpha\mu} g_{\alpha\nu}v^\mu_{(0)}v^\nu_{(0)}}{\sqrt{- g_{\rho\sigma}v^\rho_{(0)} v^\sigma_{(0)}}} \frac{\delta^3(x^i-z^i_{(0)})}{\sqrt{-g}},\label{eq:Tm0}\\
    T^{(m,1)}_{\alpha \beta} &=\left(z^i_{(1)}\frac{\partial}{\partial z^i_{(0)}} + v^i_{(1)}\frac{\partial}{\partial v^i_{(0)}} +h^{\rm R(1)}_{\mu\nu}\frac{\partial}{\partial  g_{\mu\nu}}\right)T^{(m,0)}_{\alpha \beta},\label{eq:Tm1}
\end{align}
where we note $\frac{\partial \sqrt{- g}}{\partial g_{\mu\nu}} = \frac{1}{2}\sqrt{- g}\,g^{\mu\nu}$.

Similarly, the dipole term~\eqref{eq:SEdipole} is expanded as 
\begin{equation}
T^{(d)}_{\alpha\beta}=\e^2T^{(d,0)}_{\alpha\beta}(\mathring\psi^i,\mathring\pi_i,\mathring\psi_s,x^i)+\mathcal{O}(\e^3) 
\end{equation}
with
\begin{equation}
    T^{(d,0)}_{\alpha \beta} = (m_2)^2\,  g_{\alpha\mu}g_{\beta\nu}\nabla_{\rho} \left(\frac{\delta^3\bigl(x^i- z^i_{(0)}\bigr)}{\sqrt{-g}} v^{(\mu}_{(0)}S^{\nu)\rho}\right). \label{eq:Td0}
\end{equation}
The total stress-energy tensor is hence
\begin{equation}
    T_{\alpha\beta} = \e T^{(1)}_{\alpha\beta}(\mathring\psi^i,\mathring\pi_i,x^i) + \e^2 T^{(2)}_{\alpha\beta}(\mathring\psi^i,\mathring\varpi_i,\mathring\psi_s,x^i) + \mathcal{O}(\e^3),
\end{equation}
where
\begin{align}
 T_{\alpha \beta}^{(1)} &= T^{(m,0)}_{\alpha \beta},\\
 T_{\alpha \beta}^{(2)} &= T^{(m,1)}_{\alpha \beta} + T^{(d,0)}_{\alpha \beta}.
\end{align}
These are decomposed into Fourier modes using
\begin{equation}
    T_{\alpha \beta}^{(n,\bm{k},q)} = \frac{1}{(2\pi)^4}\oint T^{(n)}_{\alpha \beta}e^{i\mathring\psi_{\bm{k}}+iq\mathring\psi_s} d^4\mathring\psi.\label{eq:Tnkq}
\end{equation}

Given the expansions of the Einstein tensor and stress-energy tensor, the Einstein equation~\eqref{eq:EFE expanded} divides into a hierarchical set of first- and second-order equations for the Fourier mode coefficients $\mathring h^{(1,\bm{k})}_{\alpha\beta}$ and $\mathring h^{(2,\bm{k},q)}_{\alpha\beta}$: 
\begin{align}
    \delta G^{(0,\bm{k})}_{\mu\nu}(\mathring h^{(1,\bm{k})}) &= 8\pi T^{(1,\bm{k})}_{\mu\nu},\label{eq:EFE1 multiscale}\\
    \delta G^{(0,\bm{k})}_{\mu\nu}(\mathring h^{(2,\bm{k},q)}) &=
    8\pi T^{(2,\bm{k},q)}_{\mu\nu}
    - \delta^2 G^{(0,\bm{k})}_{\mu\nu}(\mathring h^{(1)},\mathring h^{(1)})\nonumber\\
    &\quad - \delta G^{(1,\bm{k})}_{\mu\nu}(\mathring h^{(1,\bm{k})}),\label{eq:EFE2 multiscale}
\end{align}
where 
\begin{equation}
\delta^2 G^{(0,\bm{k})}_{\mu\nu}(\mathring h^{(1)},\mathring h^{(1)})=\sum_{\bm{k}'\in\mathbb{Z}^3}\delta^2 G^{(0,\bm{k},\bm{k}')}_{\mu\nu}(\mathring h^{(1,\bm{k}')},\mathring h^{(1,\bm{k}-\bm{k}')}).
\end{equation}
The field equations~\eqref{eq:EFE1 multiscale} and \eqref{eq:EFE2 multiscale} are (elliptic) partial differential equations in $x^i$ for each Fourier coefficient. They can be converted to Teukolsky equations as described in Refs.~\cite{Spiers:2023cip,Spiers:2023mor}. The first-order equation, Eq.~\eqref{eq:EFE1 multiscale}, is identical to the standard frequency-domain Einstein equation for a point particle on a geodesic orbit (though we once again emphasize that the leading-order trajectory $z^i_{(0)}$ is not a geodesic).

Equations~\eqref{eq:EFE1 multiscale} and \eqref{eq:EFE2 multiscale} can be solved on a grid of $\mathring\pi_i$ (and background spin) values. Equation~\eqref{eq:EFE1 multiscale} is first solved for $\mathring h^{(1,\bm{k})}_{\alpha\beta}$. The solutions $\mathring h^{(1,\bm{k})}_{\alpha\beta}$ are then used in the calculation of the source terms in Eq.~\eqref{eq:EFE2 multiscale}, both directly, through the sources $\delta^2 G^{(0,\bm{k})}_{\mu\nu}$ and $\delta G^{(1,\bm{k})}_{\mu\nu}$, and indirectly through the calculation of the forcing functions $F^{(0)}_i$ and frequency corrections $\Omega^i_{(1)}$ (which enter into $\delta G^{(1,\bm{k})}_{\mu\nu}$ and $T^{(2,\bm{k},0)}_{\alpha\beta}$). From $\mathring h^{(1,\bm{k})}_{\alpha\beta}$ and $\mathring h^{(2,\bm{k},q)}_{\alpha\beta}$, one can then compute the 1PA forcing functions $F^{(1)}_i$. Finally, the waveform amplitudes are read off the asymptotic values of $\mathring h^{(1,\bm{k})}_{\alpha\beta}$ and $\mathring h^{(2,\bm{k},q)}_{\alpha\beta}$. Using this precomputed data, Eqs.~\eqref{eq:ringed equations} can be solved to evolve the system through parameter space and generate a waveform, as described in the Introduction. We return to the waveform generation scheme in Sec.~\ref{sec:waveform generation}.

Finally, we note that the expansion~\eqref{eq:acceleration expansion} of the acceleration used in Sec.~\ref{sec:orbital motion} assumes the force is given in the form $a^i(\psi^j,\varpi_I,\tilde\psi_s,\e)$. If we directly substitute the multiscale expansion~\eqref{eq:h multiscale} into Eq.~\eqref{eq:selfforceorbit full}, we instead obtain the force in the form $a^i(\psi^j,\pi_j,\tilde\psi_s,\e;\mathring\psi^j,\mathring\varpi_j,\mathring\psi_s)$, where the dependence on $(\psi^j,\pi_j,\tilde\psi_s)$ arises from evaluating $\mathring h^{{\rm R}(n)}_{\alpha\beta}(\mathring\psi^i,\mathring\varpi_i,\mathring\psi_s,x^i)$ (and other fields appearing in $a^i$) at $x^i=z^i$, as well as from evaluating the four-velocity and spin in terms of the osculating elements and phases. The arguments after the semicolon, on the other hand, arise from the first three arguments in $\mathring h^{{\rm R}(n)}_{\alpha\beta}$.

To accommodate this mixed form of the acceleration, one can treat the expansion~\eqref{eq:acceleration expansion} as an expansion of $a^i(\psi^j,\pi_j,\tilde\psi_s,\e;\mathring\psi^j,\mathring\varpi_j,\mathring\psi_s)$ holding all arguments (other than $\e$) fixed. When calculating derivatives such as $\partial f^{(0)}_i/\partial\mathring\pi_j$ in Eq.~\eqref{eq:F1}, one must note that these terms arise from substituting the expansions~\eqref{eq:averaging transformation} into the first three arguments of $a^i(\psi^j,\pi_j,\tilde\psi_s,\e;\mathring\psi^j,\mathring\varpi_j,\mathring\psi_s)$. The derivative  $\partial f^{(0)}_i/\partial\mathring\pi_j$ in Eq.~\eqref{eq:F1} then involves $\frac{\partial}{\partial\mathring\pi_j}a^i_{(1)}(\psi^i_{(0)},\mathring\pi_i,\tilde\psi_{s(0)};\mathring\psi^i,\mathring\varpi_i,\mathring\psi_s)$, where the derivative only acts on the second argument.

\subsection{Linear-in-spin effects}

In Secs.~\ref{sec:orbital motion} and \ref{sec:field equations}, we have kept the treatment generic, without highlighting spin contributions or explicitly dropping quadratic-in-spin terms. We now isolate the linear spin terms.

We first define $a^i_{(n\text{-}\chi_2)}$, the secondary spin's linear contribution to the coefficients $a^i_{(n)}$ in the expansion~\eqref{eq:acceleration expansion}. Referring to the complete acceleration~\eqref{eq:selfforceorbit full}, we read off 
\begingroup
\allowdisplaybreaks
\begin{align}
a^\mu_{(1\text{-}\chi_2)} &= -\frac{m_2}{2}R^{\mu}{}_{\alpha \beta \gamma}u^{\alpha}  S^{\beta \gamma},\\
a^\mu_{(2\text{-}\chi_2)} &= -\frac{1}{2} P^{\mu\nu}\left(2 \mathring h_{\nu \rho ; \sigma}^{\mathrm{R}(2\text{-}\chi_2)}-\mathring h_{\rho \sigma ; \nu}^{\mathrm{R}(2\text{-}\chi_2)}\right) u^{\rho} u^{\sigma}\nonumber\\*
&\quad +\frac{m_2}{4}R^{\mu}{}_{\alpha \beta \gamma}\mathring h^{\mathrm{R}(1)}_{\rho\sigma}u^\rho u^\sigma u^{\alpha}  S^{\beta \gamma} \nonumber\\*
&\quad -\frac{m_2}{2}R^{\mu}{}_{\alpha \beta \gamma}u^{\alpha} \delta S^{\beta \gamma} \nonumber\\*
&\quad +\frac{m_2}{2} P^{\mu\nu}\left(2\mathring h^{\mathrm{R}(1)}_{\nu(\alpha;\beta)\gamma}-\mathring h^{\mathrm{R}(1)}_{\alpha\beta;\nu\gamma}\right)u^{\alpha} S^{\beta \gamma}\label{eq:a2-chi}
\end{align}
\endgroup
where all fields are evaluated at $z^i$ (rather than $z^i_{(0)}$, for example). Here we have used the `mixed' form of the acceleration described at the end of the previous section.

As discussed below Eq.~\eqref{eq:angle average}, these accelerations do not contribute to any of the leading terms ($\Omega^i_{(0)}$, $F^{(0)}_i$, and $\Omega_{s(0)}$) in the evolution equations~\eqref{eq:dpsitilde/dt}--\eqref{eq:dpsisttilde/dt}. From Eqs.~\eqref{eq:Omega1}--\eqref{eq:Omegas1}, we can read off the linear spin contributions to the subleading terms in the evolution equations: 
\begin{align}
    \Omega^i_{(1\text{-}\chi_2)} &= \left\langle A^i_{\ j} a^j_{(1\text{-}\chi_2)}\right\rangle \nonumber\\
    &\quad + \left\langle\delta \pi^{(\chi_2)}_j\frac{\partial\omega^i_{(0)}}{\partial\mathring \pi_j}+
    \delta \psi^j_{(\chi_2)} \frac{\partial\omega^i_{(0)}}{\partial\psi^j_{(0)}}\right\rangle,\label{eq:Omega1-chi}\\
    F_i^{(1\text{-}\chi_2)} &= \left\langle B_{ij}a^j_{(2\text{-}\chi_2)}\right\rangle - F^{(0)}_j\frac{\partial\left\langle \delta\pi_i^{(\chi_2)}\right\rangle}{\partial\mathring\pi_j}\nonumber\\
    &\quad +\left\langle\delta \pi^{(\chi_2)}_j\frac{\partial f_i^{(0\text{-1SF})}}{\partial\mathring \pi_j}+
    \delta \psi^j_{(\chi_2)} \frac{\partial f_i^{(0\text{-1SF})}}{\partial\psi^j_{(0)}}\right\rangle \nonumber\\
    &\quad + \left\langle\delta \pi^{(\rm 1SF)}_j\frac{\partial f_i^{(0\text{-}\chi_2)}}{\partial\mathring \pi_j}+\delta \psi^j_{(\rm 1SF)} \frac{\partial f_i^{(0\text{-}\chi_2)}}{\partial\psi^j_{(0)}}\right\rangle\label{eq:F1-chi2}.
\end{align}
Here we have used Eqs.~\eqref{eq:omega1=Aa} and~\eqref{eq:fn=Ba} to make the forces explicit, and we observed that the linear spin contribution to $\Omega_{s(1)}$ can be neglected as an overall $\mathcal{O}(s^2)$ effect in the dynamics, analogous to the linear spin contribution to $\delta\omega_{AB}$, which we neglected in Eq.~\eqref{eq:domegaAB}. The quantity $\delta\pi^{(\chi_2)}_i$ in Eqs.~\eqref{eq:Omega1-chi} and \eqref{eq:F1-chi2} is extracted from the linear spin terms in Eq.~\eqref{eq:dpi}, 
and $\delta\psi_{(\chi_2)}^i$ can be extracted from the linear spin terms in either Eq.~\eqref{eq:dpsi eqn} or Eq.~\eqref{eq:dpsi soln}. Quantities labeled with `1SF' are calculated from the first-order self-force.

There are several immediate takeaways from these formulas:
\begin{enumerate}
    \item Since the expressions are linear in the secondary spin, the average over the precession phase entirely eliminates contributions from the orthogonal, precessing components of the spin. In other words, in calculating the right-hand sides of the evolution equations~\eqref{eq:ringed equations}, one can replace $S^\alpha$ and $\delta S^\alpha$ with the precession-averaged spin vectors $\langle S^\alpha\rangle_{\tilde\psi_s}$ and $\langle\delta S^\alpha\rangle_{\tilde\psi_s}$ given in Eq.~\eqref{eq:<S>}. This irrelevance of the precession phase in the 1PA orbital dynamics has been pointed out many times previously (e.g., \cite{Witzany:2019nml,Skoupy:2023lih}). However, one can still consistently include the precession's modulation effect on the waveform at this order. We return to this last point at the end of this section and in Sec.~\ref{sec:waveform generation}.
    \item The correction $\Omega^i_{(1\text{-}\chi_2)}$ to the orbital frequency is solely due to the precession-averaged MPD force,
    \begin{equation}
        \langle a^\mu_{(1\text{-}\chi_2)}\rangle_{\tilde\psi_s} = -\frac{m_2}{2}R^{\mu}{}_{\alpha \beta \gamma}u^{\alpha} \langle S^{\beta \gamma}\rangle_{\tilde\psi_s},
    \end{equation}
    and to gauge freedom. Using Eqs.~\eqref{eq:dOmega0} and \eqref{eq:Omega1 alt}, we can also write this frequency correction as
    \begin{align}
        \Omega^i_{(1\text{-}\chi_2)} &= \Bigl\langle \delta\pi_j^{(\chi_2)}\Bigr\rangle\frac{\partial\Omega^i_{(0)}}{\partial\mathring\pi_j} \nonumber\\
        &\quad +\left\langle\frac{\partial\mathring\psi^i_{(0)}}{\partial\psi^j_{(0)}}A^j_{\ k}a^k_{(1\text{-}\chi_2)}\right\rangle \nonumber\\
        &\quad +\left\langle\frac{\partial\mathring\psi^i_{(0)}}{\partial\mathring\pi_j}B_{jk}a^k_{(1\text{-}\chi_2)}\right\rangle,
    \end{align}
    where we recall that $\Bigl\langle \delta\pi_j^{(\chi_2)}\Bigr\rangle$ is freely specifiable. This frequency correction was first computed in Ref.~\cite{Witzany:2019nml}. It is given in closed, analytical form (for a specific gauge choice) in Ref.~\cite{Witzany:2024ttz} in agreement with the numerical calculations of Ref.~\cite{Drummond:2022efc}. We return to this frequency correction in Secs.~\ref{sec:gauge choices}, \ref{sec:waveform gauge invariance}, and \ref{sec:flux balance}.
    \item The spin nutation only contributes to $F^{(1\text{-}\chi_2)}_i$, through the force
    \begin{equation}
    a^\mu_{(2\text{-}\delta S)}=-\frac{m_2}{2}R^{\mu}{}_{\alpha \beta \gamma}u^{\alpha} \langle \delta S^{\beta \gamma}\rangle_{\tilde\psi_s}
    \end{equation}
    that enters in the first term on the right-hand side of Eq.~\eqref{eq:F1-chi2}. In  Appendix~\ref{sec:localforces}, we show that in fact, the effects of this force in the 1PA orbital dynamics can be computed without solving the nutation equations.
    \item Each term in $F^{(1\text{-}\chi_2)}_i$ arises from an interaction between the secondary spin and the first-order regular field at the particle. However, as alluded to in the Introduction, none of these local terms need to be evaluated to calculate $F^{(1\text{-}\chi_2)}_i$ in practice; thanks to recent results in Refs.~\cite{Grant:2024ivt,Witzany:2024ttz}, Eq.~\eqref{eq:F1-chi2} can be replaced with an expression in terms of asymptotic fluxes. We summarize this in Sec.~\ref{sec:flux balance}.
\end{enumerate}

To complete the summary of first-order spin effects, we now turn to the field equations. The spin only enters the field equations~\eqref{eq:EFE1 multiscale} and \eqref{eq:EFE2 multiscale} in two simple ways: (i) through the spin-dipole stress-energy tensor $T^{(d,0)}_{\alpha\beta}$ in Eq.~\eqref{eq:Td0}; (ii) through the spin's contribution to $T^{(m,1)}_{\alpha\beta}$ in Eq.~\eqref{eq:Tm1}, which arises from the MPD spin force's contribution to $z^i_{(1)}$ and $v^i_{(1)}$ in Eqs.~\eqref{eq:z1} and \eqref{eq:v1}; and (iii) through contributions $\Omega^i_{(1\text{-}\chi_2)}$ and $\Omega_{s(1\text{-}\chi_2)}$ to Eq.~\eqref{eq:chain rule}, which contributes a term $\delta G^{(1\text{-}\chi_2,\bm{k})}_{\mu\nu}(\mathring h^{(1,\bm{k})})$ to Eq.~\eqref{eq:EFE2 multiscale}. 

We write the spin's total contribution to the stress-energy tensor as
\begin{equation}
    T^{(2\text{-}\chi_2)}_{\alpha\beta} = T^{(d,0)}_{\alpha\beta}+T^{(m,1\text{-}\chi_2)}_{\alpha\beta}
\end{equation}
with
\begin{equation}
    T^{(m,1\text{-}\chi_2)}_{\alpha \beta} =\left(z^i_{(1\text{-}\chi_2)}\frac{\partial}{\partial z^i_{(0)}} + v^i_{(1\text{-}\chi_2)}\frac{\partial}{\partial v^i_{(0)}}\right)T^{(m,0)}_{\alpha \beta}.
\end{equation}
The only term involving $\chi_2$ in the field equation is thus
\begin{multline}\label{eq:EFE2 spin terms}
    \delta G^{(0,\bm{k})}_{\mu\nu}\bigl(\mathring h^{(2\text{-}\chi_2,\bm{k},q)}\bigr) =
    8\pi T^{(2\text{-}\chi_2,\bm{k},q)}_{\mu\nu} \\- \delta G^{(1\text{-}\chi_2,\bm{k})}_{\mu\nu}(\mathring h^{(1,\bm{k})}),
\end{multline}
where $h^{(2\text{-}\chi_2)}_{\alpha\beta}$ is the metric perturbation sourced by the spin. The first source term on the right-hand side is confined to the libration region containing the particle's orbit. The second source term, arising from $\Omega^i_{(1\text{-}\chi_2)}$ and $\Omega_{s(1\text{-}\chi_2)}$ and Eq.~\eqref{eq:chain rule}, is distributed over the entire spacetime. As highlighted in Ref.~\cite{Mathews:2021rod} and discussed in the next section, there is considerable advantage in choosing a (phase-space) gauge that eliminates this non-compact term.

The solution to Eq.~\eqref{eq:EFE2 spin terms} enters the asymptotic waveform in two ways. First, it contributes to the first term on the right-hand side of the local force~\eqref{eq:a2-chi}, thereby contributing to the first term on the right-hand side of the 1PA forcing function~\eqref{eq:F1-chi2}. Second, it contributes directly to the second-order waveform mode amplitudes. We return to these two contributions in Secs.~\ref{sec:waveform generation} and \ref{sec:flux balance}.

\subsection{Gauge choices}
\label{sec:gauge choices}

As noted in Sec.~\ref{sec:orbital motion}, our multiscale expansion admits a residual gauge freedom on the orbital phase space, corresponding to the choice of non-oscillatory terms $\langle \Delta\psi^i\rangle$, $\langle \delta\pi_i\rangle$, and $\langle \Delta\psi_s\rangle$ in the transformations~\eqref{eq:averaging transformation}. Under different choices of these functions, subleading terms in the evolution equations change according to Eq.~\eqref{eq:1PA gauge transformation}. 

Different choices of these functions correspond to different choices of what we hold fixed when we vary $\e$. In this section, we describe three convenient gauge choices. Previous discussions along these lines can be found in Refs.~\cite{VanDeMeent:2018cgn,Drummond:2022efc,Drummond:2023wqc,Piovano:2024yks}, for example.

In Sec.~\ref{sec:waveform gauge invariance}, we explain how the final 1PA waveform is invariant under such choices. We also emphasize that these choices can nevertheless affect the accuracy of the waveform model.

\subsubsection{Fixed frequencies and fixed turning points}

We first consider the gauge choice adopted in Ref.~\cite{Pound:2021qin} for generic accelerated orbits. This gauge choice is defined by the properties 
\begin{enumerate}
    \item the phases $\mathring\psi^i$, which are angular coordinates on a torus in phase space, share an origin with the phases $\psi^i$. Concretely, $\mathring\psi^i(\psi^j=0,\mathring\pi_j)=0$, meaning $\mathring{\psi}^r=0$ corresponds to periapsis, and $\mathring\psi^\theta=0$ corresponds to maximum inclination (for orbits with fixed $\mathring\pi_i$). 
    \item $\mathring\pi_i$ is geodesically related to the physical frequencies 
    \begin{align}
    \Omega^i&\equiv \frac{d\mathring\psi^i}{dt}.
    \end{align}
    In other words, in this gauge, $\Omega^i=\Omega^i_{(0)}(\mathring \pi_i)$.
    \item \emph{either} $\mathring\pi_i$ is geodesically related to $\Omega_s\equiv \frac{d\mathring\psi_s}{dt}$, making $\Omega_s=\Omega_{s(0)}(\mathring \pi_i)$, \emph{or} $\mathring\psi_s$ shares an origin with $\tilde\psi_s$.
\end{enumerate}
We stress that our condition of `fixed turning points' is unrelated to the condition with the same name in Ref.~\cite{Drummond:2022efc}; ours is a condition on the values of the radial and polar phases $\mathring\psi^r$ and $\mathring\psi^\theta$ at turning points (periapsis and maximum inclination), while Ref.~\cite{Drummond:2022efc}'s is a condition on the values of Boyer-Lindquist radius $r$ and polar angle $\theta$ at turning points.

The first condition represents a choice of origin for the phases on the tori of constant $\mathring\pi_i$. Recalling Eq.~\eqref{eq:psi0}, we see the condition implies 
\begin{equation}\label{eq:fixed turning points}
\Delta\psi^i(0,\mathring\pi_i) = 0.
\end{equation}
To turn that into a condition on $\langle \Delta\psi^i\rangle$, we divide $\Delta\psi^i$ into its average $\langle \Delta\psi^i\rangle$ and a purely oscillatory part, $\Delta\psi^i_{\rm osc} \equiv \Delta\psi^i - \langle \Delta\psi^i\rangle$. Equation~\eqref{eq:fixed turning points} then implies
\begin{equation}
\langle \Delta\psi^i\rangle = -\Delta\psi^i_{\rm osc}(0,\mathring\pi_i).
\end{equation}

Next, to enforce the condition $\Omega^i=\Omega^i_{(0)}(\mathring \pi_i)$, we choose $\langle \delta \pi_i\rangle$ to eliminate the frequency corrections $\Omega^i_{(1)}$. From Eq.~\eqref{eq:Omega1}, this requires
\begin{multline}\label{eq:dpi fixed Omega}
    \langle\delta \pi_j\rangle\frac{\partial\Omega^i_{(0)}}{\partial\mathring \pi_j} =-\left\langle\omega^i_{(1)}\right\rangle 
     + F^{(0)}_j\frac{\partial\left\langle \Delta\psi^i\right\rangle}{\partial\mathring\pi_j}\\
     - \left\langle\delta \pi^{\rm osc}_j\frac{\partial\omega^i_{(0)}}{\partial\mathring \pi_j} + \delta \psi^j_{\rm osc} \frac{\partial\omega^i_{(0)}}{\partial\psi^j_{(0)}}\right\rangle,    
\end{multline}
where we have split $\delta\pi_i$ and $\delta\psi^i$ into averaged and oscillatory pieces, noted $\langle\omega^i_{(0)}\rangle=\Omega^i_{(0)}$, and further noted that the product of an oscillatory function with a non-oscillatory one averages to zero. The linear spin term in Eq.~\eqref{eq:dpi fixed Omega} is
\begin{align}\label{eq:dpi-chi2 fixed Omega}
    \left\langle\delta \pi^{(\chi_2)}_j\right\rangle\frac{\partial\Omega^i_{(0)}}{\partial\mathring \pi_j} &=-\left\langle\omega^i_{(1\text{-}\chi_2)}\right\rangle \nonumber\\
     &\quad - \left\langle\delta \pi^{(\chi_2)}_j\frac{\partial\omega^i_{(0)}}{\partial\mathring \pi_j}+\delta \psi^j_{(\chi_2)} \frac{\partial\omega^i_{(0)}}{\partial\psi^j_{(0)}}\right\rangle.    
\end{align}

Equation~\eqref{eq:dpi fixed Omega} has a unique solution so long as the matrix $\frac{\partial\Omega^i_{(0)}}{\partial\mathring \pi_j}$ is invertible. This means the fixed-frequencies gauge breaks down at (measure-zero) degenerate surfaces in parameter space where the frequencies fail to be good coordinates~\cite{Warburton:2013yj}.

Having fixed $\langle\delta\pi_i\rangle$ and $\langle\tilde\Delta\psi^i\rangle$, we are only left with $\langle\Delta\psi^i\rangle$. This means we do not have the freedom to simultaneously enforce $\Omega_{s(1)}=0$ and $\Delta\tilde\psi_s(0,\mathring\pi_i)=0$ (unless the oscillatory part of $\Delta\tilde \psi_s$ is an odd function of $\psi^i$, in which case it automatically vanishes at $\psi^i=0$). If we choose to eliminate $\Omega_{s(1)}$, then from Eq.~\eqref{eq:Omegas1} we see that $\left\langle \Delta\tilde\psi_s\right\rangle$ must satisfy the following partial differential equation:
\begin{multline}
    F^{(0)}_j\frac{\partial\bigl\langle \Delta\tilde\psi_s\bigr\rangle}{\partial\mathring\pi_j} = \left\langle\omega_{s(1)}^{(\rm 1SF)}\right\rangle + \left\langle\delta \pi^{(\rm 1SF)}_j\frac{\partial\omega_{s(0)}}{\partial\mathring \pi_j}\right\rangle\\
    +\left\langle\delta \psi^j_{(\rm 1SF)}\frac{\partial\omega_{s(0)}}{\partial\mathring\psi^j}\right\rangle.
\end{multline}
Here we have emphasized that these terms come solely from the first-order regular field because we neglect linear spin corrections to the spin precession.

Assuming we eliminate $\Omega_{s(1)}$, the evolution equations away from the degenerate surfaces become
\begin{subequations}
\begin{align}
    \frac{d\mathring\psi^i}{dt} &= \Omega^i_{(0)}(\mathring \pi_j),\\
    \frac{d\mathring \pi_i}{dt} &= \e F^{(0)}_i(\mathring \pi_j) + \e^2 F^{(1)}_i(\mathring \varpi_J)+{\cal O}(\e^3),\\    
    \frac{d{\mathring{\psi}}_s}{dt} &= \Omega_{s(0)}(\mathring \pi_j).
\end{align}
\end{subequations}
The forcing function $F^{(1)}_i$ is now
\begin{multline}\label{eq:F1 fixed frequencies}
    F_i^{(1)} = \left\langle f_i^{(1)}\right\rangle +\langle\delta \pi_j\rangle\frac{\partial F_i^{(0)}}{\partial\mathring \pi_j} 
    - F^{(0)}_j\frac{\partial\left\langle \delta\pi_i\right\rangle}{\partial\mathring\pi_j}\\
    + \left\langle\delta \pi^{\rm osc}_j\frac{\partial f_i^{(0)}}{\partial\mathring \pi_j} + \delta \psi^j_{\rm osc} \frac{\partial f_i^{(0)}}{\partial\mathring\psi^j}\right\rangle, 
\end{multline}
with $\langle\delta \pi_j\rangle$ given by the solution to Eq.~\eqref{eq:dpi fixed Omega}. 

This is the gauge used in the Introduction, where we used it to replace $\Omega^i_{(0)}$ with $\Omega^i$. It has several advantages: 
\begin{enumerate}
    \item It eliminates $\Omega^i_{(1)}$ terms from the second-order field equations~\eqref{eq:EFE2 multiscale} because Eq.~\eqref{eq:ddt modes} reduces to
    \begin{multline}
        \hspace{15pt}\left(\frac{\partial}{\partial t}\right)_{\!\bm{k},q} 
        = -i\bigl(k_i\Omega^i_{(0)} +q\,\Omega_{s(0)}\bigr)\\
        + \e F^{(0)}_I \frac{\partial}{\partial\mathring\varpi_I} + \mathcal{O}(\e^2).
    \end{multline}
    As highlighted in Ref.~\cite{Mathews:2021rod}, this is particularly advantageous in the case of the secondary-spin contribution to the field equations because it eliminates the noncompact source term proportional to $\Omega^i_{(1)}$ in Eq.~\eqref{eq:EFE2 spin terms}. On the other hand, the correction to $\Omega_{s(0)}$ would only enter the \emph{third}-order field equations, since the precession phase first enters the field equations at second order.
    \item It naturally yields observable quantities as functions of the physical, observable frequencies. Such functional relationships have historically been the basis for translating invariant information between different approaches to the two-body problem~\cite{LeTiec:2014oez,Barack:2018yvs}.
    \item There is some indication that it might yield more accurate waveforms than other gauge choices, though this evidence is limited to the quasicircular case~\cite{Wardell:2021fyy}, where the degenerate surfaces do not exist.
\end{enumerate}

This gauge also has a disadvantage due to the degenerate surfaces. At these surfaces, $\delta\pi_i$ diverges, causing the forcing function~\eqref{eq:F1 fixed frequencies} to diverge. A complete evolution scheme using this gauge would require a method of evolving across these singular surfaces. It is not yet clear how much of an obstacle this represents, particularly since the surfaces are deep in the strong field, near the separatrix where the multiscale expansion breaks down and the particle transitions into a plunge~\cite{Apte:2019txp,Kuchler:2024esj,Becker:2024xdi,Lhost:2024jmw}. It is also possible to avoid this problem by only eliminating corrections to some, but not all of the frequencies.

\subsubsection{Fixed frequencies and fixed emissions}

We next consider a gauge that eliminates all post-adiabatic terms in the evolution equations. This requires eliminating the forcing functions $F^{(1)}_i$ as well as the frequency corrections $\Omega_{(1)}^i$ and $\Omega_{s(1)}$. The elimination of post-adiabatic forcing functions (`fixed emissions') has been considered previously in Ref.~\cite{VanDeMeent:2018cgn}, for example.

We see from Eq.~\eqref{eq:F1} that the forcing function $F^{(1)}_i$ can be eliminated with a choice of $\langle\delta \pi_j\rangle$ satisfying the partial differential equation 
\begin{multline}
    \langle\delta \pi_j\rangle\frac{\partial F_i^{(0)}}{\partial\mathring \pi_j} - F^{(0)}_j\frac{\partial\left\langle \delta\pi_i\right\rangle}{\partial\mathring\pi_j} 
    = -\left\langle f_i^{(1)}\right\rangle  
    - \left\langle\delta \pi^{\rm osc}_j\frac{\partial f_i^{(0)}}{\partial\mathring \pi_j}\right\rangle \\
    + \left\langle\delta \psi^j_{\rm osc} \frac{\partial f_i^{(0)}}{\partial\mathring\psi^j}\right\rangle.\label{eq:fixed emissions}    
\end{multline}
Similarly, we can see from Eq.~\eqref{eq:Omega1} that $\Omega^i_{(1)}$ can be eliminated with a choice of $\langle\Delta\psi^i\rangle$ satisfying
\begin{align}
    F^{(0)}_j\frac{\partial\left\langle \Delta\psi^i\right\rangle}{\partial\mathring\pi_j} &= \left\langle\omega^i_{(1)}\right\rangle + \left\langle\delta \pi_j\frac{\partial\omega^i_{(0)}}{\partial\mathring \pi_j}+\delta \psi^j \frac{\partial\omega^i_{(0)}}{\partial\mathring\psi^j}\right\rangle.
\end{align}

The evolution equations in this gauge are simply the 0PA ones to all orders:
\begin{subequations}
\begin{align}
    \frac{d\mathring\psi^i}{dt} &= \Omega^i_{(0)}(\mathring \pi_j),\\
    \frac{d\mathring \pi_i}{dt} &= \e F^{(0)}_i(\mathring \pi_j),\\    
    \frac{d{\mathring{\psi}}_s}{dt} &= \Omega_{s(0)}(\mathring \pi_j).
\end{align}
\end{subequations}

This gauge might appear to be impossibly advantageous in that it superficially avoids the need to calculate subleading terms when generating waveforms. However, this appearance is misleading because the term $\langle f^{(1)}_i\rangle$ in Eq.~\eqref{eq:fixed emissions} is proportional to the second-order force $a^i_{(2)}$. Calculating $a^i_{(2)}$ requires the solution to the second-order field equation~\eqref{eq:EFE2 multiscale}, which in turn involves $\delta\pi_i$ in the source term~\eqref{eq:Tm1}. Therefore Eq.~\eqref{eq:fixed emissions} actually represents an extremely complicated integro-differential equation for $\delta\pi_i$, which might not even have a solution.

We can consider a more practical alternative that eliminates all contributions to $F^{(1)}_i$ \emph{except} those coming from $a^i_{(2)}$. Equation~\eqref{eq:fixed emissions} is then replaced with the condition
\begin{multline}\label{eq:almost fixed emissions}
    \langle\delta \pi_j\rangle\frac{\partial F_i^{(0)}}{\partial\mathring \pi_j} - F^{(0)}_j\frac{\partial\left\langle \delta\pi_i\right\rangle}{\partial\mathring\pi_j} 
    = 
    - \left\langle\delta \pi^{\rm osc}_j\frac{\partial f_i^{(0)}}{\partial\mathring \pi_j}\right\rangle \\
    + \left\langle\delta \psi^j_{\rm osc} \frac{\partial f_i^{(0)}}{\partial\mathring\psi^j}\right\rangle.   
\end{multline}
The evolution equation for $\mathring\pi_i$ then becomes
\begin{equation}
\frac{d\mathring \pi_i}{dt} = \e \langle B_{ij} a_{(1)}^j\rangle +\e^2 \langle B_{ij} a_{(2)}^j\rangle +\mathcal{O}(\e^3),
\end{equation}
where we have used Eq.~\eqref{eq:fn=Ba}. However, one should note that this choice can complicate the second-order Einstein equations because the solution to Eq.~\eqref{eq:almost fixed emissions} contributes to the source term~\eqref{eq:Tm1}.

\subsubsection{Fixed constants of motion}

As a final option, we consider a gauge in which the constants of motion take fixed values as we vary $\e$. This is the gauge used in Ref.~\cite{Witzany:2024ttz}, a fact that will play an important role in Sec.~\ref{sec:flux balance}.

To understand this gauge choice, we must first recall the constants of motion for spinning test particles (i.e., spinning particles that do not source a metric perturbation or experience a self-force). In addition to the particle's mass $m_2$ and spin components $\chi_\parallel$ and $\chi_\perp$, the conserved quantities are spin-corrected versions of the geodesic energy, angular momentum, and Carter constant: $P_i = (E,L_z,K)$.\footnote{The conserved quantities are more often described as $P_i$ together with the mass, spin magnitude, and Rüdiger's constant $C_Y=l_\alpha S^\alpha$; see, e.g., Refs.~\cite{Compere:2021kjz,Skoupy:2023lih,Ramond:2024sfp}. Here $l_\alpha$ is the `orbital angular momentum' vector defined in Eq.~\eqref{eq:l def}. In the test-particle limit, $C_Y$ is equivalent (up to a factor) to our $\chi_\parallel$.} In any spacetime with a Killing vector $\xi^\alpha$, the quantity
\begin{equation}\label{eq:Xi}
    \Xi_\xi=\xi^\alpha u_\alpha +\frac{m_2}{2}S^{\alpha \beta}\nabla_\alpha \xi_\beta
\end{equation}
is conserved along solutions to the MPD equations~\cite{Dixon:1970zza}. The energy and angular momentum are the quantities $\Xi_\xi$ associated with Kerr's timelike and axial Killing vectors: 
\begin{equation}\label{eq:E and L}
E \equiv -\Xi_t\quad \text{and} \quad L_z \equiv \Xi_\phi. 
\end{equation}
The spin-corrected Carter constant, introduced in Ref.~\cite{doi:10.1098/rspa.1981.0046}, is
\begin{equation}
\label{eq:CarterWspin}
    K=K^{\alpha \beta} u_\alpha u_\beta +m_2 L_{\alpha \beta \gamma} S^{\alpha \beta}u^\gamma,
\end{equation}
in which we have defined~\cite{doi:10.1098/rspa.1981.0046, Witzany:2019nml}
\begin{equation}
    L_{\alpha\beta\gamma}\equiv-2\left(Y^\delta{}_\beta \nabla_\delta Y_{\gamma \alpha} -  Y_\gamma{}^\delta \nabla_\delta Y_{\alpha \beta}\right).
\end{equation} 
We refer to Ref.~\cite{Compere:2021kjz}, for example, for a review of the construction of these conserved quantities.

Each of the three conserved quantities take the form of a geodesic term plus a correction proportional to the spin. Hence, in terms of our osculating elements $\pi_i$ and phases $\psi^i$, we can write them as
\begin{equation}\label{eq:osculating P_i}
    P_i = P_i^{(0)}(\pi_j) + \e \delta P^{(\chi_2)}_i(\psi^j,\pi_j,S^{\alpha\beta}).
\end{equation}
Here 
\begin{equation}
P_i^{(0)} = (E^{(0)},L_z^{(0)},K^{(0)}) = (-u_t, u_\phi, K^{\alpha\beta}u_\alpha u_\beta)
\end{equation}
are the geodesic conserved quantities, which are given as functions of $\pi_i$ in Eqs.~(222)--(224) of Ref.~\cite{Pound:2021qin}, where the quantity $Q^{(0)}$ in Eq.~(224) of Ref.~\cite{Pound:2021qin} is related to $K^{(0)}$ by $Q^{(0)}=K^{(0)}-\bigl(L^{(0)}_z-m_1^{(0)}\chi_1^{(0)} E^{(0)}\bigr)^2$. Along the accelerated orbit, the geodesic term varies with time, regardless of whether we consider test-particle orbits or self-accelerated ones. Upon substitution of the expansions~\eqref{eq:psi to psitilde} and \eqref{eq: pi to pitilde}, Eq.~\eqref{eq:osculating P_i} becomes
\begin{multline}
    P_i = P_i^{(0)}(\mathring \pi_j) - \e \delta \pi_j \frac{\partial P_i^{(0)}}{\partial\mathring \pi_j} \\
    + \e \delta P^{(\chi_2)}_i(\psi^j_{(0)},\mathring \pi_j,S^{\alpha\beta}) + \mathcal{O}(\e^2).
\end{multline}
It will also be necessary to involve the angle-averaged version,
\begin{multline}
    \langle P_i\rangle  = P_i^{(0)}(\mathring \pi_j) - \e \langle\delta \pi_j\rangle \frac{\partial P_i^{(0)}}{\partial\mathring \pi_j} \\
    + \e \left\langle\delta P^{(\chi_2)}_i(\psi^j_{(0)},\mathring \pi_j,S^{\alpha\beta})\right\rangle + \mathcal{O}(\e^2).\label{eq:<Pi>}
\end{multline}

The gauge condition we consider now is
\begin{equation}
    \langle P_i\rangle  = P_i^{(0)}(\mathring \pi_j).
\end{equation}
In other words, we choose $\mathring \pi_j$ to be geodesically related to the (averaged) spinning-particle constants of motion. From Eq.~\eqref{eq:<Pi>}, we see this condition is enforced with the choice 
\begin{equation}
    \langle \delta \pi_i\rangle = \frac{\partial P_i^{(0)}}{\partial\mathring \pi_j}\langle \delta P^{}_i\rangle.
\end{equation}

This gauge is convenient because it enables us to immediately find the evolution of $\mathring \pi_i$ from the evolution of~$\langle P_i\rangle$:
\begin{equation}\label{eq:tbguageflux}
    \frac{d\mathring \pi_i}{dt} = \frac{\partial \mathring \pi_i}{\partial P^{(0)}_k} \frac{d\langle P_k\rangle}{dt},
\end{equation}
where $\partial \mathring \pi_i/\partial P^{(0)}_k$ denotes the inverse of the (geodesic) Jacobian $\partial P^{(0)}_k/\partial \mathring \pi_i$. This formula holds even for the spin-independent part of the dynamics. However, since we do not have a useful formula for the 1PA spin-independent piece of $d\langle P_k\rangle/dt$, it is not (at the moment) particularly useful for computing that piece of the 1PA dynamics. Instead, the formula becomes especially useful, as we explain in Sec.~\ref{sec:flux balance}, when calculating the linear-in-spin 1PA piece of $d\mathring \pi_i/dt$.

Note that in this gauge we are still left with the freedom to choose $\langle\Delta\psi^i\rangle$ and $\langle\Delta\tilde\psi_s\rangle$.

\section{Waveform}\label{sec:waveform generation}

We now summarize the waveform-generation scheme that results from the multiscale expansion. We then explain how the waveform is invariant under the gauge freedom discussed in the preceding section. Finally, we discuss the impact that the secondary spin's precession has on the waveform.

\subsection{Waveform generation scheme}

The gravitational waveform is extracted from the metric perturbation at future null infinity,
\begin{equation}
    h = h_+-ih_\times = \lim_{r\to\infty}\bigl(r h_{\alpha\beta}\bar m^{\alpha}\bar m^{\beta}\bigr),
\end{equation}
where the limit is taken at fixed retarded time $u$, and $\bar m^\alpha$ is the standard Newman-Penrose complex basis vector on the celestial sphere~\cite{Pound:2021qin}. Given the form of the metric perturbation~\eqref{eq:h1 Fourier}--\eqref{eq:h2 Fourier}, the waveform through second order in $\e$ can be written as
\begin{align}
    h &= \sum_{\bm{k}\in\mathbb{Z}^3}\Bigl[\e \mathring h^{(1)}_{\bm{k}}(\mathring \pi_i,\theta,\phi)
     +\e^2 \mathring h^{(2\text{-}\bcancel{\chi_2})}_{\bm{k}}(\mathring \varpi_i,\theta,\phi)\nonumber \\
    &\qquad +\e^2 \chi_\parallel \mathring h^{(2\text{-}\chi_\parallel)}_{\bm{k}}(\mathring \pi_i,\theta,\phi) \nonumber\\
    &\qquad + \e^2 \chi_\perp\sum_{q=\pm1} \mathring h^{(2\text{-}\chi_\perp)}_{\bm{k}q}(\mathring \pi_i,\theta,\phi)e^{-iq\mathring\psi_s}\Bigr]e^{-i\mathring\psi_{\bm{k}}},\label{eq:multiscale waveform with spin}
\end{align}
with $\mathring h^{(1)}_{\bm{k}}=\lim_{r\to\infty}\Bigl(r h^{(1,\bm{k})}_{\alpha\beta}\bar m^{\alpha}\bar m^{\beta}\Bigr)$, for example. Equation~\eqref{eq:multiscale waveform with spin} extends Eq.~\eqref{multiscale waveform} to include the secondary spin. Here $\mathring h^{(2\text{-}\bcancel{\chi_2})}_{\alpha\beta}$ denotes the $\chi_2$-independent part of the second-order metric perturbation, identical to the second-order term in Eq.~\eqref{multiscale waveform}. We have also divided the linear spin contribution---the solution to Eq.~\eqref{eq:EFE2 spin terms}---into pieces proportional to $\chi_\parallel$ and $\chi_\perp$, respectively. 

The waveform's time dependence is governed by Eqs.~\eqref{eq:ringed equations}, where $t$ is re-interpreted as retarded time $u$ along future null infinity. This identification between time along the particle's worldline and time at future null infinity is achieved using hyperboloidal slicing, for example~\cite{Miller:2020bft,Miller:2023ers}. Even with such slicing, there is considerable subtlety in extracting the second-order waveform, as detailed in Ref.~\cite{Cunningham:2024dog}. However, this complexity only arises from nonlinearity in the field equations. It does not affect the first-order and linear-in-spin pieces of the waveform, which can be extracted from the Fourier amplitudes $\mathring h^{(1,\bm{k})}_{\alpha\beta}$ and $\mathring h^{(2\text{-}\chi_2,\bm{k})}_{\alpha\beta}$ as standard in linear perturbation theory~\cite{Pound:2021qin}.

\subsection{Gauge invariance}\label{sec:waveform gauge invariance}

The residual gauge freedom discussed in Sec.~\ref{sec:orbital motion} and \ref{sec:gauge choices} comprises transformations 
\begin{subequations}\label{eq:gauge freedom}
\begin{align}
\mathring\pi_i&\to \mathring\pi'_i=\mathring\pi_i-\e\langle\delta\pi_i\rangle, \\
\mathring\psi^i&\to\mathring\psi'^i=\mathring\psi^i-\langle\Delta\psi^i\rangle,\\ 
\mathring\psi_s&\to\mathring\psi'_s=\mathring\psi_s-\langle\Delta\tilde\psi_s\rangle. 
\end{align}
\end{subequations}
To understand the minus sign, note that Eq.~\eqref{eq:averaging transformation} expresses  old variables in terms of new ones, while Eq.~\eqref{eq:gauge freedom} represents the inverse: new in terms of old. 

Under the phase-space gauge transformation~\eqref{eq:gauge freedom}, the orbital evolution equations~\eqref{eq:ringed equations} change according to Eqs.~\eqref{eq:1PA gauge transformation}. We now show that the amplitudes in the waveform~\eqref{eq:multiscale waveform with spin} change under this transformation in a way that precisely compensates the changes in the orbital evolution equations, leaving the waveform invariant. 

The metric perturbation~\eqref{eq:h multiscale} in terms of the new phase-space variables is
\begin{equation}\label{eq:h multiscale primed}
    h_{\alpha\beta} = \e \mathring h'^{(1)}_{\alpha\beta}(\mathring\psi'^i,\mathring \varpi'_I,x^i)+\e^2 \mathring h'^{(2)}_{\alpha\beta}(\mathring\psi'^i,\mathring \varpi'_I,\mathring\psi'_s,x^i)+\ldots,
\end{equation}
where $(\mathring\psi'^i,\mathring\pi'_i)$ are determined from the ordinary differential equations
\begin{subequations}\label{eq:primed evolution equations}
\begin{align}
    \frac{d\mathring\psi'^i}{dt} &= \Omega^i_{(0)}(\mathring \pi'_j)+\e\Omega'^i_{(1)}(\mathring \varpi'_J)+{\cal O}(\e^2),\label{eq:dpsi'/dt}\\
    \frac{d\mathring \pi'_i}{dt} &= \e F^{(0)}_i(\mathring \pi'_j) + \e^2 F'^{(1)}_i(\mathring \varpi'_J)+{\cal O}(\e^3),\label{eq:dpi'/dt}\\
    \frac{d\mathring{\psi}'_s}{dt} &= \Omega_{s(0)}(\mathring \pi'_j)+\e\Omega'_{s(1)}(\mathring \varpi'_J)+{\cal O}(\e^2).
\end{align}
\end{subequations}
For simplicity, we assume $\delta m'_1=\delta m_1$ and $\delta\chi'_1=\delta\chi_1$, and we exclude transformations involving the perturbations of the primary's mass and spin.

We first find the relationship between $h'^{(n)}_{\alpha\beta}$ and $h^{(n)}_{\alpha\beta}$. To do so, we  start from the metric perturbation~\eqref{eq:h multiscale} in the unprimed gauge. After expressing $(\mathring\psi^i,\mathring\pi_i)$ in terms of $(\mathring\psi'^i,\mathring\pi'_i)$ using Eq.~\eqref{eq:gauge freedom}, we then re-expand Eq.~\eqref{eq:h multiscale} in powers of $\e$ at fixed $(\mathring\psi'^i,\mathring\pi'_i)$. Since the result and Eq.~\eqref{eq:h multiscale primed} both represent expansions of $h_{\alpha\beta}$ at fixed phase-space coordinates $(\mathring\psi'^i,\mathring\pi'_i)$, the coefficients must agree at each order in $\e$. Making that identification between coefficients in the two expressions, we find
\begin{align}
    \mathring h'^{(1)}_{\alpha\beta}(\mathring\psi'^i,\mathring \varpi'_I) &= \mathring h^{(1)}_{\alpha\beta}(\mathring\psi'^i+\langle\Delta\psi^i\rangle,\mathring \varpi'_I),\label{eq:h1' to h1}\\
    \mathring h'^{(2)}_{\alpha\beta}(\mathring\psi'^i,\mathring \varpi'_I,\mathring\psi'_s) &= \mathring h^{(2)}_{\alpha\beta}\bigl(\mathring\psi'^i+\langle\Delta\psi^i\rangle,\mathring \varpi'_I,\mathring\psi'_s+\langle\Delta\tilde\psi_s\rangle\bigr)\nonumber\\
    &\quad + \langle\delta\pi_i\rangle\frac{\partial}{\partial\mathring\pi'_i}\mathring h^{(1)}_{\alpha\beta}(\mathring\psi'^i+\langle\Delta\psi^i\rangle,\mathring\pi'_i),\label{eq:h2' to h2}
\end{align}
where we have suppressed the dependence on $x^i$. Note that the $\mathring\pi'_i$ derivative in the last line acts only on the second argument of $\mathring h^{(1)}_{\alpha\beta}$, not on the $\langle\Delta\psi^i\rangle(\mathring\pi'_i)$ appearing in the first argument.

Since the waveform~\eqref{eq:multiscale waveform with spin} is expressed in terms of Fourier modes, we next Fourier expand the left- and right-hand sides of Eqs.~\eqref{eq:h1' to h1} and \eqref{eq:h2' to h2} with respect to the functions' first and third arguments. This immediately yields relationships between mode coefficients,
\begin{align}
    \mathring h'^{(1,\bm{k})}_{\alpha\beta} &= \mathring h^{(1,\bm{k})}_{\alpha\beta}e^{-i\langle\Delta\psi_{\bm{k}}\rangle},\\
    \mathring h'^{(2\text{-}\bcancel{\chi_2},\bm{k})}_{\alpha\beta} &= \left[\mathring h^{(2\text{-}\bcancel{\chi_2})}_{\alpha\beta} + \left\langle\delta\pi_i^{(\bcancel{\chi_2})}\right\rangle\frac{\partial}{\partial\mathring\pi'_i}\mathring h^{(1)}_{\alpha\beta}\right]e^{-i\langle\Delta\psi_{\bm{k}}\rangle},\\  
    \mathring h'^{(2\text{-}\chi_\parallel,\bm{k})}_{\alpha\beta} &= \left[\mathring h^{(2\text{-}\chi_\parallel)}_{\alpha\beta} + \left\langle\delta\pi_i^{(\chi_\parallel)}\right\rangle\frac{\partial}{\partial\mathring\pi'_i}\mathring h^{(1)}_{\alpha\beta}\right]e^{-i\langle\Delta\psi_{\bm{k}}\rangle},\\
    \mathring h'^{(2\text{-}\chi_\perp,\bm{k},q)}_{\alpha\beta} &= \mathring h^{(2\text{-}\chi_\perp,\bm{k},q)}_{\alpha\beta}e^{-i\langle\Delta\psi_{\bm{k}}\rangle - iq\langle\Delta\tilde\psi_{s}\rangle}.
\end{align}
Here we have divided $\langle\delta\pi_i\rangle$ into pieces independent of $\chi_2$ and proportional to $\chi_\parallel$. We exclude a term proportional to $\chi_\perp$ in $\langle\delta\pi_i\rangle$ because such a term would introduce a $\mathring\psi_s$-independent contribution to $\mathring h^{(2\text{-}\chi_\perp)}_{\alpha\beta}$. 

The relationships between the Fourier mode amplitudes in the two gauges imply that the waveform in the primed gauge can be written as follows in terms of the amplitudes in the unprimed gauge: 
\begin{align}
    h &= \sum_{\bm{k}\in\mathbb{Z}^3}\Bigl\{\e \mathring h^{(1)}_{\bm{k}}(\mathring \pi'_i)
     +\e^2\Bigl[ \mathring h^{(2\text{-}\bcancel{\chi_2})}_{\bm{k}}(\mathring \varpi'_i) \nonumber \\
    &\quad + \chi_\parallel \mathring h^{(2\text{-}\chi_\parallel)}_{\bm{k}}(\mathring \pi'_i) +\left\langle\delta\pi_i\right\rangle\frac{\partial}{\partial\mathring\pi'_i}\mathring h^{(1)}_{\bm{k}}(\mathring\pi_i')\nonumber\\
    &\quad + \chi_\perp\sum_{q=\pm1} \mathring h^{(2\text{-}\chi_\perp)}_{\bm{k}q}(\mathring\pi_i')e^{-iq(\mathring\psi'_s+\langle\Delta\tilde\psi_s\rangle)}\Bigr]\Bigr\}e^{-i(\mathring\psi'_{\bm{k}}+\langle\Delta\psi_{\bm{k}}\rangle)}.\label{eq:multiscale waveform primed}
\end{align}

We now appeal to the equations of motion~\eqref{eq:primed evolution equations} to show that the above waveform agrees with the one in the unprimed gauge, Eq.~\eqref{eq:multiscale waveform with spin}. Examining Eqs.~\eqref{eq:multiscale waveform with spin} and~\eqref{eq:multiscale waveform primed}, we see that the two expressions become equivalent, up to $\mathcal{O}(\e^3)$ differences, if the equations of motion imply the relationships~\eqref{eq:gauge freedom}. Since the transformation laws~\eqref{eq:1PA gauge transformation} for the equations of motion were derived from those same relationships, it is clear that the equations of motion must be compatible with them. But doing the converse, obtaining the desired relationships from the equations of motion, is a worthwhile exercise.

Consider the equation of motion~\eqref{eq:dpsi'/dt} for the orbital phases. Appealing to Eq.~\eqref{eq:1PA gauge transformation}, we write it as
\begin{multline}
    \frac{d\mathring\psi'^i}{dt} = \Omega^i_{(0)}(\mathring \pi'_j) + \e\biggl[\Omega^i_{(1)}(\mathring \varpi'_J) + \langle \delta\pi_j\rangle\frac{\partial\Omega^i_{(0)}}{\partial\mathring\pi'_j} \\
    -F^{(0)}_j\frac{\partial\langle\Delta\psi^i\rangle}{\partial\mathring\pi'_j}\biggr] + {\cal O}(\e^2).    
\end{multline}
We can immediately rewrite this as
\begin{subequations}\label{eq:dpsi'/dt to dpsi/dt}
\begin{align}
    \frac{d\mathring\psi'^i}{dt} &= \Omega^i_{(0)}\bigl(\mathring \pi'_j + \e\langle\delta\pi_j\rangle\bigr) + \e \Omega^i_{(1)}\bigl(\mathring \varpi'_J+ \e\langle\delta\varpi_J\rangle\bigr) 
    \nonumber\\
    &\quad -\frac{d\langle\Delta\psi^i\rangle}{dt} + {\cal O}(\e^2),\\
    &= \frac{d\mathring\psi^i}{dt} -\frac{d\langle\Delta\psi^i\rangle}{dt} + {\cal O}(\e^2),
\end{align}
\end{subequations}
defining $\delta\varpi_J=(\delta\pi_i,0,0)$. Hence, we find the required relationship $\mathring\psi'^i = \mathring\psi^i-\Delta\psi^i + \mathcal{O}(\e)$ so long as their initial conditions agree. The difference is of order $\e$, rather than the $\e^2$ in Eq.~\eqref{eq:dpsi'/dt to dpsi/dt}, because the integration is over a time scale of order $1/\e$.

Similarly, the equation of motion~\eqref{eq:dpsi'/dt}, given Eq.~\eqref{eq:1PA gauge transformation}, can be written as
\begin{multline}
    \frac{d\mathring \pi'_i}{dt} = \e F^{(0)}_i(\mathring \pi'_j) + \e^2\biggl[ F^{(1)}_i(\mathring \varpi'_J) +\langle \delta\pi_j\rangle\frac{\partial F^i_{(0)}}{\partial\mathring\pi'_i} \\
    -F^{(0)}_j\frac{\partial\langle\delta\pi_i\rangle}{\partial\mathring\pi'_j}\biggr] + {\cal O}(\e^3).
\end{multline}
This implies
\begin{subequations}
\begin{align}
    \frac{d\mathring\pi'_i}{dt} &= \e F_i^{(0)}\bigl(\mathring \pi'_j + \langle\delta\pi_j\rangle\bigr) + \e^2 F_i^{(1)}\bigl(\mathring \varpi'_J + \langle\delta\varpi_J\rangle\bigr) 
    \nonumber\\
    &\quad -\e\frac{d\langle\delta\pi_i\rangle}{dt} + {\cal O}(\e^3),\\
    &= \frac{d\mathring\pi_i}{dt} -\e\frac{d\langle\delta\pi_i\rangle}{dt} + {\cal O}(\e^3),
\end{align}
\end{subequations}
and we find the required relationship $\mathring\pi'_i = \mathring\pi_i-\e\delta\pi_i + \mathcal{O}(\e^2)$ (again, so long as their initial conditions agree).

Therefore, as promised, the transformation of the evolution equations counterbalances the transformation of the waveform amplitudes, such that the waveform is invariant under the residual gauge freedom we consider. This invariance might appear trivial when shown in this way. However, its significance becomes clearer when we consider the many manifestations of the choice of gauge in solving the Einstein equation and calculating forcing functions:
\begin{enumerate}
    \item The choice of $\langle\Delta\psi^i\rangle$ affects the Fourier mode decomposition of the stress-energy tensor and of all the Einstein equations.
    \item The choice of $\langle\delta\pi_i\rangle$ affects the first-order terms $z^i_{(1)}$ and $v^i_{(1)}$ in the parametrization of the particle's Boyer-Lindquist trajectory; see Eqs.~\eqref{eq:z1} and \eqref{eq:v1}.
    \item The changes in $z^i_{(1)}$ and $v^i_{(1)}$ affect the term $T^{(m,1)}_{\alpha\beta}$ in the stress-energy tensor; see Eq.~\eqref{eq:Tm1}.
    \item The change in $T^{(m,1)}_{\alpha\beta}$ affects the piece of $h^{(2)}_{\alpha\beta}$ that $T^{(m,1)}_{\alpha\beta}$ sources; see Eq.~\eqref{eq:EFE2 multiscale}.
    \item The choice of $\langle\Delta\psi^i\rangle$ and $\langle\delta\pi_i\rangle$ affect the value of $\Omega^i_{(1)}$ and therefore the source term $\delta G^{(1,\bm{k})}_{\mu\nu}$ in the field equation~\eqref{eq:EFE2 multiscale}.
    \item The change in $\delta G^{(1,\bm{k})}_{\mu\nu}$ affects the piece of $h^{(2)}_{\alpha\beta}$ that $\delta G^{(1,\bm{k})}_{\mu\nu}$ sources.
    \item The changes in $h^{(2)}_{\alpha\beta}$ affect the second-order force $a^i_{(2)}$.
\end{enumerate}

Since the final waveform is invariant, one can choose whichever gauge is deemed most convenient for these calculations. As we discuss in Sec.~\ref{sec:flux balance}, one can also choose a gauge that is convenient for solving the field equations and then transform the outputs into a gauge that is convenient for calculating forcing functions.

Before moving on, we emphasize two important facts. First, even though the waveform is invariant under the gauge transformations we consider, that does not mean waveforms computed in two different gauges will be identical (given identical initial conditions). The invariance is a limiting statement, in that the two waveforms agree up to nonzero $\mathcal{O}(\e^3)$ differences, and only in the sense that they agree when one is re-expanded at fixed values of the variables used in the other (i.e., when they are compared in precisely the \emph{same} limit $\ee\to0$). In other words, they will decidedly \emph{not} be numerically identical functions of time for a given, finite value of $\ee$. Different gauges move information between different terms in the expansion, and different gauges will change the magnitude of omitted higher-order terms. This in turn can mean that one choice of gauge can yield more accurate waveforms than another choice. 

The second important point is that the 1PA frequency correction $\Omega^i_{(1)}$, forcing function $F^{(1)}_i$, and leading-order amplitudes $\mathring{h}^{(1)}_{\bm{k}}$ are all affected by the choice of $\langle\Delta\psi^i\rangle$. These three quantities work together in unison, and all three must be computed with a single, consistent choice of gauge---or else transformed into a common gauge as a post-processing step before using them as inputs in a 1PA waveform model. Recognizing this is especially important because different choices for $\langle\Delta\psi^i\rangle$ are already in use in 0PA waveform generation, as previously discussed in the supplemental material of Ref.~\cite{Isoyama:2021jjd}. 

\subsection{Relevance of the secondary spin's precession}

So far, we have not specifically highlighted the impact of the secondary spin's precession, except to reaffirm that it does not contribute to the 1PA orbital evolution. We now investigate its role in more detail. 

To make the assessment, we write the waveform~\eqref{eq:multiscale waveform with spin} as 
\begin{subequations}
\begin{align}
h&=\sum_{\bm{k}\in\mathbb{Z}^3}h_{\bm{k}}(\mathring\psi_{\bm{k}},\mathring\pi_i,\theta,\phi,\e) \\
&= \sum_{\bm{k}\in\mathbb{Z}^3}\sum_{n\geq0}\e^n h^{(n)}_{\bm{k}}(\mathring\psi_{\bm{k}},\mathring\pi_i,\theta,\phi).
\end{align}
\end{subequations}
We then decompose each mode into a real amplitude and a complex phase factor, 
\begin{equation}\label{eq:real amplitudes}
h_{\bm{k}} = A_{\bm{k}}e^{-i \Phi_{\bm{k}}},
\end{equation}
where $A_{\bm{k}}=|h_{\bm{k}}|$ and $\Phi_{\bm{k}} = -\arg(h_{\bm{k}})$. The phase $\Phi_{\bm{k}}$ characterizes the total waveform's phase evolution in an invariant way. This contrasts with phases such as $\mathring\psi_{\bm{k}}$, which are associated with complex-valued amplitudes such as $\mathring h^{(1)}_{\bm{k}}$. In a product such as $\mathring h^{(1)}_{\bm{k}}e^{-i\mathring\psi_{\bm{k}}}$, we can freely move  phases between the complex amplitude and the complex exponential, making the phase non-unique, as we saw when considering gauge transformations in the previous section. The decomposition~\eqref{eq:real amplitudes} avoids this ambiguity.

Decomposing Eq.~\eqref{eq:multiscale waveform with spin} into the form~\eqref{eq:real amplitudes}, we find
\begin{multline}\label{eq:waveform phase}
    \Phi_{\bm{k}} = \mathring\psi_{\bm{k}} -\e^0\arg\bigl(h^{(1)}_{\bm{k}}\bigr)
    +\frac{\e}{\bigl|h^{(1)}_{\bm{k}}\bigr|^2}\Bigl\{d^{\,0}_{\bm{k}}
  +\chi_\perp d^{\,\rm s}_{\bm{k}}\sin\bigl(\mathring\psi_s\bigr)\\
 +\chi_\perp d^{\,\rm c}_{\bm{k}}\cos\bigl(\mathring\psi_s\bigr)\Bigr\}
 + {\cal O}(\e^2),
\end{multline}
with
\begin{subequations}
\begin{align}
    d^{\,0}_{\bm{k}} &= -{\rm Re}\,h^{(1)}_{\bm{k}} {\rm Im}\,h^{(2)}_{\bm{k},0} + {\rm Im}\,h^{(1)}_{\bm{k}}{\rm Re}\,h^{(2)}_{\bm{k},0},\\
    d^{\,\rm s}_{\bm{k}} &= \Bigl({\rm Im}\,h^{(2\text{-}\chi_\perp)}_{\bm{k},+1}-{\rm Im}\,h^{(2\text{-}\chi_\perp)}_{\bm{k},-1}\Bigr)\,{\rm Im}\,h^{(1)}_{\bm{k}}\nonumber\\
    &\quad + \Bigl({\rm Re}\, h^{(2\text{-}\chi_\perp)}_{\bm{k},+1}-{\rm Re}\, h^{(2\text{-}\chi_\perp)}_{\bm{k},-1}\Bigr)\,{\rm Re}\, h^{(1)}_{\bm{k}},\\
    d^{\,\rm c}_{\bm{k}} &= \Bigl({\rm Re}\,h^{(2\text{-}\chi_\perp)}_{\bm{k},+1}+{\rm Re}\,h^{(2\text{-}\chi_\perp)}_{\bm{k},-1}\Bigr)\,{\rm Im}\,h^{(1)}_{\bm{k}}\nonumber\\
    &\quad - \Bigl({\rm Im}\, h^{(2\text{-}\chi_\perp)}_{\bm{k},+1}+{\rm Im}\, h^{(2\text{-}\chi_\perp)}_{\bm{k},-1}\Bigr)\,{\rm Re}\, h^{(1)}_{\bm{k}}.
\end{align}
\end{subequations}
Here $h^{(2)}_{\bm{k},0}=\Bigl(h^{(2\text{-}\bcancel{\chi_2})}_{\bm{k}} + \chi_\parallel h^{(2\text{-}\chi_\parallel)}_{\bm{k}}\Bigr)e^{-i\bm{\psi_k}}$ is the total second-order $\bm{k}$ mode excluding the secondary spin precession.

Similarly,
\begin{multline}
    A_{\bm{k}} = \e |h^{(1)}_{\bm{k}}| + \frac{\e^2}{\bigl|h^{(1)}_{\bm{k}}\bigr|}\Bigl\{a^0_{\bm{k}}
  + \chi_\perp a^{\rm s}_{\bm{k}} \sin\bigl(\mathring\psi_s\bigr)\\
 +\chi_\perp a^{\rm c}_{\bm{k}}\cos\bigl(\mathring\psi_s\bigr)\Bigr\}
 + {\cal O}(\e^3),
\end{multline}
with
\begin{subequations}
\begin{align}
    a^{\,0}_{\bm{k}} &= {\rm Re}\,h^{(1)}_{\bm{k}} {\rm Re}\,h^{(2)}_{\bm{k},0} + {\rm Im}\,h^{(1)}_{\bm{k}}{\rm Im}\,h^{(2)}_{\bm{k},0},\\
    a^{\,\rm s}_{\bm{k}} &= -\Bigl({\rm Re}\,h^{(2\text{-}\chi_\perp)}_{\bm{k},+1}-{\rm Re}\,h^{(2\text{-}\chi_\perp)}_{\bm{k},-1}\Bigr)\,{\rm Im}\,h^{(1)}_{\bm{k}}\nonumber\\
    &\quad +\Bigl({\rm Im}\, h^{(2\text{-}\chi_\perp)}_{\bm{k},+1}-{\rm Im}\, h^{(2\text{-}\chi_\perp)}_{\bm{k},-1}\Bigr)\,{\rm Re}\, h^{(1)}_{\bm{k}},\\
    a^{\,\rm c}_{\bm{k}} &= \Bigl({\rm Re}\,h^{(2\text{-}\chi_\perp)}_{\bm{k},+1}+{\rm Re}\,h^{(2\text{-}\chi_\perp)}_{\bm{k},-1}\Bigr)\,{\rm Re}\,h^{(1)}_{\bm{k}}\nonumber\\
    &\quad + \Bigl({\rm Im}\, h^{(2\text{-}\chi_\perp)}_{\bm{k},+1}+{\rm Im}\, h^{(2\text{-}\chi_\perp)}_{\bm{k},-1}\Bigr)\,{\rm Im}\, h^{(1)}_{\bm{k}}.
\end{align}
\end{subequations}

Now, to compare these terms to the usual $n$PA counting, we note that the form of the evolution equations~\eqref{eq:dpsitilde/dt}--\eqref{eq:dpsisttilde/dt} immediately implies that $\mathring{\psi}_{\bm{k}}$ can be expanded as 
\begin{equation}
    \mathring{\psi}_{\bm{k}}(\e t,\e) = \frac{1}{\e}\left[\mathring{\psi}^{(0)}_{\bm{k}}(\e t)+\e\mathring{\psi}^{(1)}_{\bm{k}}(\e t)+\mathcal{O}(\e^2)\right]
\end{equation}
in terms of the `slow time' $\e t$; cf. Eq.~\eqref{phases}. Here $\mathring{\psi}^{(0)}_{\bm{k}}/\e$ is the 0PA term and subsequent orders are $n$PA. Comparing this to Eq.~\eqref{eq:waveform phase}, we see that the precession of the secondary spin contributes to the waveform phase at the same order in $\e$ as a 2PA effect.

One might hastily conclude two things: (i) if we are justified in neglecting 2PA terms when modeling some class of binaries (such as EMRIs), then we are equally well justified in neglecting the secondary spin precession; and (ii) if we think we should not include 1PA first-order conservative SF effects in a waveform model until we have also included 1PA second-order dissipative SF effects, then we should likewise not include the orthogonal component of the secondary spin unless we include all 2PA terms. 

However, both of these conclusions might be \emph{too} hasty. About the first, we note that the effect of the spin precession is qualitatively different than the effect of a 2PA forcing function: unlike a 2PA term, the precession terms (in both the phase and the amplitude) oscillate on a fast time scale rather than only varying on the slow, radiation-reaction time scale of the inspiral. About the second conclusion, we note that first-order conservative and second-order dissipative effects have the same qualitative effect on the waveform, and the division between them is gauge dependent. There is hence no reason to expect that including just one of them will improve the fidelity of a waveform model. This contrasts with the secondary spin precession, which has a qualitatively different signature than other effects that contribute to the waveform phase at the same order in $\e$. Ultimately, data analysis studies should assess the relevance of the secondary spin precession for gravitational-wave science. In Ref.~\cite{Mathews:2025txc}, we highlight that $a^{\,\rm c}_{\bm{k}}$, $d^{\,\rm c}_{\bm{k}}$, $a^{\,\rm s}_{\bm{k}}$ and $d^{\,\rm s}_{\bm{k}}$ are numerically small compared to $a^{\,0}_{\bm{k}}$ and $d^{\,0}_{\bm{k}}$ for quasicircular, approximately equatorial inspirals into slowly spinning primaries. Their relative contribution to the waveform for generic configurations warrants further investigation.

\section{Evolution with asymptotic fluxes}
\label{sec:flux balance}

As we have presented, and as summarized in the Introduction, the secondary spin contributes to the 1PA waveform in three ways: (i) through a correction $h^{(2\text{-}\chi_2)}_{\bm{k}q}$ to the waveform amplitudes, obtained by solving Eq.~\eqref{eq:EFE2 spin terms}, (ii) through a correction $\Omega^i_{(1\text{-}\chi_2)}$ to the orbital frequencies, given analytically in Eq.~\eqref{eq:Omega1-chi} in terms of the MPD force (and an arbitrary gauge choice), and (iii) through a correction $F^{(1\text{-}\chi_2)}_i$ to the 1PA forcing function.

The first two of these ingredients can be obtained independently from all other 1PA calculations; they do not require the solution to the first-order field equation~\eqref{eq:EFE1 multiscale}\footnote{The exception to this statement is if the fixed-frequencies gauge is not used. In that case, the source term $\delta G^{(1\text{-}\chi_2,\bm{k})}_{\alpha\beta}$ in Eq.~\eqref{eq:EFE2 spin terms} requires the solution to the first-order field equation as input. The analogous statement applies if solving the analogous Teukolsky equation.} or any local calculations of regular fields at the particle. However, in the form~\eqref{eq:F1-chi2} the forcing function $F^{(1\text{-}\chi_2)}_i$ requires as inputs the local force generated by the regular part of the solution $h^{(2\text{-}\chi_2,\bm{k},0)}_{\alpha\beta}$ to Eq.~\eqref{eq:EFE2 spin terms}; the regular field and first-order self-force extracted from the solution to the first-order field equation~\eqref{eq:EFE1 multiscale}; and the spin perturbation $\langle\delta S^\alpha\rangle_{\tilde\psi_s}$ in Eq.~\eqref{eq:<S>}, which in turn seems to require the solution to the nutation equations~\eqref{eq:dtheta_c eqn} and \eqref{eq:dtheta_s eqn} (though Appendix~\ref{sec:localforces} shows that the nutation equations do not need to be explicitly solved even when constructing $F^{(1\text{-}\chi_2)}_i$ from local forces and torques).

Remarkably, a flux-balance law recently derived by Grant~\cite{Grant:2024ivt} allows us to entirely bypass this complexity and instead calculate  $F^{(1\text{-}\chi_2)}_i$ solely from the values of $h^{(1,\bm{k})}_{\alpha\beta}$ and $h^{(2\text{-}\chi_2,\bm{k},0)}_{\alpha\beta}$ (or equivalent Teukolsky mode amplitudes) at future null infinity and at the primary's future horizon. Working in the pseudo-Hamiltonian framework of Ref.~\cite{Blanco:2023jxf}, Grant considered small corrections to spinning test-body motion due to the linear metric perturbation that the body produces, also ultimately linearizing in the spin. Such a framework is conceptually different than our own, but since the spin effects we are interested in are linear in the perturbation produced by the body, we can directly import Grant's result for those effects. 

We state Grant's flux-balance law as\footnote{See Eq.~(5.44) in Ref.~\cite{Grant:2024ivt}. Note that our $\mathcal{F}\bigl[h,\partial_{\vartheta^\alpha} h\bigr]$ corresponds to $(\partial_{\vartheta_\beta})^A\mathcal{F}_A$ in that equation, and we have used Eq.~(5.12) therein with $P_\beta$ replaced by $J_\beta$. Equation~(5.44) is directly for the constants of motion $P_\beta$, while we find it conceptually clearer to start with balance law for the actions.}
\begin{equation}\label{eq:AlexFluxBal J}
    \left\langle\frac{d J_\alpha}{d\tau}\right\rangle_{\!\tau} = -\left\langle\mathcal{F}\bigl[h,\partial_{\vartheta^\alpha_0} h\bigr]\right\rangle_\tau +\mathcal{O}(\e^2\bcancel{s},\e^3).
\end{equation}
Here we use $\mathcal{O}(\e^2\bcancel{s})$ to denote spin-independent 1PA terms. The variables $(\vartheta^\alpha,J_\alpha)$ are action-angle variables for the linearised MPD system in Kerr spacetime, and $\vartheta^\alpha_0$ is the angle value at the reference time where we wish to compute the average. On the left, $\langle\rangle_\tau$ is an average over proper time around the reference time. On the right, the bilinear operator $\left\langle\mathcal{F}\bigl[h,\partial_{\vartheta^\alpha_0} h\bigr]\right\rangle_\tau$ is a certain $\tau$-averaged flux to future null infinity and down the black hole horizon, constructed from the linear perturbation $h_{\alpha\beta}$ due to a spinning particle.

Later in this section, we describe how Eq.~\eqref{eq:AlexFluxBal J} can be recast as
\begin{equation}\label{eq:AlexFluxBal J2}
    \left\langle\frac{d J_\alpha}{dt}\right\rangle=-{\cal F}_\alpha +\mathcal{O}(\e^2 \bcancel{s},\e^3),
\end{equation}
where $\langle\cdot\rangle$ is our angle average, and (adapting the notation of Ref.~\cite{Isoyama:2018sib}) the flux is\footnote{For comparison with Ref.~\cite{Isoyama:2018sib}, we note that the mode numbers there are written as $k_i=(n,k,m)$.}
\begin{equation}\label{eq:fluxes}
{\cal F}_\alpha =\sum_{\ell\bm{k}}\frac{\varepsilon_\alpha}{4 \pi \omega_{\bm{k}}^3}\left( |\mathcal{Z}^{\text{out}}_{\ell\bm{k}}|^2+\frac{\omega_{\bm{k}}}{\mathcal{P}_{\bm{k}}}|\mathcal{Z}^{\text{down}}_{\ell \bm{k}}|^2\right).
\end{equation}
Here 
\begin{equation}\label{eq:flux frequencies}
\omega_{\bm{k}}\equiv k_i\Omega^i=k_i\Bigl(\Omega^i_{(0)}+\e \Omega^i_{(1\text{-}\chi_2)}\Bigr),
\end{equation}
$\varepsilon_{\alpha}\equiv(-\omega_{\bm{k}},k_r,k_\theta,k_\phi)$, $\mathcal{P}_{\bm{k}}\equiv \omega_{\bm{k}}-k_\phi \chi^{(0)}_1/(2r_+)$, and $r_+$ is the usual outer horizon radius. 
The quantities 
\begin{equation}\label{eq:flux amplitudes}
\mathcal{Z}^{\text{out/down}}_{\ell \bm{k}} = \mathcal{Z}^{(1)\text{out/down}}_{\ell \bm{k}} + \e \chi_\parallel\mathcal{Z}^{(2\text{-}\chi_\parallel)\text{out/down}}_{\ell \bm{k}}    
\end{equation}
are the usual Teukolsky mode amplitudes at the horizon (`down') and at future null infinity (`out'), which here are constructed from the solution to the (linear) Teukolsky equation with a spinning-particle source, corresponding to the sum of the solutions $h^{(1,\bm{k})}_{\alpha\beta}$ and $h^{(2\text{-}\chi_2,\bm{k},0)}_{\alpha\beta}$ to Eqs.~\eqref{eq:EFE1 multiscale} and~\eqref{eq:EFE2 spin terms}. As in the previous sections, there is no contribution from $\chi_\perp$ at this order.

Equation~\eqref{eq:AlexFluxBal J2} also implies evolution equations for the spin-corrected constants of motion $P_i=(E,L_z,K)$ defined in Eqs.~\eqref{eq:E and L} and \eqref{eq:CarterWspin}:
\begin{equation}
\label{eq:AlexFluxBal}
    \left\langle \frac{d P_i}{d t}\right\rangle=-\frac{\partial P_i}{\partial J_\alpha}\mathcal{F}_\alpha + \mathcal{O}(\e^2 \bcancel{s},\e^3),
\end{equation}
where it is understood that the Jacobian $\partial P_i/\partial J_\alpha$ is calculated for a test particle to linear order in spin. The form~\eqref{eq:AlexFluxBal} requires the relationships $P_i(J_\alpha)$, whose inverse relations $J_\alpha(P_i)$ were recently calculated in closed form by Witzany and collaborators~\cite{Witzany:2024ttz}. In particular, the linear spin corrections to the Jacobian $\partial J_\alpha/ \partial P_i$ can be obtained analytically from Eqs.~(B.9) and~(B.15) of Ref.~\cite{Witzany:2024ttz} (and $\partial P_i/ \partial J_\alpha$ from its matrix inverse).

Equation~\eqref{eq:AlexFluxBal} represents a complete, practical description of dissipation at 0PA and of the secondary spin's contribution to dissipation at 1PA order. At 0PA it reduces to the standard flux-driven evolution equations~\cite{Sago:2005fn,Isoyama:2018sib}. At 1PA, it recovers the spinning-body energy and angular momentum flux balance formulae of Ref.~\cite{Akcay:2019bvk} (also see Appendix~\ref{sec:localforces}), and it provides a formula for $dK/dt$ that completes the description. 

In Sec.~\ref{sec:combo} below, we describe how Eq.~\eqref{eq:AlexFluxBal J} is translated into the forms~\eqref{eq:AlexFluxBal J2} and \eqref{eq:AlexFluxBal}. In Sec.~\ref{sec:pragmatic} we outline how to use Eq.~\eqref{eq:AlexFluxBal} to compute the forcing functions $F^{(1\text{-}\chi_2)}_i$ in the 1PA waveform-generation scheme of Sec.~\ref{sec:waveform generation}. Readers uninterested in the technical details can skip directly to Sec.~\ref{sec:pragmatic}.

\subsection{Importing the results of Grant, Witzany et al., and Isoyama et al.}
\label{sec:combo}

We consider each of the three ingredients in Eq.~\eqref{eq:AlexFluxBal} in turn: actions, averages, and fluxes.

\subsubsection{Actions}

Grant's result~\eqref{eq:AlexFluxBal J} is valid for any set of phase-space coordinates $(\vartheta^{\cal A},J_{\cal A})$ that behave as action-angle variables in the test-body limit, by which we mean $d\vartheta^{\cal A}/d\tau=\partial H/\partial J_{\cal A}=\nu_{\cal A}(J_{\cal B})$ and $dJ_{\cal A}/d\tau=0$ when $h^{\rm R}_{\alpha\beta}\to0$. Here calligraphic indices denote coordinates on the 10D phase space for the test-body MPD dynamics at linear order in spin, and $H$ is a suitable Hamiltonian for the MPD dynamics. To relate this setting to ours, note the 10D phase space has  (noncanonical) coordinates $(x^{\cal A},p_{\cal A})$ with $x^{\cal A}=(x^\alpha,\psi_s)$ and $p_{\cal A}=(p_\alpha,p_{\psi_s})$, where $p_{\psi_s}=\chi_\parallel-\chi_2$~\cite{Witzany:2019nml}. The bulk of our paper works instead on the physical 7D submanifold defined by the mass-shell condition $\sqrt{-g^{\alpha\beta}p_\alpha p_\beta}=m_2$ and $p_{\psi_s}=\text{constant}$, reducing~$t$ to a parameter rather than a coordinate.

Fortunately, Witzany et al.~\cite{Witzany:2019nml,Witzany:2024ttz} have recently provided action-angle coordinates that are appropriate for use in Eq.~\eqref{eq:AlexFluxBal J}, based on a Hamilton-Jacobi formulation of the test-body MPD equations (through linear order in $s$ and assuming the Tulczyjew-Dixon SSC). The actions are given by
\begin{equation}\label{eq:JtJphiJpsi}
J_t=-E,\quad J_\phi=L_z,\quad J_{\psi_s}=\chi_\parallel-\chi_2,     
\end{equation}
and
\begin{equation}\label{eq:Jy}
    J_y(P_{\cal B})\equiv \frac{1}{2\pi m_2}\oint_{\gamma_y} \Pi_{\cal A} dx^{\cal A},
\end{equation}
for $y=r,\theta$. Here $\gamma_y$ are any two homotopically inequivalent closed radial and polar contours on the torus of constant $P_{\cal A}=(m_2,E,L_z,K,\chi_\parallel)$ (and any constant $t$), and we have introduced a factor $1/m_2$ to work with actions that are $m_2$-independent at leading order. The quantities $\Pi_{\cal B}=(\Pi_\mu,\Pi_{\psi_s})$ are the momenta conjugate to $x^{\mathcal{A}}=\{x^\alpha, \psi_s\}$~\cite{Witzany:2018ahb,Witzany:2024ttz}, related to $p_{\cal A}$ by $\Pi_{\psi_s}=p_{\psi_s}$ and 
\begin{equation}
    \Pi_\mu \equiv p_\mu+\frac{1}{2}m_2^2 \bar\omega_{AB \mu}S^{AB},
\end{equation}
with $\bar \omega_{AB \mu}\equiv(\nabla_\mu \E_{A}^{\alpha}) \E_{B \alpha}$ such that $u^\mu \bar \omega_{AB \mu}=\omega_{AB}$ and $S^{AB}$ are the triad components of the (dimensionless) spin tensor.\footnote{To define the derivative $\nabla_\mu \E_{A}^{\alpha}$, we must promote the tetrad $\E_{A}^{\alpha}$ to a field in a neighborhood of the worldline. Since the tetrad is defined along a geodesic (or an accelerated curve as in Sec.~\ref{sec:spin}), it is immediately promoted to a field by considering a congruence of such curves~\cite{Witzany:2019nml,Witzany:2024ttz}.} Equation~\eqref{eq:Jy} also applies for $y=\phi,\psi_s$ with appropriate closed contours $\gamma_y$, but in those cases it immediately reduces to Eq.~\eqref{eq:JtJphiJpsi} because $\Pi_\phi=L_z$ and $\Pi_{\psi_s}=\chi_\parallel-\chi_2$ are constant on the torus. 

In Ref.~\cite{Witzany:2024ttz}, Witzany and collaborators derived closed-form analytical expressions for these action variables in terms of the test-body conserved quantities $P_{\cal A}$ by performing the loop integrals with Hadamard finite-part integration. The results are linearized in spin in the form 
\begin{equation}\label{eq:J(P) expansion}
    J_\alpha = J^{(0)}_\alpha(P_i) + \e \chi_\parallel J^{(1\text{-}\chi_2)}_\alpha(P_i) +\mathcal{O}(s^2),
\end{equation}
with no contribution from $\chi_\perp$ at linear order. 
They also provided the linear spin corrections to the Jacobian $\partial J_\alpha/\partial P_i$, meaning $\partial J^{(1\text{-}\chi_2)}_\alpha/\partial P_i$, from which we can obtain the linear spin correction to  $\partial P_i/\partial J_\alpha$ in Eq.~\eqref{eq:AlexFluxBal}. Many elements of the Jacobian are trivial by virtue of the relations~\eqref{eq:JtJphiJpsi}, and (as we explain in Sec.~\ref{sec:pragmatic}) ultimately the only elements required are $\partial J_y/\partial P_i$ for either $y=r$ or $y=\theta$. 

These results are conveniently available in a Mathematica notebook in the supplemental material of Ref.~\cite{Witzany:2024ttz}. That notebook also contains closed-form expressions for the frequency corrections due to the spin, in the form
\begin{equation}\label{eq:Omega(P) expansion}
\Omega^i=\Omega^i_{(0)}(P_j) + \e\chi_\parallel\Omega^i_{(1\text{-}\chi_2)}(P_j) +\mathcal{O}(s^2);   
\end{equation}
recall that $\chi_\perp$ cannot contribute because the frequency involves an average over the precession phase. Both this and Eq.~\eqref{eq:J(P) expansion} are expansions in $\e$ at fixed $P$. Since $P_i=\langle P_i\rangle$ for a spinning test particle, we can replace $P_i$ with $\langle P_i\rangle$ in the above expressions. The expressions then also hold true in the presence of self-force in the fixed-constants-of-motion gauge discussed in Sec.~\ref{sec:gauge choices}; otherwise, they omit $\mathcal{O}(\e \bcancel{s})$ self-force terms.

\subsubsection{Averages}

Grant's derivation is based on regular perturbation theory rather than a multiscale expansion. Specifically, he works with a spinning particle, considers the linear perturbation it sources, and finds the effect of that linear perturbation on its motion. This type of approach is well known to be ill-behaved on large time and space scales~\cite{Pound:2015wva}, but incorporating the results of such an approach into well-behaved schemes is also standard SF lore.

We consider a spacetime described by our multiscale expansion~\eqref{eq:h multiscale}. To put it in the form assumed by Grant, we expand our spacetime metric (and particle trajectory) in an ordinary power series in $\e$ near an arbitrary time $t_0$. We then define an average with respect to a time $\lambda$ as
\begin{equation}\label{eq:time average}
    \langle \cdot \rangle_\lambda \equiv \lim_{T\to\infty}\frac{1}{2T}\int_{\lambda(t_0)-T}^{\lambda(t_0)+T}\!\! \cdot\; d\lambda. 
\end{equation}
This is the average in Eq.~\eqref{eq:AlexFluxBal}, with $\lambda=\tau$.

Before mapping quantities in the regular expansion onto those in the multiscale expansion, we first convert Eq.~\eqref{eq:AlexFluxBal J} to an average over $t$. At the same time, we effectively convert to the formalism in the body of this paper: a 3+1 split in which $t$ is a parameter along trajectories rather than a coordinate. This is achieved by observing that if $J_\alpha$ is conjugate to $\vartheta^\alpha$ for a Hamiltonian $H$ that generates proper-time evolution [i.e., $d\vartheta^\alpha/d\tau=\partial H/\partial J_\alpha=\nu^\alpha(J_\beta)$], then $J_\alpha$ is also conjugate to $\mathring\psi^\alpha$ for a Hamiltonian ${\cal H}$ that generates coordinate-time evolution [i.e., $d\mathring\psi^i/dt=\partial {\cal H}/\partial J_i=\Omega^i(J_k)$ and $\mathring\psi^t=t$]~\cite{Blanco:2022mgd,Kakehi:2024bnh,Lewis:2025ydo}. Using ${\cal H}$ in place of $H$ in Grant's derivation yields
\begin{equation}\label{eq:F[h,dh]=F}
    \left\langle\frac{d J_\alpha}{dt}\right\rangle_{\!t} = -\left\langle\mathcal{F}\Bigl[h,\partial_{\mathring\psi^\alpha_0} h\Bigr]\right\rangle_t\equiv -\mathcal{F}_\alpha,
\end{equation}
with $\mathring\psi^t_0\equiv t_0$ and $\mathring\psi^i_0\equiv\mathring\psi^i(t_0)$.\footnote{When using $t$ as a time parameter, Hamilton's equations only apply for $\alpha=i$. As a consequence, Grant's derivation only yields Eq.~\eqref{eq:F[h,dh]=F} for $\alpha=i$. However, the equation for $\alpha=t$ follows from the others as an equation for the Hamiltonian itself: $\left\langle \frac{d{\cal H}}{dt}\right\rangle_t =  \left\langle \frac{\partial{\cal H}}{\partial J_i}\frac{dJ_i}{dt}\right\rangle_t = \left\langle\Omega^i  \frac{dJ_i}{dt}\right\rangle_t = -\left\langle{\cal F}\Bigl[h,\Omega^i\partial_{\mathring\psi^i_0} h\Bigr]\right\rangle_t$. This agrees with the $\alpha=t$ component of Eq.~\eqref{eq:F[h,dh]=F} because $J_t=-{\cal H}$ and, by virtue of Eq.~\eqref{eq:psi near t0}, $h_{\alpha\beta}$ depends on $\mathring\psi^\alpha_0$ in the combination $(\mathring\psi^i_0-\Omega^i t_0)$.} Here $\langle\mathcal{F}\rangle_t$ denotes the sum of fluxes through the horizon and out to future null infinity, now averaged over advanced time $v$ at the horizon and retarded time $u$ at infinity. 
Alternatively, we can obtain this result directly from Eq.~\eqref{eq:AlexFluxBal J} using  the general relation $\langle df/d\tau\rangle_\tau=\langle df/dt\rangle_t/\langle d\tau/dt\rangle_t$~\cite{Drasco:2005kz,Isoyama:2018sib} and the Jacobian $\partial\mathring\psi^\beta/\partial\vartheta^\alpha$.

To relate $\langle dJ_\alpha/dt\rangle_t$ to $\langle dJ_\alpha/dt\rangle$, we now examine the regular expansion around $t_0$. Inspecting Eqs.~\eqref{eq:ringed equations} reveals that $(\mathring\psi^i,\mathring\pi_i)$ can be written as functions $\mathring\psi^i(\e t,\e)$ and $\mathring\pi_i(\e t,\e)$, where $\mathring\psi^i(\e t,\e)$ has the expansion~\eqref{phases}, and $\mathring\pi_i(\e t,\e)$ has an expansion
\begin{equation}
    \mathring\pi_i(\e t,\e) = \mathring\pi^{(0)}_i(\e t) + \e \mathring\pi^{(1)}_i(\e t) + \mathcal{O}(\e^2).
\end{equation}
We expand these around their values at $t_0$ using $\e t = \tilde t_0 + \e \Delta t$ with $\Delta t\equiv (t-t_0)$: 
\begin{align}
    \mathring\psi^i(\e t, \e) &= \mathring\psi^i(\tilde t_0,\e) + \Delta t\,\Omega^i(\tilde t_0,\e) + \mathcal{O}(\e \Delta t^2),\label{eq:psi near t0}\\
    \mathring\pi_i(\e t,\e) &= \mathring\pi_i(\tilde t_0,\e) + \e\Delta t\frac{d}{d\tilde t_0}\mathring\pi_i(\tilde t_0,\e) + \mathcal{O}(\e^2 \Delta t^2).\label{eq:pi near t0}
\end{align}
The first term in Eq.~\eqref{eq:pi near t0} represents the constant parameters of a test-particle orbit (plus self-force contributions), 
\begin{equation}\label{eq:pi near t0 - chi}
\mathring\pi_i(\tilde t_0,\e)=\mathring\pi^{(0)}_i(\tilde t_0)+\e \mathring\pi^{(1\text{-}\chi_2)}_i(\tilde t_0)+\mathcal{O}(\e \bcancel{s}); 
\end{equation}
and the first two terms in Eq.~\eqref{eq:psi near t0} represent the test-particle orbital phases (plus self-force terms), with arbitrary initial values $\mathring\psi^i(\tilde t_0,\e)$ and constant frequencies 
\begin{equation}\label{eq:Omega near t0 - chi}
\Omega^i(\tilde t_0,\e) = \Omega^i(\tilde t_0)+\e\Omega^i_{(1\text{-}\chi_2)}(\tilde t_0)+\mathcal{O}(\e \bcancel{s}).
\end{equation}

Unlike $\mathring\pi_i$, the test-body constants of motion $P_i$, defined in Eqs.~\eqref{eq:E and L} and \eqref{eq:CarterWspin}, contain oscillatory contributions due to the self-force, and the actions $J_\alpha(P_i)$ inherit those oscillations. We divide them into averaged and oscillatory terms, 
\begin{equation}\label{eq:J split}
J_\alpha = \langle J_\alpha\rangle + J^{\rm osc}_\alpha,    
\end{equation}
as defined generically in Eq.~\eqref{eq:av osc split}. $\langle J_\alpha\rangle$ is a function of $\mathring\pi_i$ and therefore has an expansion of the form~\eqref{eq:pi near t0},
\begin{equation}\label{eq:Jav near t0}
    \langle J_\alpha\rangle = \langle J_\alpha\rangle(\tilde t_0,\e) + \e \Delta t \frac{d}{d\tilde t_0}\langle J_\alpha\rangle(\tilde t_0,\e) + \mathcal{O}(\e^2\Delta t^2).
\end{equation}
On the other hand, $J^{\rm osc}_\alpha$ is an oscillatory function of the phases and hence has an expansion
\begin{equation}\label{eq:Josc near t0}
    J^{\rm osc}_\alpha =  \e\sum_{\bm{k}\neq0}J^{(1,\bm{k})}_\alpha(\tilde t_0,\e)e^{-i\bigl(\mathring\psi^0_{\bm{k}}+\omega_{\bm{k}}\Delta t\bigr)} + \mathcal{O}(\e^2\Delta t^2)
\end{equation}
with $\mathring\psi^0_{\bm{k}}=k_i\mathring\psi^i(\tilde t_0,\e)$ and $\omega_{\bm{k}}=k_i\Omega^i(\tilde t_0,\e)$.

By substituting Eq.~\eqref{eq:J split} into the time average~\eqref{eq:time average}, we immediately find
\begin{equation}\label{eq:<dJdt>_t=d<J>/dt}
    \left\langle \frac{d J_\alpha}{d t}\right\rangle_{\!t} =  \e\frac{d\langle J_\alpha\rangle(\tilde t_0,\e)}{d\tilde t_0} +\mathcal{O}(\e^2).
\end{equation}
We also see that the $t$ average is ill defined beyond order $\e$ because of the terms quadratic (and higher order) in $\Delta t$ in Eqs.~\eqref{eq:Jav near t0} and \eqref{eq:Josc near t0}. We hence define the average to apply only at leading order, setting subleading terms to zero before averaging.

Since the equality~\eqref{eq:<dJdt>_t=d<J>/dt} applies for any $\tilde t_0$, we can rewrite it as
\begin{equation}
    \left\langle \frac{d J_\alpha}{d t}\right\rangle_{\!t} =  \frac{d\langle J_\alpha\rangle}{dt},
\end{equation}
discarding subleading terms as explained above. We also have the trivial identity
\begin{equation}\label{eq:<dfdt>=d<f>/dt}
    \left\langle \frac{d f}{d t}\right\rangle = \frac{d\langle f\rangle}{dt}
\end{equation}
for any $f$ since $\left\langle d f^{\rm osc}/d t\right\rangle=0$. Combining these, we obtain our desired identity:
\begin{equation}
    \left\langle \frac{d J_\alpha}{d t}\right\rangle_{\!t} = \left\langle \frac{d J_\alpha}{d t}\right\rangle.
\end{equation}

We similarly consider the flux ${\cal F}_\alpha$ on the right-hand side of Eq.~\eqref{eq:F[h,dh]=F}, which is an integral of products of $h_{\alpha\beta}$ and $\partial_{\mathring\psi^\alpha_0}h_{\alpha\beta}$ over cuts of the horizon and future null infinity, averaged over all time along those surfaces. Here $h_{\alpha\beta}$ is the linear perturbation sourced by a spinning particle on an orbit with constant parameters~\eqref{eq:pi near t0 - chi} and corresponding constant frequencies~\eqref{eq:Omega near t0 - chi}, discarding the $\mathcal{O}(\e \bcancel{s})$ terms in those equations. We can write this metric perturbation as  
\begin{equation}\label{eq:hModes}
h_{\alpha\beta} = \sum_{\bm{k}\in\mathbb{Z}^3}h^{(\bm{k})}_{\alpha\beta}\,e^{-i\bigl(\mathring\psi_{\bm{k}}^0+\omega_{\bm{k}}\Delta t\bigr)} + \mathcal{O}(\e^2\bcancel{s},\e^3,\chi_\perp)
\end{equation}
with 
\begin{multline}
    h^{\bm{k}}_{\alpha\beta} = \e \mathring h^{(1,\bm{k})}_{\alpha\beta}(\mathring\pi_i,x^i) + \e^2\chi_\parallel \mathring h^{(2\text{-}\chi_\parallel,\bm{k})}_{\alpha\beta}(\mathring\pi_i,x^i),
\end{multline}
in the notation of previous sections, where $\mathring\pi_i$ is evaluated at $t_0$ and we  omit the $\mathcal{O}(\e \bcancel{s})$ terms in Eqs.~\eqref{eq:pi near t0 - chi} and~\eqref{eq:Omega near t0 - chi}.  We also omit the term proportional to $\chi_\perp$ in the metric perturbation, which  can only contribute to the flux at $\mathcal{O}(s^2)$ due to its oscillatory dependence on the precession phase (just as it could only contribute to the local dynamics at that order). Given Eq.~\eqref{eq:hModes}, the derivative in $\partial_{\mathring\psi^\alpha_0}h_{\alpha\beta}$ can be replaced by 
\begin{equation}
    \frac{\partial}{\partial\mathring\psi^\alpha_0} = -i\varepsilon_\alpha,
\end{equation}    
where $\varepsilon_\alpha$ is defined below Eq.~\eqref{eq:flux frequencies} and we used $\partial_{t_0}\Delta t= -1$.

Equation~\eqref{eq:hModes} is precisely what one would obtain for the metric perturbation by substituting the expansions~\eqref{eq:psi near t0}, \eqref{eq:pi near t0 - chi}, and \eqref{eq:Omega near t0 - chi} into the multiscale expansion~\eqref{eq:h multiscale} with Eqs.~\eqref{eq:h1 Fourier} and \eqref{eq:h2 Fourier}. The average over time along the horizon and future null infinity also yields precisely the same expression in terms of Fourier mode amplitudes and frequencies as one would obtain by averaging over angles in the multiscale expansion, noting we can neglect $\delta m_1$ and $\delta \chi_1$ when calculating linear-in-spin 1PA terms.

\subsubsection{Fluxes}

Finally, we convert the fluxes $\mathcal{F}_\alpha$ into the form~\eqref{eq:fluxes}. These fluxes, as mentioned above, are defined from time averages of symplectic currents, which are products of the retarded metric perturbation evaluated on the horizon and at future infinity. They can be expressed in terms of Teukolsky amplitudes following standard methods of metric reconstruction~\cite{Grant:2020nmu}. However, we can skip that calculation by appealing to the results of Isoyama et al. for $\langle dJ_\alpha/dt\rangle$ in the case of a nonspinning particle~\cite{Isoyama:2018sib}.

The essential point is that $\mathcal{F}_\alpha\equiv \mathcal{F}\bigl[h,\partial_{\mathring\psi^\alpha_0}h\bigr]$ is an identical function of $h_{\mu\nu}$ and $\partial_{\mathring\psi^\alpha_0}h_{\mu\nu}$ regardless of whether $h_{\mu\nu}$ is sourced by a spinning or a nonspinning particle. More concretely, it is an identical function of the mode coefficients $h^{\bm{k}}_{\alpha\beta}$ and frequencies $\omega_{\bm{k}}$ in Eq.~\eqref{eq:hModes} regardless of their values. The conversion into Teukolsky amplitudes is likewise independent of the values of $h^{\bm{k}}_{\alpha\beta}$ and frequencies $\omega_{\bm{k}}$. Therefore the expression for $\mathcal{F}_\alpha$, and hence for $\langle dJ_\alpha/dt\rangle$, in terms of Teukolsky amplitudes and mode frequencies, is functionally the same for a spinning body as for a non-spinning body. It follows that $\mathcal{F}_\alpha$ for a spinning particle must be the same function of Teukolsky amplitudes and mode frequencies as Isoyama et al.'s result for $\langle dJ_\alpha/dt\rangle$ for a nonspinning particle. This is the formula reproduced in Eq.~\eqref{eq:fluxes}. 

\subsection{Pragmatic summary}
\label{sec:pragmatic}

To incorporate Eq.~\eqref{eq:AlexFluxBal} into the 1PA waveform generation scheme of Sec.~\ref{sec:waveform generation}, we only need to extract its linear-in-spin term and convert it into an expression for $d\mathring\pi_i/dt$.

We first write Eq.~\eqref{eq:AlexFluxBal} more explicitly. Noting $J_t=-E$ and $J_\phi=L_z$ along with Eq.~\eqref{eq:<dfdt>=d<f>/dt}, we have immediately
\begin{subequations}\label{eq:Edot and Ldot}
  \begin{align}
    \frac{d\langle E\rangle}{dt} &= {\cal F}_t +\mathcal{O}(\e^2 \bcancel{s}),\\
    \frac{d \langle L_z\rangle }{dt} &= -{\cal F}_\phi+\mathcal{O}(\e^2 \bcancel{s}).
\end{align}  
\end{subequations}
Next, to obtain the evolution formula for the Carter constant, rather than expanding the Jacobian in Eq.~\eqref{eq:AlexFluxBal}, we simply rearrange 
\begin{equation}
\frac{dJ_r}{dt} = \frac{\partial J_r}{\partial  P_j }\frac{d P_j}{dt}    
\end{equation}
to obtain 
\begin{align}
    \frac{dK}{dt} &= \left(\frac{\partial J_r}{\partial K}\right)^{\!-1}\left(\frac{dJ_r}{dt} - \frac{\partial J_r}{\partial E}\frac{dE}{dt} - \frac{\partial J_r}{\partial L_z}\frac{dL_z}{dt}\right).
\end{align}
Taking the average yields
\begin{equation}\label{eq:Kdot}
    \frac{d\langle K\rangle}{dt}
    = \left(\frac{\partial J_r}{\partial K}\right)^{\!-1}\left(-{\cal F}_r - \frac{\partial J_r}{\partial E}{\cal F}_t + \frac{\partial J_r}{\partial L_z}{\cal F}_\phi\right),
\end{equation}
where it is understood that in $\partial J_r/\partial P_i$ we only include the linear spin correction (neglecting the 1SF correction, which contains oscillations) and replace $P_i$ with $\langle P_i\rangle$. The same expression holds with $J_r$ replaced by $J_\theta$.

Equations~\eqref{eq:Edot and Ldot} and \eqref{eq:Kdot} can be straightforwardly linearized in $\chi_\parallel$ (noting $\chi_\perp$ does not appear). This involves substituting the frequencies~\eqref{eq:flux frequencies} and amplitudes~\eqref{eq:flux amplitudes} into the fluxes~\eqref{eq:fluxes} and substituting the radial or polar action~\eqref{eq:J(P) expansion}.

The explicit form of the result depends on the choice of phase space gauge. In principle, the evolution equations~\eqref{eq:Edot and Ldot} and \eqref{eq:Kdot} apply in any gauge. However, the actions~\eqref{eq:J(P) expansion} and frequencies~\eqref{eq:Omega(P) expansion} are expressed in the fixed-constants-of-motion gauge discussed in Sec.~\ref{sec:gauge choices}. Meanwhile, the field equation for the mode amplitudes ${\cal Z}^{(2\text{-}\chi_\parallel)\text{out/down}}_{\ell \bm{k}}$ is simplest in the fixed-frequencies gauge. In the remainder of this section, we summarize the prescription in these two gauges, which we follow Ref.~\cite{Piovano:2024yks} in labeling `FF' (fixed frequencies) and `FC' (fixed constants).

We first derive the transformation between the two gauges. We assume the gauge freedom $\langle\Delta\psi^i\rangle$ is specified in the same way in both cases. We also note that the gauge imposed on 1SF effects is independent of the gauge imposed on linear spin effects, and we only consider the latter here. 

The frequencies in the FC gauge are given by Eq.~\eqref{eq:Omega(P) expansion}, which we rewrite as
\begin{equation}\label{eq:Omega - FC}
\Omega^i = \Omega^i_{(0)}(\mathring \pi^{\rm FC}_j) + \e \chi_\parallel\Omega^i_{(1\text{-FC})}(\mathring \pi^{\rm FC}_j).
\end{equation}
Here $\mathring\pi_i$ are geodesically related to $\langle P_i\rangle$, or equivalently, $\langle P_i\rangle(\mathring\pi^{\rm FC}_j)=P^{(0)}_i(\mathring\pi^{\rm FC}_j)$. To transform to the FF gauge, we write
\begin{equation}
\mathring\pi_i^{\rm FC}=\mathring\pi_i^{\rm FF} + \e\chi_\parallel\delta\pi^{\rm FF}_i.
\end{equation}
We then substitute this into Eq.~\eqref{eq:Omega - FC} and enforce the fixed-frequencies condition
\begin{equation}
    \Omega^i = \Omega^i_{(0)}(\mathring\pi^{\rm FF}_j),
\end{equation}
which yields
\begin{equation}\label{eq:FC to FF}
    \delta\pi_j^{\rm FF} = - \frac{\partial\mathring\pi_j}{\partial\Omega^i_{(0)}}\Omega^i_{(1\text{-FC})}. 
\end{equation}
Here $\partial\mathring\pi_j/\partial\Omega^i_{(0)}$ represents the inverse of the geodesic Jacobian $\partial\Omega^i_{(0)}/\partial\mathring\pi_j$.

With Eq.~\eqref{eq:FC to FF} in hand, we now summarize the prescription in the two gauges:
\begin{enumerate}
    \item In the FC gauge, we can read off the linear spin contribution to the forcing function in the final evolution equation~\eqref{eq:ringed equations} from Eq.~\eqref{eq:tbguageflux}:
\begin{equation}
    F^{(1\text{-FC})}_i = \frac{\partial \mathring \pi_i}{\partial P^{(0)}_k} \left(\frac{d\langle P_k\rangle}{dt}\right)^{(1\text{-FC})},
\end{equation}
where $\partial \mathring \pi_i/\partial P^{(0)}_k$ is the inverse of the geodesic Jacobian $\partial P^{(0)}_k/\partial \mathring \pi_i$. $\left(\frac{d\langle P_k\rangle}{dt}\right)^{(1\text{-FC})}$ is the linear spin term in Eqs.~\eqref{eq:Edot and Ldot} and~\eqref{eq:Kdot}, extracted using the expansions of the frequencies~\eqref{eq:flux frequencies}, amplitudes~\eqref{eq:flux amplitudes}, and radial action~\eqref{eq:J(P) expansion}. The mode amplitudes ${\cal Z}^{(2\text{-}\chi_\parallel)\text{out/down}}_{\ell \bm{k}}$ are calculated from the Teukolsky analog of Eq.~\eqref{eq:EFE2 spin terms}, accounting for $\Omega^i_{(1\text{-FC})}$ terms in the noncompact source term. Alternatively, ${\cal Z}^{(2\text{-}\chi_\parallel)\text{out/down}}_{\ell \bm{k}}$ in the FC gauge can be calculated by first solving the Teukolsky analog of Eq.~\eqref{eq:EFE2 spin terms} in the FF gauge (with no noncompact source term), and then using
\begin{equation}
    {\cal Z}^{(2\text{-FC})}_{\ell \bm{k}} = {\cal Z}^{(2\text{-FF})}_{\ell \bm{k}} 
    -\chi_\parallel \delta\pi^{\rm FF}_i\partial_{\mathring\pi_i}{\cal Z}^{(1)}_{\ell \bm{k}}.
\end{equation}
    \item In the FF gauge, we have
    \begin{equation}
        \hspace{20pt}\frac{d\mathring\pi_i}{dt} = \frac{\partial\mathring\pi_i}{\partial\Omega^j_{(0)}}\left(\frac{\partial\Omega^j_{(0)}}{\partial P_k} +\e\chi_\parallel \frac{\partial\Omega^j_{(1\text{-FC})}}{\partial P_k}\right)\frac{d\langle P_k\rangle}{dt},
    \end{equation}
    where $\partial\mathring\pi_i/\partial\Omega^j_{(0)}$ is the inverse of the geodesic Jacobian. From this, we read off the linear spin term:
    \begin{multline}
        \hspace{20pt}F^{(1\text{-}FF)}_i = \frac{\partial\mathring\pi_i}{\partial P^{(0)}_k}\left(\frac{d\langle P_k\rangle}{dt}\right)^{(1\text{-FF})} \\
        + \chi_\parallel\frac{\partial\mathring\pi_i}{\partial\Omega^j_{(0)}} \frac{\partial\Omega^j_{(1\text{-FC})}}{\partial P_k}\left(\frac{d\langle P_k\rangle}{dt}\right)^{\!(0)},
    \end{multline}    
    where $\partial\mathring\pi_i/\partial P^{(0)}_k$ is the inverse of the geodesic Jacobian. In calculating $\left(\frac{d\langle P_k\rangle}{dt}\right)^{(1\text{-FF})}$, we set $\Omega^i_{(1\text{-}\chi_2)}$ to zero everywhere it appears. Specifically, it is set to zero in the fluxes~\eqref{eq:fluxes}, and there is no noncompact source term in the Teukolsky equation for the amplitudes ${\cal Z}^{(2\text{-}\chi_\parallel)\text{out/down}}_{\ell \bm{k}}$. In Eq.~\eqref{eq:Kdot} we require $J_r$ in the FF gauge. This is straightforwardly obtained from its value in the FC gauge:  
    \begin{multline}
    \hspace{20pt}J_\alpha = J^{(0)}_\alpha(\mathring \pi^{\rm FF}_i) + \e \chi_\parallel J^{(1\text{-FC})}_\alpha(\mathring \pi^{\rm FF}_i) \\
    + \e\chi_\parallel\delta\pi^{\rm FF}_i\frac{\partial P_j^{(0)}}{\partial\mathring\pi_i}\frac{\partial J^{(0)}_\alpha}{\partial P_j^{(0)}}.
    \end{multline}
\end{enumerate}

\section{Discussion and conclusions}
\label{sec:conclusions}
We conclude with a summary of this work and of progress toward generic 1PA waveform models.

\subsection{This work}

In this paper, we have extended the 1PA multiscale waveform-generation framework of Ref.~\cite{Pound:2021qin} to include a generic secondary spin. The framework accounts for all 1PA effects for generic orbits of a spinning secondary around a Kerr black hole, including, in particular, all 1PA spin-precession effects (though excluding orbital resonances). The scheme is summarized in Sec.~\ref{sec:waveform generation}, and the contribution of the secondary spin is further outlined and streamlined in Sec.~\ref{sec:flux balance}.

Our analysis began in Sec.~\ref{sec:spin} with a detailed study of the spin degrees of freedom for a spinning, gravitating secondary body with an arbitrary, self-accelerated orbital configuration and precessing spin orientation in Kerr spacetime. We have characterized both the precession and the nutation of the secondary's spin with simple parameters and corresponding evolution formulae. Our formulation has elucidated that the conserved quantity related to Rüdiger's constant ($\chi_\parallel$) is exactly constant in the MPD-Harte system, at linear order in spin, as a simple consequence of the existence of an orthonormal Fermi-Walker-transported basis in the effective metric. 

Having suitably parameterized the secondary spin, we incorporated it into the multiscale framework in Secs.~\ref{sec:generic} and \ref{sec:waveform generation}. 
In doing so, we recovered the known result that the secondary spin's precession decouples from the 1PA orbital evolution (at least away from resonances). This decoupling was easily anticipated~\cite{Witzany:2023bmq, Drummond:2023wqc, Skoupy:2023lih, Piovano:2024yks} because any precession effect at linear order in the spin is purely oscillatory, meaning it cannot survive averaging over the precession period. However, we also showed the less obvious result that the nutation equations do not need to be solved at 1PA order. This required a more careful analysis because the nutation generates a force that \emph{does} survive precession-averaging and does contribute to the 1PA orbital dynamics. 

Despite the fact that the precession does not enter the 1PA orbital evolution, we advocated in Sec.~\ref{sec:waveform generation} that its direct contribution to the waveform, through an additional $\mathcal{O}(\e^2)$ oscillatory amplitude, is worth calculating. Since it represents a qualitatively new feature in the waveform, modulating the waveform phase and amplitude, it is potentially relevant for data analysis in some regions of parameter space. On the other hand, nutation will only make a direct contribution to the waveform at a still higher order in the mass ratio, in an order-$\e^3$ term sourced by the $\delta S^\alpha$ contribution to the dipole stress-energy tensor $T^{(d)}_{\alpha\beta}$. This makes nutation's direct contribution exceedingly unlikely to be relevant (unlike its indirect contribution through its impact on the 1PA orbital dynamics).

In addition to incorporating secondary spin, we have also illuminated broader aspects of the multiscale framework, specifically streamlining the derivation of the final orbital evolution equations in Sec.~\ref{sec:orbital motion} and analysing the framework's gauge freedom in Secs.~\ref{sec:gauge choices} and \ref{sec:waveform gauge invariance}. We described various gauge choices in the 1PA waveform generation scheme. We then formalized the invariance of the resulting waveform under such choices, while emphasizing that, despite this invariance, the choices \emph{can have differing implications on waveform accuracy}. 

Finally, we have shown how to combine the results of Grant~\cite{Grant:2024ivt}, Witzany et al.~\cite{Witzany:2024ttz} and Isoyama et al.~\cite{Isoyama:2018sib} into an evolution formula for the spin-corrected Carter constant in terms of Teukolsky amplitudes. As explained in Sec.~\ref{sec:flux balance}, this formula, alongside the energy and angular momentum flux balance formulae of Ref.~\cite{Akcay:2019bvk}, can be readily incorporated into our multiscale framework, where it enables calculations of the secondary spin's complete contribution to 1PA waveforms while avoiding any evaluation of local self-forces and regular fields at the particle.
Computing the secondary spin's contribution to the 1PA forcing function $F^{(1\text{-}\chi_2)}_i$ across the parameter space hence only requires the Teukolsky amplitudes for a spinning secondary as input (and only the contribution from the nonprecessing component of the spin). These amplitudes are now available from Refs.~\cite{Skoupy:2023lih, Piovano:2024yks}. 

In Appendix~\ref{sec:localforces}, we have also presented a simplified local expression for the averaged rate of change of the Carter constant in terms of the effective metric at the particle, which may be used as a consistency check of the Teukolsky flux formula in future work.

\subsection{Path to a complete 1PA waveform model}

The first 1PA waveform model was limited to nonspinning, quasicircular binaries~\cite{Wardell:2021fyy}. The most general 1PA waveform models at the time of writing are presented in a series of upcoming companion papers that make immediate use of the  multiscale framework we presented in this paper (`Paper I'). In Ref.~\cite{Mathews:2025txc}, hereafter `Paper II', the 1PA waveform model of Ref.~\cite{Wardell:2021fyy} is extended to include a generic precessing secondary spin and a slowly spinning primary whose spin axis has a small misalignment with the orbital angular momentum. In Ref.~\cite{Honet:2025gge}, hereafter `Paper III', the model of Ref.~\cite{Wardell:2021fyy} is extended to allow for a rapidly spinning primary whose spin is (anti-)aligned with the orbital angular momentum; this is achieved by hybridizing existing SF information (0PA fluxes and 1SF conservative 1PA effects) with known post-Newtonian results for the 2SF energy fluxes (and terms beyond 1PA order). The results of Papers~II and~III are combined and extended in Ref.~\cite{Honet:2025lmk} to describe quasi-circular binaries with generic (anti-)aligned spins on both bodies. Reference~\cite{Honet:2025lmk} also demonstrates how the parameter-space coverage of 1PA waveforms can be pushed to high spins and comparable masses by leveraging the resummations in Papers II and III.

These models remain limited in their coverage of spin precession (and eccentricity). There are two major computational hurdles on the path to fully generic 1PA models: (i) the significant effort required in computing the dissipative effects of $h^{(2)}_{\alpha\beta}$ and (ii) the large intrinsic parameter space over which SF calculations must be performed. We stress that these computations are performed offline, and (when optimized) the online waveform generation is computationally inexpensive and rapid. For generic 1PA waveforms relying solely on strong-field SF results, the computational frameworks necessary to compute $h^{(2)}_{\alpha\beta}$~\cite{Pound:2014xva,Wardell:2015ada, Miller:2016hjv,Miller:2020bft,Spiers:2023mor,Miller:2023ers,Cunningham:2024dog} are still being extended to include eccentricity and a rapidly spinning primary~\cite{Toomani:2021jlo, Durkan:2022fvm, PanossoMacedo:2022fdi, Osburn:2022bby, Leather:2023dzj, Dolan:2023enf, Spiers:2023cip, Bourg:2024vre, Wardell:2024yoi, Spiers:2024src, Bourg:2024cgh, PanossoMacedo:2024pox}. More study is required to assess whether post-Newtonian results for 2SF dissipative effects (as used in Paper III) will be sufficiently accurate across the full parameter space of realistic asymmetric binaries.

At 0PA order, modeling efforts have been aided by a substantial simplification:  the forcing functions $F^{(0)}_i$ in the evolution equation~\eqref{Jdot} can be written in terms of asymptotic Teukolsky mode amplitudes at infinity and the primary black hole's horizon~\cite{Sago:2005fn}. This means all necessary inputs for a 0PA waveform model ($F^{(0)}_i$ and the waveform amplitudes $h^{(1)}_{\bm{k}}$) can be determined directly from the solution to the Teukolsky equation with a point-mass source, avoiding the need to reconstruct the complete first-order metric perturbation or to calculate the complete self-force it exerts~\cite{Barack:2018yvs}. As we have highlighted, the same shortcut is now possible for the secondary spin's contribution to 1PA waveform models. We expect the same to also hold for the $\chi_2$-independent sector of the 1PA dynamics, but such a result has not yet been established.
As 1PA waveform models are extended to include spin precession and eccentricity, transient orbital resonances between $\Omega_r$ and $\Omega_\theta$ will also require careful treatment~\cite{Flanagan:2010cd, Pound:2021qin}. There has been considerable progress to that end~\cite{Isoyama:2021jjd,Nasipak:2021qfu, Nasipak:2022xjh,Lynch:2024ohd}. 

Finally, most of the development of SF waveforms to date has focused on the inspiral stage of the waveform. Current multiscale methods employed in SF models break down as the binary transitions to the merger-ringdown regime.
While these stages likely contribute very little to the total signal-to-noise ratio of EMRI signals detected by LISA, for example, they become increasingly important in the intermediate-mass-ratio regime and for asymmetric-mass sources observable by ground-based detectors. Thus, including the merger and ringdown in SF waveform models will be a critical step toward their direct use in these cases. Fortunately, there is significant progress toward this goal~\cite{Apte:2019txp,Lim:2019xrb, Rifat:2019ltp,Compere:2021iwh, Compere:2021zfj, Kuchler:2024esj, Becker:2024xdi}, building on the pioneering work of Refs.~\cite{Ori:2000zn, Buonanno:2000ef}.

Finally we remark that, even before accurate 1PA SF models are extended across the entire parameter space (and into the merger-ringdown regime), their intermediate results may be used in calibrating effective models with more extensive coverage. Such calibration has a long history~\cite{Barack:2018yvs}, with Refs.~\cite{Nagar:2022fep,vandeMeent:2023ols,Albertini:2024rrs,Leather:2025nhu} standing as recent examples. The results presented in Refs.~\cite{Mathews:2025txc,Honet:2025gge,Honet:2025lmk} could be used to further calibrate these models.

\begin{acknowledgments}

We thank Soichiro Isoyama, Paul Ramond, Mostafizur Rahman, and Vojtech Witzany for helpful discussions. AP particularly thanks Alex Grant for discussions about the evolution of the spin-corrected Carter constant. 
JM acknowledges support from the Irish Research Council under grant GOIPG/2018/448 and by the NUS Faculty of Science, under the research grant 22-5478-A0001. AP acknowledges the support of a Royal Society University Research Fellowship and the ERC Consolidator/UKRI Frontier Research Grant GWModels (selected by the ERC and funded by UKRI [grant number EP/Y008251/1]). This work has made use of the xAct tensor algebra package~\cite{Brizuela:2008ra, Martin-Garcia:2008ysv, xTensorOnline}.

\end{acknowledgments}

\appendix

\section{Solution to the nutation equations}
\label{sec:NutationSol}
The secondary spin vector's nutation is governed by Eqs.~\eqref{eq:dtheta_c eqn} and~\eqref{eq:dtheta_s eqn}, which we restate here for convenience:
\begin{align}
    \frac{d\delta\vartheta_c}{d\tau} - \omega_{12}\delta\vartheta_s &= -\delta\omega_{23},\\
    \frac{d\delta\vartheta_s}{d\tau} + \omega_{12}\delta\vartheta_c &= \delta\omega_{13}.   
\end{align}
In this appendix we describe how to solve these equations.

We first decouple the first-order coupled system in favor of two independent second-order differential equations:
\begin{align}
   \frac{d}{d \tau}\left(\frac{1}{\omega_{12}}\frac{d\delta\vartheta_c}{d\tau}\right)+\omega_{12} \delta\vartheta_c &= -\frac{d}{d \tau }\left(\frac{\delta\omega_{23}}{\omega_{12}}\right)+\delta \omega_{13},\\
    \frac{d}{d\tau}\left(\frac{1}{\omega_{12}}\frac{d\delta\vartheta_s}{d\tau}\right) +\omega_{12} \delta\vartheta_s &=\frac{d}{d \tau}\left( \frac{\delta\omega_{13}}{\omega_{12}}\right)+\delta\omega_{23},  
\end{align}
or switching time variable,
\begin{align}
   \frac{d}{d t}\left(\frac{u^t}{\omega_{12} }\frac{d\delta\vartheta_c}{dt}\right)+\frac{\omega_{12}}{u^t}\delta\vartheta_c &= -\frac{d}{d t }\left(\frac{\delta\omega_{23}}{\omega_{12}}\right)+\frac{\delta\omega_{13}}{u^t},\label{eq:decoupled nutation c}\\
    \frac{d}{dt}\left(\frac{u^t}{\omega_{12}}\frac{d\delta\vartheta_s}{dt}\right) +\frac{\omega_{12}}{u^t}\delta\vartheta_s &=\frac{d}{d t}\left( \frac{\delta\omega_{13}}{\omega_{12}}\right)+\frac{\delta\omega_{23}}{u^t}.   \label{eq:decoupled nutation s}
\end{align}
Recall that $\frac{\omega_{12}}{u^t}=-\omega_{s(0)}$; see Eq.~\eqref{eq:omegas1}.

Next we note that in the multiscale analysis, $\omega_{AB}$, $u^t$, and $\delta\omega_{AB}$ are all functions of the form $\omega_{12}=\omega^{(0)}_{12}(\mathring\psi^i,\mathring \pi_i)+\mathcal{O}(\e,s)$, $\delta\omega_{AB}=\delta\omega_{AB}(\mathring\psi^i,\mathring\varpi_i)+\mathcal{O}(\e^2,s)$, etc. We therefore adopt the ansatz $\delta\vartheta_c=\delta\vartheta_{c(0)}(\mathring\psi^i,\mathring \varpi_I) +\mathcal{O}(\e)$ and analogously for $\delta\vartheta_s$. At leading order, the equations~\eqref{eq:decoupled nutation c} and \eqref{eq:decoupled nutation s} depend only on time derivatives of the phases $\mathring\psi^i$ since time derivatives of $\mathring\varpi_i$ are subleading order. The first equation hence becomes
\begin{multline}
   \Omega_{(0)}^i\Omega_{(0)}^j\frac{\partial}{\partial \mathring \psi^i}\left(\frac{1}{\omega_{s(0)} }\frac{\partial\delta\vartheta_{c(0)}}{\partial \mathring \psi^j}\right)+\omega_{s(0)}\delta\vartheta_{c(0)} =\\
   \Omega_{(0)}^i\frac{\partial}{\partial \mathring \psi^i}\left(\frac{\delta\omega^{(0)}_{23}}{\omega^{(0)}_{12}}\right)-\frac{\delta\omega^{(0)}_{13}}{u^t_{(0)}},
\end{multline}
with a similar expression for the second equation. By substituting the Fourier series anzatz
\begin{equation}
    \delta\vartheta^{(0)}= \sum_{\bm{k}}\delta\vartheta^{(0)}_{\bm{k}}(\mathring \pi_i)e^{-i \mathring \psi_{\bm{k}}},
\end{equation}
for both $\delta\vartheta_{c(0)}$ and $\delta\vartheta_{s(0)}$ along with the Fourier series for the four-velocity, $\omega^{(0)}_{12}$ and $\delta\omega^{(0)}_{AB}$, one obtains an algebraic equation with coupled modes.

\section{Conservative and dissipative forces}
\label{app:conservative/dissipative}

The analysis of orbital motion in Sec.~\ref{sec:orbital motion} makes use of the fact that certain forces are purely conservative and therefore cannot contribute to the 0PA forcing function~\eqref{eq:F0}. Here we explain this fact by recalling some basic features of the problem.

We first define dissipative and conservative forces according to their behavior under time reversal $(t,\psi^i,\tilde\psi_s)\mapsto(-t,-\psi^i,-\tilde\psi_s)$:
\begin{align}
    a^\alpha_{\rm diss}(\psi^i,\pi_i,\tilde\psi_s) &\equiv \frac{1}{2}a^\alpha(\psi^i,\pi_i,\tilde\psi_s)\nonumber\\
    &\quad -\frac{1}{2}\varepsilon^\alpha a^\alpha(-\psi^i,\pi_i,-\tilde\psi_s),\\
    a^\alpha_{\rm cons}(\psi^i,\pi_i,\tilde\psi_s) &\equiv \frac{1}{2}a^\alpha(\psi^i,\pi_i,\tilde\psi_s) \nonumber\\
    &\quad +\frac{1}{2}\varepsilon^\alpha a^\alpha(-\psi^i,\pi_i,-\tilde\psi_s),
\end{align}
where $\varepsilon^\alpha=(-1,1,1,-1)$ in Boyer-Lindquist coordinates and there is no summation over $\alpha$. Under this definition, the first-order MPD force is purely conservative, as is the linear force due to the mass and spin perturbations $\delta m_1$ and $\delta\chi_1$. This fact can be straightforwardly verified by explicit computation.

Next, we note the key symmetry properties of the matrices $A^i_{\ j}(\psi^i,\pi_i)$ and $B_{ij}(\psi^i,\pi_i)$ appearing in Eqs.~\eqref{eq:omega1=Aa} and \eqref{eq:fn=Ba}:
\begin{align}
    A^i_{\ y}(-\psi^i,\pi_i) &= A^i_{\ y}(\psi^i,\pi_i),\\
    A^i_{\ \phi}(-\psi^i,\pi_i) &= -A^i_{\ \phi}(\psi^i,\pi_i),\\
    B_{i y}(-\psi^i,\pi_i) &= -B_{i y}(\psi^i,\pi_i),\\
    B_{i \phi}(-\psi^i,\pi_i) &= B_{i \phi}(\psi^i,\pi_i),    
\end{align}
where $y$ denotes either $r$ or $\theta$. These properties are easily verified by inspection of Eqs.~(289)--(295) in Ref.~\cite{Pound:2021qin}.

Finally, we note that $\langle f\rangle=0$ for any odd function of $(\psi^i,\tilde\psi_s)$. This is trivial if we average over $(\psi^i,\tilde\psi_s)$ but slightly less obvious for our average over $(\mathring\psi^i,\mathring\psi_s)$. To see that it holds for the average over $(\mathring\psi^i,\mathring\psi_s)$, consider that the ringed variables are odd functions of the unringed ones (and vice versa) if we choose them to have the same origin, implying the $\mathring\psi$-average vanishes for an odd function of the unringed phases. Since the functions of interest here are $2\pi$-periodic in both sets of phases, it follows that the $\mathring\psi$-average also vanishes even if the ringed and unringed phases do not have a common origin.

Combining all the above properties, we find
\begin{align}
    \langle A^i_{\ j}a^j_{\rm diss}\rangle &= 0, \\
    \langle B_{ij}a^j_{\rm cons}\rangle &= 0.    
\end{align}
In other words, under an average, $A^i_{\ j}$ projects out dissipative pieces of the force, and $B_{ij}$ projects out conservative contributions. The promised conclusion follows: the first-order MPD force and the linear forces due to $\delta m_1$ and $\delta\chi_1$ cannot contribute to the 0PA forcing function~\eqref{eq:F0}.

\section{Osculating action angles}
\label{app:action-angles}

In the body of the paper we formulate the orbital motion in terms of quasi-Keplerian phase-space coordinates $(\psi^i,\pi_i)$. It is also possible to begin with geodesic action angles $\mathring\psi^i_{(0)}$ in place of the quasi-Keplerian phases~$\psi^i$. Here we outline that approach and also show how~$\mathring\psi^i_{(0)}$ serves as a helper function for the approach in the body of the paper.

The geodesic action angles $\mathring\psi^i_{(0)}$ satisfy Eq.~\eqref{eq:geodesic action-angles} in the case of a geodesic motion. They can be defined from a type-2 canonical transformation, but here it will be more useful to define them (for generic, accelerated or geodesic orbits) as the solution to the leading-order part of the transformation~\eqref{eq:psi to psitilde}:
\begin{equation}
    \psi^i = \mathring\psi^i_{(0)} + \Delta\psi^i(\mathring\psi^j_{(0)},\pi_i).
\end{equation}
As neatly summarized in Ref.~\cite{Lynch:2024ohd}, the solution $\mathring\psi^i_{(0)}(\psi^j,\pi_j)$ can be written analytically as
\begin{equation}\label{eq:geodesic action angle func}
    \mathring\psi^i_{(0)}(\psi^j,\pi_j) = q^i(\psi^j,\pi_j) + \Omega^i_{(0)}(\pi_j)\delta t(\psi^j,\pi_j)
\end{equation}
with
\begin{equation}
    \delta t(\psi^j,\pi_j) = t^r(q^r(\psi^r,\pi_i),\pi_i) + t^\theta(q^\theta(\psi^\theta,\pi_i),\pi_i). 
\end{equation}
The functions $q^r(\psi^r,\pi_i)$ and $q^\theta(\psi^\theta,\pi_i)$ are given in Eqs.~(20b) and~(21b) of Ref.~\cite{Lynch:2024ohd}. The functions $t_r(q_r,\pi_i)$ and $t_\theta(q_r,\pi_i)$ are given in Eqs.~(28) and~(39) of Ref.~\cite{Fujita:2009bp}, with the quantities `$\lambda^{(r)}$' and `$\lambda^{(\theta)}$' therein replaced by $q^r/\Upsilon^r(\pi_i)$ and $q^\theta/\Upsilon^\theta(\pi_i)$, respectively. Here $\Upsilon^\alpha$ are the `Mino time' orbital frequencies~\cite{Fujita:2009bp,Pound:2021qin}.

If we apply the osculating-geodesics approach of Sec.~\ref{sec:orbital motion}, then in place of Eq.~\eqref{eq:dpsi/dt} we arrive at equations of the form
\begin{align}
    \frac{d\mathring\psi^i_{(0)}}{dt} &= \Omega^i_{(0)}(\pi_j) + \e \delta\Omega^i_{(0)}(\mathring\psi^j_{(0)},\varpi_j,\tilde\psi_s) +\mathcal{O}(\e^2),\label{eq:dpsiring0/dt}\\
    \frac{d\pi_i}{dt} &= \e g_i^{(0)}(\mathring\psi^j_{(0)},\varpi_j,\tilde\psi_s)\nonumber \\
    &\qquad\quad + \e^2 g_i^{(1)}(\mathring\psi^j_{(0)},\varpi_j,\tilde\psi_s) +\mathcal{O}(\e^3).
\end{align}
The functions $g^{(n)}_i$ are trivially related to the functions $f_i^{(n)}$ in Eq.~\eqref{eq:dpi/dt} by 
\begin{equation}
g_i^{(n)}\bigl(\mathring\psi^j_{(0)}\bigr)=f_i^{(n)}\bigl(\mathring\psi^j_{(0)} + \Delta\psi^j(\mathring\psi^k_{(0)},\pi_k)\bigr),
\end{equation}
suppressing the arguments $\varpi_j$ and $\tilde\psi_s$. The function $\delta\Omega^i_{(0)}$ can be related to those in Eqs.~\eqref{eq:dpsi/dt} and \eqref{eq:dpi/dt} by differentiating $\mathring\psi^i_{(0)}(\psi^j,\pi_j)$ with respect to $t$, substituting Eqs.~\eqref{eq:dpsi/dt} and \eqref{eq:dpi/dt}, and comparing to Eq.~\eqref{eq:dpsiring0/dt}. The result is
\begin{equation}\label{eq:dOmega0}
    \delta\Omega^i_{(0)} = \frac{\partial\mathring\psi^i_{(0)}}{\partial\psi^j}\omega^j_{(1)} +\frac{\partial\mathring\psi^i_{(0)}}{\partial\pi_j}f^{(0)}_j
\end{equation}
as well as the geodesic expression
\begin{equation}
    \Omega^i_{(0)}(\pi_j) = \frac{\partial\mathring\psi^i_{(0)}}{\partial\psi^j}\omega^i_{(0)}(\psi^k,\pi_k).
\end{equation}

We can transform to the variables $(\mathring\psi^i,\mathring\varpi_I,\mathring\psi_s)$ used in the multiscale expansion using a near-identity averaging transformation,
\begin{align}
    \mathring\psi^i_{(0)} &= \mathring\psi^i + \e \delta\mathring\psi^i_{(0)}(\mathring\psi^j,\mathring \varpi_J,{\mathring\psi}_s) + {\cal O}(\e^2).\label{eq:psiring0 to psitilde}
\end{align}
Following the same steps as in Sec.~\ref{sec:orbital motion}, one finds the forcing functions $F^{(n)}_i$, frequency correction $\Omega^i_{(1)}$, and the quantities $\delta\pi_i$ and $\delta\mathring\psi^i_{(0)}$ in the transformations~\eqref{eq:psiring0 to psitilde} and~\eqref{eq: pi to pitilde}. This process is substantially simplified by the fact that the leading term in Eq.~\eqref{eq:dpsiring0/dt} is non-oscillatory, and hence no $\mathcal{O}(\e^0)$ oscillatory term is needed in the transformation~\eqref{eq:psiring0 to psitilde}.

A disadvantage of this approach is the complicated relationship between $\mathring\psi^i_{(0)}$ and the Boyer-Lindquist trajectory. This makes the expansion of the stress-energy tensor in Sec.~\ref{sec:field equations} more complicated, for example. However, its advantages can outweigh this disadvantage.

As alluded to above, this approach also provides useful helper functions for the approach taken in the body of the paper. In particular, we can use it to find the function $\delta\psi^i$ in the transformation~\eqref{eq:psi to psitilde} without directly solving Eq.~\eqref{eq:dpsi eqn}. To achieve this, we substitute the expansions~\eqref{eq:averaging transformation} into $\mathring\psi^i_{(0)}(\psi^j,\pi_j)$, yielding
\begin{multline}
    \hspace{-10pt}\mathring\psi^i_{(0)}(\psi^j,\pi_j) = \mathring\psi^i_{(0)}(\psi^j_{(0)},\mathring\pi_j) + \e\left(\delta\psi^j \frac{\partial\mathring\psi^i_{(0)}}{\partial\psi^j_{(0)}}+\delta\pi_j \frac{\partial\mathring\psi^i_{(0)}}{\partial\mathring\pi_j}\right) \\
    + \mathcal{O}(\e^2).
\end{multline}
Identifying this with Eq.~\eqref{eq:psiring0 to psitilde} and rearranging, we find $\mathring\psi^i_{(0)}(\psi^j_{(0)},\mathring\pi_j)=\mathring\psi^i$ and
\begin{equation}\label{eq:dpsi soln}
    \delta\psi^j\frac{\partial\mathring\psi^i_{(0)}}{\partial\psi^j_{(0)}} = \delta\mathring\psi^i_{(0)}-\delta\pi_j \frac{\partial\mathring\psi^i_{(0)}}{\partial\mathring\pi_j}.
\end{equation}
Since the matrix $\partial\mathring\psi^i_{(0)}/\partial\psi^j_{(0)}$ is invertible, this determines $\delta\psi^j$ in terms of $\delta\mathring\psi^i_{(0)}$.

$\delta\mathring\psi^i_{(0)}$, in turn, is determined (up to an arbitrary non-oscillatory part) by the analogue of Eq.~\eqref{eq:dpsi eqn}. That analogue is the $\mathcal{O}(\e)$ term in Eq.~\eqref{eq:dpsiring0/dt} after substituting the transformations~\eqref{eq:psiring0 to psitilde} and \eqref{eq: pi to pitilde}:
\begin{equation}\label{eq:dpsiring0 eqn0}
    \Omega^i_{(1)} + \Omega^j_{(0)}\frac{\partial\delta\mathring\psi^i_{(0)}}{\partial\mathring\psi^j} + \Omega_{s(0)}\frac{\partial\delta\mathring\psi^i_{(0)}}{\partial\mathring\psi_s} = \delta\pi_j\frac{\partial\Omega^i_{(0)}}{\partial\mathring\pi_j} +\delta\Omega^i_{(0)}.
\end{equation}
Averaging this equation yields an alternative expression for $\Omega^i_{(1)}$,
\begin{equation}\label{eq:Omega1 alt}
        \Omega^i_{(1)} = \langle \delta\pi_j\rangle\frac{\partial\Omega^i_{(0)}}{\partial\mathring\pi_j} +\langle\delta\Omega^i_{(0)}\rangle.
\end{equation}
Substituting this formula for $\Omega^i_{(1)}$ back into Eq.~\eqref{eq:dpsiring0 eqn0}, we obtain
\begin{equation}
    \Omega^j_{(0)}\frac{\partial\delta\mathring\psi^i_{(0)}}{\partial\mathring\psi^j} + \Omega_{s(0)}\frac{\partial\delta\mathring\psi^i_{(0)}}{\partial\mathring\psi_s} = \delta\pi^{\rm osc}_j\frac{\partial\Omega^i_{(0)}}{\partial\mathring\pi_j} +\delta\Omega^i_{(0)\rm osc},\label{eq:dpsiring0 eqn}
\end{equation}
where the oscillatory part of a function $f$ is defined as
\begin{equation}\label{eq:av osc split}
    f_{\rm osc} \equiv f - \langle f\rangle.
\end{equation}

The simple equation~\eqref{eq:dpsiring0 eqn} contrasts with the complicated equation~\eqref{eq:dpsi eqn} for $\delta\psi^i$. Equation~\eqref{eq:dpsiring0 eqn} can be immediately solved for $\delta\mathring\psi^i_{(0)}$ by, for example, expanding $\delta\mathring\psi^i_{(0)}$, $\delta\pi^{\rm osc}_j$, and $\delta\Omega^i_{(0)\rm osc}$ in Fourier series.  Substituting the solution for $\delta\mathring\psi^i_{(0)}$ into Eq.~\eqref{eq:dpsi soln} then determines $\delta\psi^i$.

\section{Local expressions for 1PA forcing functions}
\label{sec:localforces}

Here we derive local expressions for $\langle dP_i/dt\rangle$ in terms of $S^{\alpha\beta}$ and $h^{{\rm R}(1)}_{\alpha\beta}$. One important outcome is that the correction term $\delta S^{\alpha}$ in Eq.~\eqref{eq:S and dS} does not need to be explicitly computed in order to calculate these rates of change. Hence, one need not calculate the nutation angles at 1PA order even if one uses local evolution equations.

Our overarching approach is to express $\langle dP_i/dt\rangle$ in terms of quantities that would manifestly vanish if the Killing vectors and tensor of Kerr were also Killing in the effective metric.

\subsection{Evolution of energy and angular momentum}

Reference~\cite{Akcay:2019bvk} derived convenient local expressions for the averaged rates of change of $E$ and $L_z$. However, that derivation involved errors that compensated earlier errors in Ref.~\cite{Akcay:2019bvk}'s expressions for the local forces and torques; see Ref.~\cite{Mathews:2021rod}. Here for completeness we provide a corrected and streamlined derivation of Ref.~\cite{Akcay:2019bvk}'s formula.

For a generic vector $\hat v^\alpha$ (whose indices are raised and lowered with the effective metric), we introduce the quantities 
\begin{subequations}
\begin{align}
\hat\Sigma&=\hat g_{\alpha \beta} \hat v^\alpha \hat u^\alpha,\\
\delta\hat\Sigma&= \hat S^{\beta\gamma}\hat\nabla_\beta \hat v_\gamma,
\end{align}
\end{subequations}
with derivatives
\begin{subequations}
\begin{align}
\frac{d\hat\Sigma}{d\hat \tau}&=\frac{1}{2}\mathcal{L}_{\hat v}\hat g_{\alpha \beta} \hat u^\alpha \hat u^\beta +v_\alpha \hat a^\alpha+\mathcal{O}(s^2),\\
\frac{d \delta\hat\Sigma}{d\hat\tau}&= \hat S^{\beta\gamma} \hat u^\lambda \hat\nabla_\lambda\hat\nabla_\beta \hat v_\gamma+\mathcal{O}(s^2)\nonumber\\
&=-2\hat v_\alpha \hat a^\alpha  +\hat S^{\beta\gamma} \hat u^\lambda\left( \hat \nabla_\beta \hat \nabla_\lambda \hat v_\gamma -2 \hat \nabla_{[\gamma}\hat\nabla_{\lambda]}\hat v_\beta\right) \nonumber\\
&\quad+\mathcal{O}(s^2).\label{eq:deltaSigmaderiv}
\end{align}
\end{subequations}
In going from the second line to the last line, we have used the Riemann symmetries inside $\hat a^\alpha$ and the definition of the Riemann tensor. Next we split up the symmetric and anti-symmetric terms in $\gamma\leftrightarrow\lambda$, using
\begin{equation}
\hat\nabla_\lambda \hat v_\gamma =\hat \nabla_{[\lambda} \hat v_{\gamma]}+\hat \nabla_{(\lambda} \hat v_{\gamma)}=\hat \nabla_{[\lambda} \hat v_{\gamma]}+\frac{1}{2}\mathcal{L}_{\hat v}\hat g_{\lambda \gamma},
\end{equation}
and the antisymmetry of the spin tensor. We find Eq.~\eqref{eq:deltaSigmaderiv} reduces to
\begin{equation}
 \frac{d \delta\hat\Sigma}{d\hat\tau}=\hat S^{\beta\gamma} \hat u^\lambda \hat \nabla_\beta\mathcal{L}_{\hat v}\hat g_{\lambda \gamma} -2\hat v_\alpha \hat a^\alpha+\mathcal{O}(s^2),
\label{eq:deltaSigmaderivsimp}   
\end{equation}
where we have eliminated terms proportional to $\hat\nabla_{[\lambda}\hat\nabla_\gamma v_{\beta]}$ by virtue of the first Bianchi identity. Putting these two terms together we have
\begin{multline}
\label{eq:Sigmahatdot}
\frac{d\hat\Sigma}{d\hat \tau}+\frac{m_2}{2}\frac{d\delta\hat\Sigma}{d\hat \tau}=\frac{1}{2}\mathcal{L}_{\hat v}\hat g_{\alpha \beta} \hat u^\alpha \hat u^\beta\\
+\hat S^{\beta\gamma} \hat u^\lambda \hat \nabla_\beta\mathcal{L}_{\hat v}\hat g_{\lambda \gamma}+\mathcal{O}(s^2).
\end{multline}

Now consider the quantity
\begin{equation}
    \hat \Xi=\hat g_{\alpha\beta}\xi^\alpha \hat u^\beta +\frac{m_2}{2}\hat S^{\gamma \delta}\hat g_{\delta \beta}\hat \nabla_\gamma\xi^\beta,
\end{equation}
which corresponds to $\hat\Sigma+\frac{m_2}{2}\delta\hat\Sigma$ with $\hat v^\alpha = \xi^\alpha$. 
If $\xi^\alpha$ satisfied Killing's equation in the effective metric, then this quantity would be conserved~\cite{doi:10.1098/rspa.1981.0046}, which we have made obvious in Eq.~\eqref{eq:Sigmahatdot}. We can still use the fact that $\mathcal{L}_\xi g_{\alpha\beta}=0$, which immediately tells us, with Eq.~\eqref{eq:Sigmahatdot}, that
\begin{equation}
    \frac{d\hat \Xi}{dt}= \frac{1}{2} \dot z^\alpha  \hat u^\beta \mathcal{L}_{\xi} h^{\mathrm{R}}_{\alpha\beta}- \frac{m_2}{2}\hat S^{\gamma \delta} \dot z^\lambda  \hat \nabla_\delta \mathcal{L}_\xi h^{\mathrm{R}}_{\gamma \lambda}+\mathcal{O}(s^2),
    \label{eq:Xihatderivexpand}
\end{equation}
where we used $\frac{d\hat\tau}{dt}\hat u^\alpha=\dot z^\alpha$. 
Consider now the test-body quantity, $\Xi$, which we can relate to the hatted quantity via $\hat \Xi=\Xi+\delta \Xi$. We have
\begin{equation}
    \left\langle \frac{d \Xi}{ dt} \right\rangle = \left\langle\frac{d\hat \Xi}{ dt}\right\rangle -\left\langle\frac{d \delta \Xi}{ dt}\right\rangle.
\end{equation}
Using $d/dt = \Omega^i_{(0)}(\mathring\pi_j)\partial/\partial\mathring\psi^i + \mathcal{O}(\e)$, we see the latter term is too high order to consider:
\begin{equation}
    \left\langle \frac{d\delta \Xi}{ dt}\right\rangle=\mathcal{O}(\e^3).
\end{equation}
Therefore the average rate of change of the test-body conserved quantity $\Xi$ is
\begin{multline}
    \left \langle\frac{d\Xi}{dt}\right\rangle=\frac{1}{2}\left\langle \dot z^\alpha \hat u^\beta \mathcal{L}_{\xi}h^{\mathrm{R}}_{\alpha\beta}-m_2 S^{\alpha \beta}\dot z^\gamma \nabla_\beta \mathcal{L}_\xi h^{\mathrm{R}}_{\alpha \gamma}\right\rangle\\
    +\mathcal{O}(\e^3,s^2).
\end{multline}
If we neglect terms quadratic in $h^{\rm R}_{\alpha\beta}$ and introduce a proper-time average~\cite{Drasco:2005kz,Isoyama:2018sib} 
\begin{equation}
\langle\cdot\rangle_\tau\equiv \frac{\left\langle\frac{d\tau}{dt}\cdot\right\rangle}{\left\langle \frac{d\tau}{dt}\right\rangle}, 
\end{equation}
then this result takes the more symmetrical form
\begin{multline}
    \left \langle\frac{d\Xi}{d\tau}\right\rangle_{\!\tau} = \frac{1}{2}\left\langle u^\alpha u^\beta \mathcal{L}_{\xi}h^{\mathrm{R}}_{\alpha\beta}-m_2 S^{\alpha \beta}u^\gamma \nabla_\beta \mathcal{L}_\xi h^{\mathrm{R}}_{\alpha \gamma}\right\rangle_\tau\\
    +\mathcal{O}(\e^2,s^2)
    \label{eq:E/AMderivAve}.    
\end{multline}

Reference~\cite{Akcay:2019bvk} showed that  the above local evolution satisfies a balance law with the asymptotic flux of energy (angular momentum), taking $\xi^\alpha=-t^\alpha  (\phi^\alpha)$ and using the approach of Gal'tsov~\cite{Galtsov:1982hwm} and Mino~\cite{Mino:2003yg}.

As foreshadowed at the beginning of this appendix, the result for $\left \langle\frac{d\Xi}{dt}\right\rangle$ does not depend on the spin correction $\delta S^\alpha$. The underlying reason for this could perhaps be lost in our streamlined derivation, but we can explain it with a more pedestrian approach. If we consider 
\begin{equation}
    \Xi = g_{\alpha\beta}\xi^\alpha u^\beta +\frac{m_2}{2} S^{\gamma \delta} g_{\delta\beta}\nabla_\gamma\xi^\beta,
\end{equation}
then we first comment that replacing $S^{\gamma \delta}$ with $\hat S^{\gamma \delta}$ only changes $\Xi$ at order $\e^2$, hence changing $\langle dP_i/dt\rangle$ at order $\e^3$, meaning 2PA. The direct contribution of $\delta S^{\alpha}$ is therefore not relevant to the averaged rate of change at 1PA order. Next consider
\begin{align}\label{eq:Xidot - acceleration}
    \frac{d\Xi}{d\tau} = \xi_\beta a^\beta + \frac{m_2}{2}u^\alpha\nabla_\alpha\left(S^{\gamma \delta} \nabla_\gamma\xi_\delta\right).
\end{align}
$\delta S^{\alpha}$ contributes to the first term through the acceleration~\eqref{eq:a2-chi}. Appealing to the identity $\xi^\mu R_{\mu\alpha\beta\gamma}=\nabla_\alpha \nabla_\beta \xi_\gamma$ for a Killing vector, we can write that contribution to Eq.~\eqref{eq:Xidot - acceleration} as
\begin{subequations}
\begin{align}
    \xi_\beta a^\beta_{(\delta S)} &= -\frac{m_2}{2}\xi^\mu R_{\mu\alpha\beta\gamma}u^\alpha\delta S^{\beta\gamma}\\
    &= -\frac{d}{d\tau}\left(\frac{m_2}{2} \nabla_\beta\xi_\gamma\,\delta S^{\beta\gamma}\right) + \frac{m_2}{2} \nabla_\beta\xi_\gamma \frac{D\delta S^{\beta\gamma}}{d\tau}.
\end{align}
\end{subequations}
The first term can be neglected under averaging because it is a total derivative of an order-$\e^2$ quantity. The second term does contribute to $\langle d\Xi/dt\rangle$, but only through the derivative of $\delta S^{\alpha\beta}$ rather than through  $\delta S^{\alpha\beta}$ on its own. That derivative can be written directly in terms of the first-order regular field using
\begin{equation}
    \frac{D \delta S^{\alpha\beta}}{d\tau} = \frac{D \hat S^{\alpha\beta}}{d\tau}-\frac{D S^{\alpha\beta}}{d\tau} ,
\end{equation}
with Eqs.~\eqref{eq:selfforcespin}, \eqref{eq:<S> leading} (after precession averaging for simplicity), and \eqref{eq:DE3dtau}. Hence, although $\delta S^{\alpha\beta}$ does contribute to the 1PA orbital evolution, it and the nutation angles it involves do not need to be explicitly computed.

\subsection{Evolution of the spin-corrected Carter constant}

Next we consider the quantity
\begin{equation}
\label{eq:CarterWspinhat}
    \hat K= K_{\alpha \beta} \hat u^\alpha \hat u^\beta +m_2 L_{\alpha \beta \gamma} \hat S^{\alpha \beta}\hat u^\gamma,
\end{equation}
in which we have defined
\begin{equation}
\label{eq:L}
     L_{\alpha\beta\gamma}\equiv -2  g^{\delta \lambda}\left(Y_{\lambda\beta}  \nabla_{\delta} Y_{\gamma\alpha} -  Y_{\gamma \lambda}  \nabla_\delta Y_{\alpha \beta}\right).
\end{equation} 
The quantity $\hat K$ reduces to the spin-corrected Carter constant in the test-body limit. Its rate of change is
\begin{equation}
    \frac{d \hat K}{d\hat \tau}=m_2 \hat V_{\alpha \beta \gamma \delta} \hat S^{\alpha \beta}\hat u^\gamma \hat u^\delta+\hat u^\alpha \hat u^\beta \hat u^\gamma\hat U_{\alpha \beta \gamma}+\mathcal{O}(s^2),   
\end{equation}
where we have defined
\begin{subequations}\label{eq:U and V}
\begin{align}
    \hat U_{\alpha \beta \gamma}&\equiv \hat \nabla_\gamma K_{\alpha \beta},\\
    \hat V_{\alpha \beta \gamma \delta}&\equiv \hat \nabla_\delta  L_{\alpha \beta \gamma}- K_{\gamma \rho}\hat R^{\rho}{}_{\delta \alpha \beta}
    .\label{eq:Vhat}    
\end{align}
\end{subequations}
Using the fact that $\hat S^{\alpha \beta}=  - \hat \epsilon^{\alpha \beta \mu \nu}\hat{S}_{\mu}\hat u_{\nu}$, we can write this as
\begin{multline}\label{eq:dKhat/dtauhat}
    \frac{d \hat K}{d\hat \tau}=-2m_2  {}^*\hat V_{\alpha (\beta \gamma \delta)} \hat S^{\alpha} \hat u^{\beta}\hat u^\gamma \hat u^\delta\\
    +\hat u^\alpha \hat u^\beta \hat u^\gamma\hat U_{(\alpha \beta \gamma)}+\mathcal{O}(s^2), 
\end{multline}
having used the notation for the left-handed dual 
\begin{equation}\label{eq:V dual}
{}^*\hat V_{\alpha \beta \gamma \delta}= \frac{1}{2}\hat \epsilon_{\alpha \beta}{}^{\mu \nu}\hat V_{\mu \nu \gamma \delta}.  
\end{equation}

Now, to utilize the background Killing and Killing-Yano symmetries, we expand out the dependence on the effective metric such that
\begin{subequations}
    \begin{align}
    {}^*\hat V_{\alpha \beta \gamma \delta}&={}^* V_{\alpha \beta \gamma \delta} +\delta{}^* V_{\alpha \beta \gamma \delta},\\
    \hat U_{\alpha \beta \gamma}&=U_{\alpha \beta \gamma}+\delta U_{\alpha \beta \gamma}.
    \end{align}
\end{subequations}
The background quantities are obtained by removing all the hats in the definitions~\eqref{eq:U and V} and \eqref{eq:V dual}, and the $\delta$ terms are linear in the regular field $h^{\rm R}_{\alpha\beta}$. 

The background terms $U_{\alpha \beta \gamma}$ and $V_{\alpha \beta \gamma}$ drop out of Eq.~\eqref{eq:dKhat/dtauhat} by virtue of the identities
\begin{subequations}
 \begin{align}
      U_{(\alpha \beta \gamma)}&=0,\\
^{*}V_{\alpha (\beta \gamma \delta)}&=0\label{eq:hodgeVbackground},
 \end{align}  
\end{subequations}
the former of which is simply the statement that $K_{\alpha\beta}$ is a Killing tensor of the background metric, and the latter of which is proved in Ref.~\cite{Compere:2021kjz, c839080c-5d3d-34d9-a247-38e3c109da14}. In addition to using these identities, we also allow ourselves to discard terms quadratic in the regular field, since we are interested specifically in terms that are linear in spin. 

Combining these simplifications, we reduce the evolution equation~\eqref{eq:dKhat/dtauhat} to
\begin{equation}
    \frac{d \hat K}{d \tau}= V_{1} + V_{2}+\mathcal{O}(\e^2\bcancel{s},\e^3, s^2),
\end{equation}
having defined the `symmetry violation' terms
\begin{subequations}%
\begin{align}
    V_1&\equiv  u^\alpha u^\beta  u^\gamma\delta U_{(\alpha \beta \gamma)} ,\\
    V_2&\equiv -2m_2  \delta{}^* V_{\alpha (\beta \gamma \delta)}  S^{\alpha}  u^{\beta}u^\gamma  u^\delta \label{eq:V2}
\end{align}
\end{subequations}
and used $\mathcal{O}(\e^2\bcancel{s})$ to denote spin-independent $\mathcal{O}(\e^2)$ terms. 
At this stage we can see the evolution equation has no explicit dependence on $\delta S^\alpha$ because it would contribute $\mathcal{O}(\e^3)$ to $V_2$. Like in the analysis of the previous section, the effect of the force generated by $\delta S^\alpha$ has been put in a form that does not require an explicit calculation of the nutation angles.

Finally, as in the previous subsection, the spin-corrected Carter constant will evolve according to
\begin{align}
\left \langle\frac{d  K}{d\tau}\right\rangle_{\!\tau}&=\biggl \langle\frac{d  \hat K}{d\tau}\biggr\rangle_{\!\tau}+\mathcal{O}(\e^3)\nonumber\\
&=\left\langle V_1+V_2 \right \rangle_\tau+\mathcal{O}(\e^2 \bcancel{s},\e^3,s^2)\label{eq:Carterevolutionsimp},
\end{align}
since the difference between the two is a total derivative that can be neglected on average. We also stress that the average immediately eliminates the precessing component of the spin, such that $S^\alpha$ can be replaced with the parallel component~\eqref{eq:<S> leading}. In the remainder of the section, we derive more explicit expressions for the quantities $V_1$ and $V_2$.

The first violation term is straightforward to compute explicitly in terms of the metric perturbation:
\begin{align}
V_1&= -2 u^\alpha  u^\beta  u^\gamma \delta \Gamma^{\delta}{}_{\gamma \alpha} K_{\delta \beta}\nonumber\\
&= -u^\alpha  u^\gamma \left(2 h^{\mathrm{R}}_{\delta \alpha;\gamma}-h^{\mathrm{R}}_{\alpha \gamma;\delta} \right) n^{\delta}+\mathcal{O}(\e^2\bcancel{s}),\label{eq:V1terms}
\end{align}
where we have defined $n^\alpha\equiv K^{\alpha}{}_{\beta}u^\beta$ and introduced the perturbation to the Christoffel symbol,
\begin{equation}
    \delta\Gamma^{\alpha}_{\beta \gamma}\equiv \frac{1}{2}g^{\alpha \delta}\left(2 h^{\mathrm{R}}_{\delta (\beta;\gamma)}-h^{\mathrm{R}}_{\beta\gamma;\delta} \right)+\mathcal{O}(\e^2\bcancel{s}).
\end{equation}

To compute $V_2$ we use
\begin{equation}
\label{eq:deltavstar}
    \delta{}^* V_{\alpha \beta \gamma \delta}=\frac{1}{2}\epsilon_{\alpha \beta}{}^{\rho \sigma}\delta V_{\rho \sigma \gamma \delta}+\frac{1}{2}\delta \epsilon_{\alpha \beta}{}^{\rho \sigma}V_{\rho \sigma \gamma \delta}+\mathcal{O}(\e^2\bcancel{s})
\end{equation}
with
\begin{align}
    \delta V_{\alpha \beta \gamma \delta}=&-\delta \Gamma^{\rho}_{\delta\alpha}L_{\rho \beta\gamma}-\delta \Gamma^{\rho}_{\delta\beta}L_{\alpha \rho \gamma}\nonumber\\
    &-\delta \Gamma^{\rho}_{\delta\gamma}L_{\alpha \beta \rho}
     -2 K_{\gamma \rho}\delta \Gamma^{\rho}{}_{\delta [\beta; \alpha]}+\mathcal{O}(\e^2\bcancel{s}),
\end{align}
and 
\begin{equation}
    \delta\epsilon_{\alpha \beta}{}^{\rho \sigma}=\frac{1}{2}\epsilon_{\alpha \beta}{}^{\rho \sigma} g^{\mu\nu}h^{\mathrm{R}}_{\mu\nu} + 2 \epsilon_{\alpha \beta}{}^{\mu [\rho}h^{\mathrm{R}\, \sigma]}{}_{\mu}+\mathcal{O}(\e^2\bcancel{s}).
\end{equation} 

Substituting Eq.~\eqref{eq:deltavstar} into Eq.~\eqref{eq:V2}, we obtain
\begin{multline}
\label{eq:V2terms}
    V_2=-m_2 \Bigl( 
    4 \epsilon_{\alpha \beta}{}^{\mu [\rho}h^{\mathrm{R}\, \sigma]}{}_{\mu}V_{\rho \sigma \gamma \delta}\\
    +\epsilon_{\alpha \beta}{}^{\rho \sigma}\delta V_{\rho \sigma \gamma \delta}\Bigr)  S^{\alpha}  u^{\beta}u^\gamma  u^\delta+ \mathcal{O}(\e^2\bcancel{s}),
\end{multline}
having eliminated a term with Eq.~\eqref{eq:hodgeVbackground}.

We can now extract the linear-in-spin 1PA terms in Eq.~\eqref{eq:Carterevolutionsimp}. First, 
\begin{align}
    V^{(1\text{-}\chi_2)}_1 &= -u^\alpha_{(0)}  u^\gamma_{(0)} \left(2 \mathring h^{\mathrm{R}(2\text{-}\chi_2)}_{\delta \alpha;\gamma}-\mathring h^{\mathrm{R}(2\text{-}\chi_2)}_{\alpha \gamma;\delta} \right) n^{\delta}_{(0)} \nonumber\\
    &\quad +\left(\delta\psi^i_{(\chi_2)}\frac{\partial }{\partial \psi^i_{(0)}} + \delta\pi_i^{(\chi_2)}\frac{\partial }{\partial \mathring\pi_i}\right)V^{(0)}_1,
\end{align}
where 
$V^{(0)}_1 = -u^\alpha_{(0)}  u^\gamma_{(0)} \left(2 \mathring h^{\mathrm{R}(1)}_{\delta \alpha;\gamma}-\mathring h^{\mathrm{R}(1)}_{\alpha \gamma;\delta} \right) n^{\delta}_{(0)}$ and we follow the notation of Sec.~\ref{sec:field equations}. In $V_2$, we simply replace $h^{\rm R}_{\alpha\beta}$ and  $u^\alpha$ with $\mathring h^{\rm R(1)}_{\alpha\beta}$  and $u^\alpha_{(0)}$ in Eq.~\eqref{eq:V2terms}.

In the case of a nonspinning secondary (such that $V_2=0$), and at leading order in $\e$, Sago et al. showed that Eq.~\eqref{eq:Carterevolutionsimp} can be rewritten as a practical flux-like formula in terms of Teukolsky mode amplitudes~\cite{Sago:2005fn}. (See also Refs.~\cite{Hughes:2005qb,Isoyama:2018sib}. Note that the formula requires additional information from the local dynamics at/near orbital resonances, as shown in Ref.~\cite{Isoyama:2018sib}, for example.) Their derivation began from Mino's result that $\langle dK/d\tau\rangle_\tau$, at 0PA order, only depends on the self-force generated by the radiative part of the metric perturbation~\cite{Mino:2003yg}. They then followed Gal'tsov's method~\cite{Galtsov:1982hwm} of reconstructing the radiative metric perturbation in terms of Teukolsky mode amplitudes. It might be possible to extend this analysis to Eq.~\eqref{eq:Carterevolutionsimp} in full.
As explained in Sec.~\ref{sec:flux balance}, Grant's result~\cite{Grant:2024ivt} bypasses the need for such an analysis, but it would provide a healthy consistency check.

\bibliography{WaveformsSpin}

\end{document}